\documentclass[a4paper,11pt]{article}
\pdfoutput=1 %
\usepackage{jcappub} 

\usepackage[T1]{fontenc} 

\usepackage[utf8]{inputenc}

\usepackage{hyperref} 
\usepackage{url}
\usepackage{graphicx}
\usepackage{esvect}
\usepackage{graphics}
\usepackage{natbib}
\usepackage{mathrsfs}
\usepackage{blindtext}

\usepackage{comment}

\usepackage{lineno}

\usepackage[table]{xcolor}
\usepackage{multirow}

\usepackage[normalem]{ulem}
\usepackage{color}

\usepackage{soul}

\usepackage{supertabular}
\usepackage{tabularx}

\def\units#1{~\hbox{$\,{\rm #1}$}}

\title{\fontsize{22}{20} FLUKA cross sections for cosmic-ray interactions with the DRAGON2 code}

\author[a]{P.~De~La~Torre~Luque}
\emailAdd{pedro.delatorreluque@fysik.su.se}
\author[b]{M.~N.~Mazziotta}
\emailAdd{mazziotta@ba.infn.it}
\author[c]{A.~Ferrari}
\author[b, d]{F.~Loparco}
\author[e]{P.~R.~Sala}
\author[b]{D.~Serini}

\affiliation[a]{The Oskar Klein Centre, Department of Physics, Stockholm University, AlbaNova\\
  SE-10691 Stockholm, Sweden}
\affiliation[b]{Istituto Nazionale di Fisica Nucleare, Sezione di Bari, via Orabona 4, I-70126 Bari, Italy}
\affiliation[c]{Institute of Astroparticle Physics, Karlsruhe Institute of Technology, Campus North, Bldg. 401 Postfach 3640, D-76021 Karlsruhe, Germany}
\affiliation[d]{Dipartimento di Fisica ``M. Merlin" dell'Universit\`a e del Politecnico di Bari, via Amendola 173, I-70126 Bari, Italy}
\affiliation[e]{Istituto Nazionale di Fisica Nucleare, Sezione di Milano, Via Celoria,16, 20133 Milano, Italy}

\date{\today}

\abstract{Secondary particles produced in spallation reactions of cosmic rays with the interstellar gas provide valuable information that allow us to investigate the injection and transport of charged particles in the Galaxy. A good understanding of the cross sections of production of these particles is crucial to correctly interpret our models, although the existing experimental data is very scarce and uncertain. We have developed a new set of cross sections, both inelastic and inclusive, computed with the {\tt FLUKA} Monte Carlo nuclear code and tested its compatibility with CR data. Inelastic and inclusive cross sections have been compared to the most up-to-date data and parameterisations finding a general good agreement. 

Then, these cross sections have been implemented in the {\tt DRAGON2} code to characterize the spectra of CR nuclei up to $Z=26$ and the secondary-to-primary ratios of B, Be and Li. 
Interestingly, we find that the FLUKA cross sections allow us to predict an energy-dependence of the B, Be and Li flux ratios which is compatible with AMS-02 data and to reproduce simultaneously these flux ratios with a scaling lower than $20\%$. 
Finally, we implement the cross sections of production of gamma rays, calculated with {\tt FLUKA}, in the {\tt Gammasky} code and compute diffuse gamma-ray sky maps and the local HI emissivity spectrum, finding a very good agreement with Fermi Large Area Telescope data.}

\begin{document}
\maketitle
\flushbottom

\section{Introduction}
\label{sec:intro}
Galactic cosmic rays (CRs) are known to play an important role in the Galaxy, regulating its evolution, shape, mass-metallicity relation, magnetic fields and other processes~\cite{GalFormation, Hopkins, CR-Galaxy, Farber_2018}. Understanding their propagation may help reach a better interpretation of the gamma-ray emissions throughout the Milky Way, the characterisation of environmental plasmas or even in indirect searches of dark matter. 

The widely accepted standard scenario is that collision-less scatterings of CRs with magneto-hydrodynamic (MHD) plasma waves cause them to follow an erratic motion that can be approximated as a diffusive transport~\cite{Gabici:2019jvz, Ginz&Syr, berezinskii1990astrophysics}. These interactions extend several kiloparsecs above and below the galactic disk~\cite{Luque:2021joz, CarmeloBeB, Weinrich_halo, biswas2018constraining, moskalenko2000diffuse, bringmann2012radio}, constituting the so-called magnetised halo~\cite{stephens1998cosmic}. During their journey, CRs repeatedly cross the galactic disk undergoing hadronic interactions (spallation reactions) with the gas in the interstellar medium (ISM) and produce secondary particles that can be secondary CRs, gamma rays or neutrinos~\cite{tjus2020closing}. These secondary particles are crucial since they bring valuable information of the CR transport and interactions. In particular, the flux ratios of secondary-to-primary CRs (considering as primary CRs those mainly accelerated in astrophysical sources) are commonly used to constrain the propagation parameters defining their transport~\cite{reinert2018precision, derome2019fitting}.

Recently, many works have claimed the need of improving cross sections measurements on spallation reactions in order to reduce the uncertainty related to the determination of the propagation parameters from the secondary-to-primary flux ratios~\cite{Luque:2021joz, Weinrich_combined, Korsmeier:2021brc, ICPPA_Pedro, GONDOLO2014175, GenoliniRanking, Tomassetti:2015nha}. Due to the lack of experimental data in most of the channels and its uncertainty, there have been several attempts meant to develop the cross sections network purely from physical principles, usually relying on Monte Carlo simulations and event generators. However, as they are mainly adjusted to reproduce accelerator data, they have demonstrated to be less accurate in the region between $100$~MeV to $\sim$ hundreds GeV than the current dedicated parameterisations~\cite{Mashnik:2010xh, Mashnik:2016dmf, Kerby:2015qaa, GenoliniRanking}. 
However, these simulation codes (as, for example, GEANT4~\cite{PSHENICHNOV2010604}, PHITS~\cite{PHIS, SIHVER2010892}, SHIELD-HIT~\cite{Hultqvist_2012}) have experienced a positive boost in the last years, usually driven by radiological and medical applications, which need to accurately describe the transport of ions in different materials and their interactions.

This paper aims at computing the full cross sections network (including inelastic and inclusive cross sections) needed to be implemented in CR propagation codes using one of the most recently updated Monte Carlo nuclear codes {\tt FLUKA}~\cite{Ferrari:2005zk, FLUKA1, Arico:2019pcz} and demonstrating its compatibility with data. {\tt FLUKA} is a general purpose tool that can be used to transport particles in arbitrarily complex geometries, including magnetic fields, through the FLUKA combinatorial geometry. Its application in astroparticle physics studies is one of the main goals of {\tt FLUKA}, as it is demonstrated by a variety of recent works~\cite{Battistoni:2008hga, Andersenarticle, Heinbockel:2011zz, Tusnski:2019rpd, FlukaSun}, raising high expectations in the CR community. This work updates the cross sections calculated by Ref.~\cite{Mazziot}, extend its calculations to isotopes up to Z=26 and explore in detail the predicted cross sections for B, Be and Li production.

These cross sections are implemented in the {\tt DRAGON2} code\footnote{\label{note1} Available at \url{https://github.com/cosmicrays/DRAGON2-Beta\_version}}~\cite{DRAGON2-1, DRAGON2-2} in order to demonstrate that they can be used to reproduce the new and accurate CR data at a similar level as common analytical and semi-analytical cross sections parameterisations such as the DRAGON2~\cite{Luque:2021joz, Evoli:2019wwu} and GALPROP~\cite{GALPROPXS, GALPROPXS1} cross sections, focusing on the study of the secondary CRs B, Be and Li. We employ a Markov chain Monte Carlo (MCMC) analysis, already used in earlier works~\cite{Luque_MCMC, Luque_Ap}, to determine the propagation parameters from the most recent data from the AMS-02 collaboration~\cite{AMS_gen, Aguilar:2015ooa, Aguilar:2018keu, Aguilar:2020ohx, aguilar2017observation, aguilar2018observation, aguilar2019towards, AMS_NNaAl, AMS_Fe, AMS_Pos2019, AMS_F}.
Furthermore, we derive the cross sections for production of leptons (electrons and positrons) and gamma rays from CR collisions with the interstellar medium and implement them in the {\tt Gammasky} code~\cite{Cirelli_2014, GSky_diffuse, Tavakoli:2011wz} in order to compute gamma-ray sky maps and the local HI emissivity spectrum (see Refs.~\cite{Casandjian:2015hja, Orlando:2017mvd} for a discussion on these measurements and predictions). The FLUKA cross sections for secondary lepton production will be discussed in a separated paper, more dedicated to the positron spectrum. Also the production of isotopes around the so-called iron group will be discussed in a separated work.

This paper is organized as follows: in section~\ref{sec:FXSecs} we discuss the main aspects of the cross sections computations performed and show them in comparison with relevant parameterisations and data. Then, in section~\ref{sec:FDRAGON}, we demonstrate that the FLUKA cross sections can be successfully used in a CR propagation code like the {\tt DRAGON2} code and report the results of the MCMC analysis for the flux ratios of B, Be and Li over C and O. Section~\ref{sec:FGamma} is dedicated to show the computations of the galactic diffuse gamma-ray emission with the FLUKA cross sections. Finally, the main highlights are discussed in section~\ref{sec:Fconc}.

\section{FLUKA cross sections for CR spallation interactions}
\label{sec:FXSecs}

One of the main advantages of using nuclear codes for the computation of cross sections of CR interactions is that it allows us to calculate them in a broad range of energies and species. The possible energy range covered for hadron-hadron and hadron-nucleus interactions in {\tt FLUKA} extends from a threshold ($\sim 1 \units{MeV}$) up to $10^4 \units{TeV}$, while electromagnetic and muon interactions can be treated from $1 \units{keV}$ ($100 \units{eV}$ for photons) up to $10^4 \units{TeV}$. Nucleus-nucleus interactions are also supported from $1 \units{MeV/n}$ up to $10^4 \units{TeV/n}$. We refer the reader to Ref.~\cite{Mazziot} for a brief description on the interaction models that {\tt FLUKA} employs for these interactions in different energy ranges. Moreover, full information on the different models used by the code and its related publications and references can be found in the FLUKA webpage \footnote{\url{http://www.fluka.org/fluka.php?id=publications&mm2=3}}.

\begin{figure*}[!tbh]
\begin{center}
\footnotesize{a)}\includegraphics[width=0.45\textwidth,height=0.24\textheight,clip] {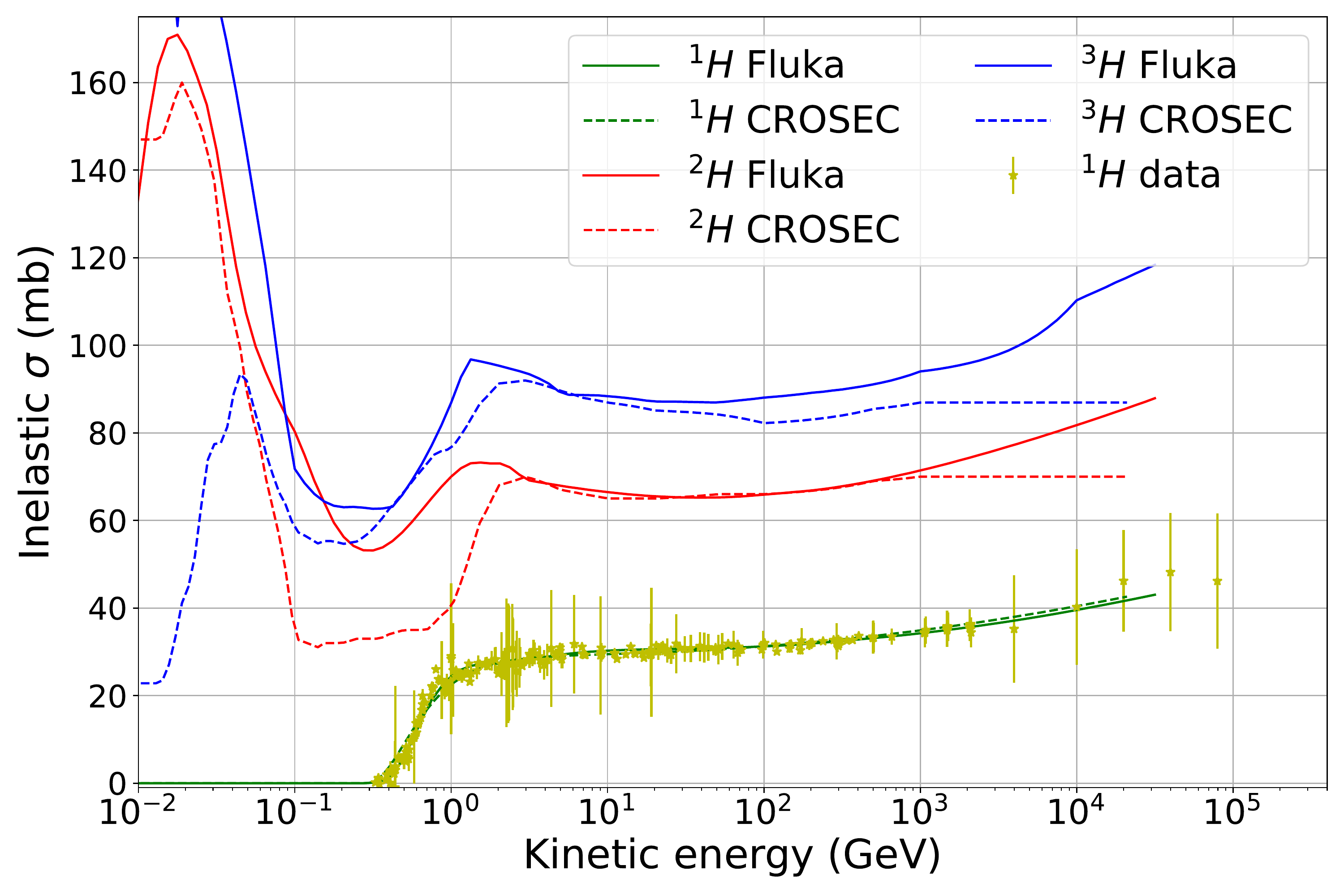} \hspace{0.3cm}
\footnotesize{b)}\includegraphics[width=0.45\textwidth,height=0.24\textheight,clip] {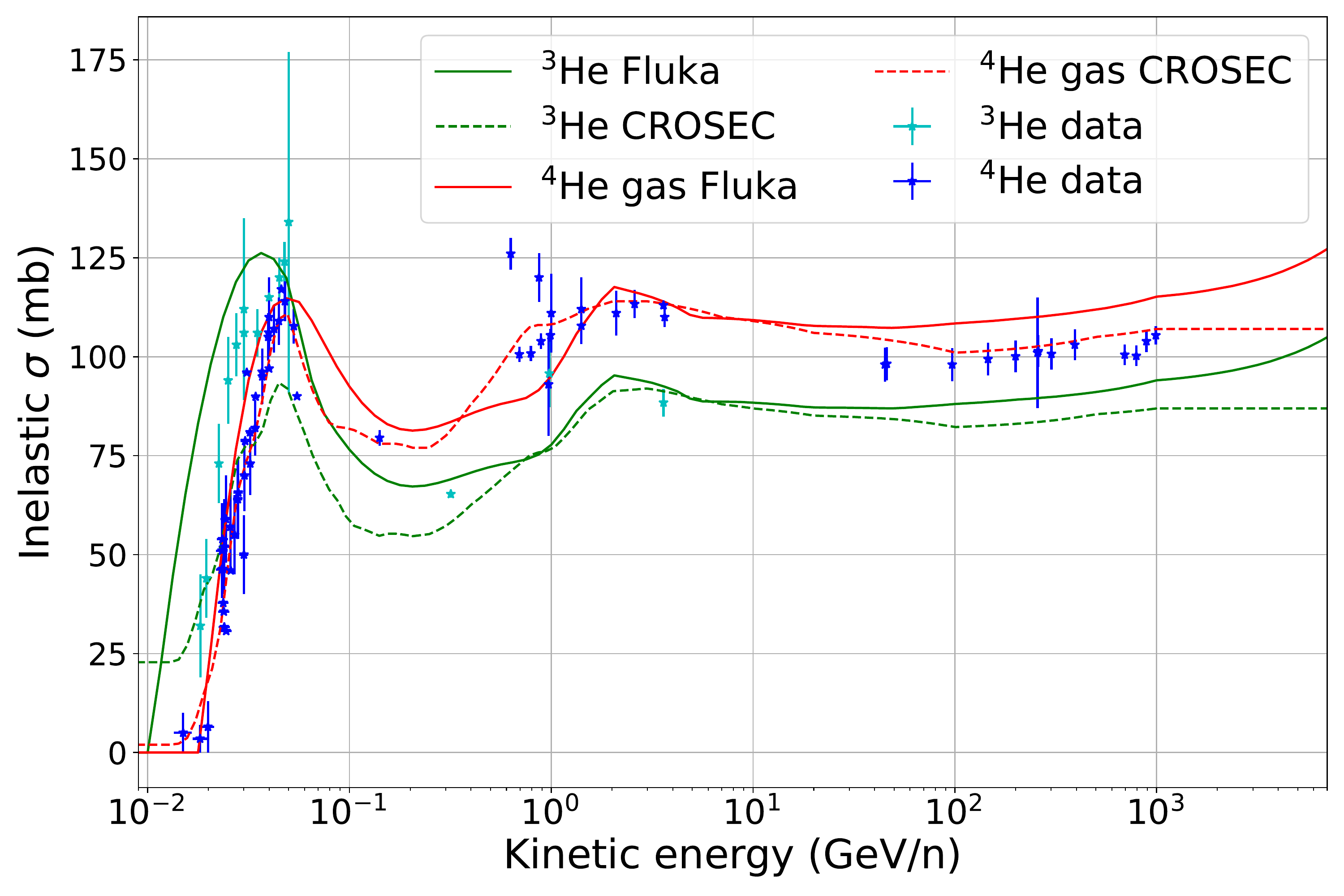}

\footnotesize{c)}\includegraphics[width=0.45\textwidth,height=0.24\textheight,clip] {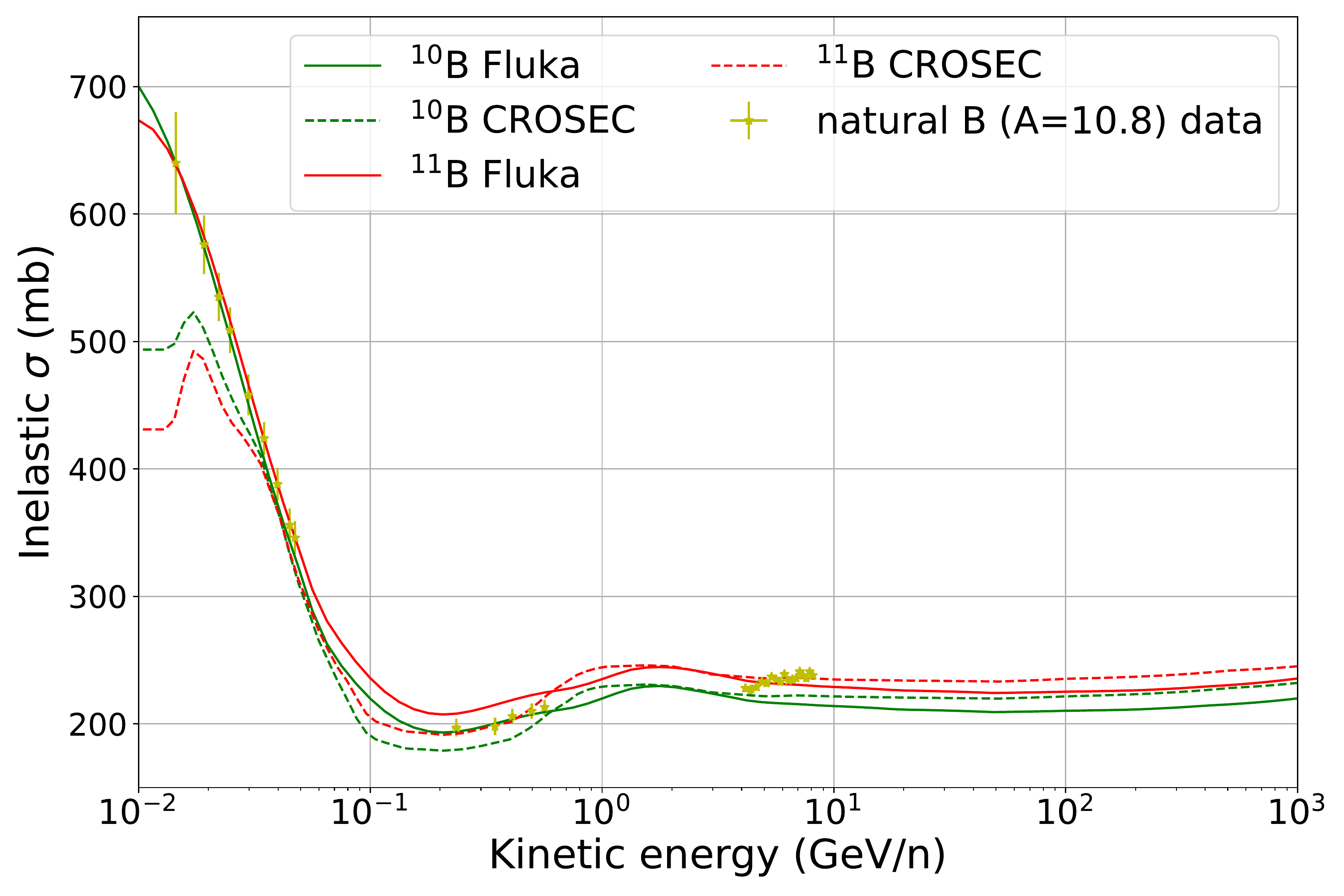} \hspace{0.3cm}
\footnotesize{d)}\includegraphics[width=0.45\textwidth,height=0.24\textheight,clip] {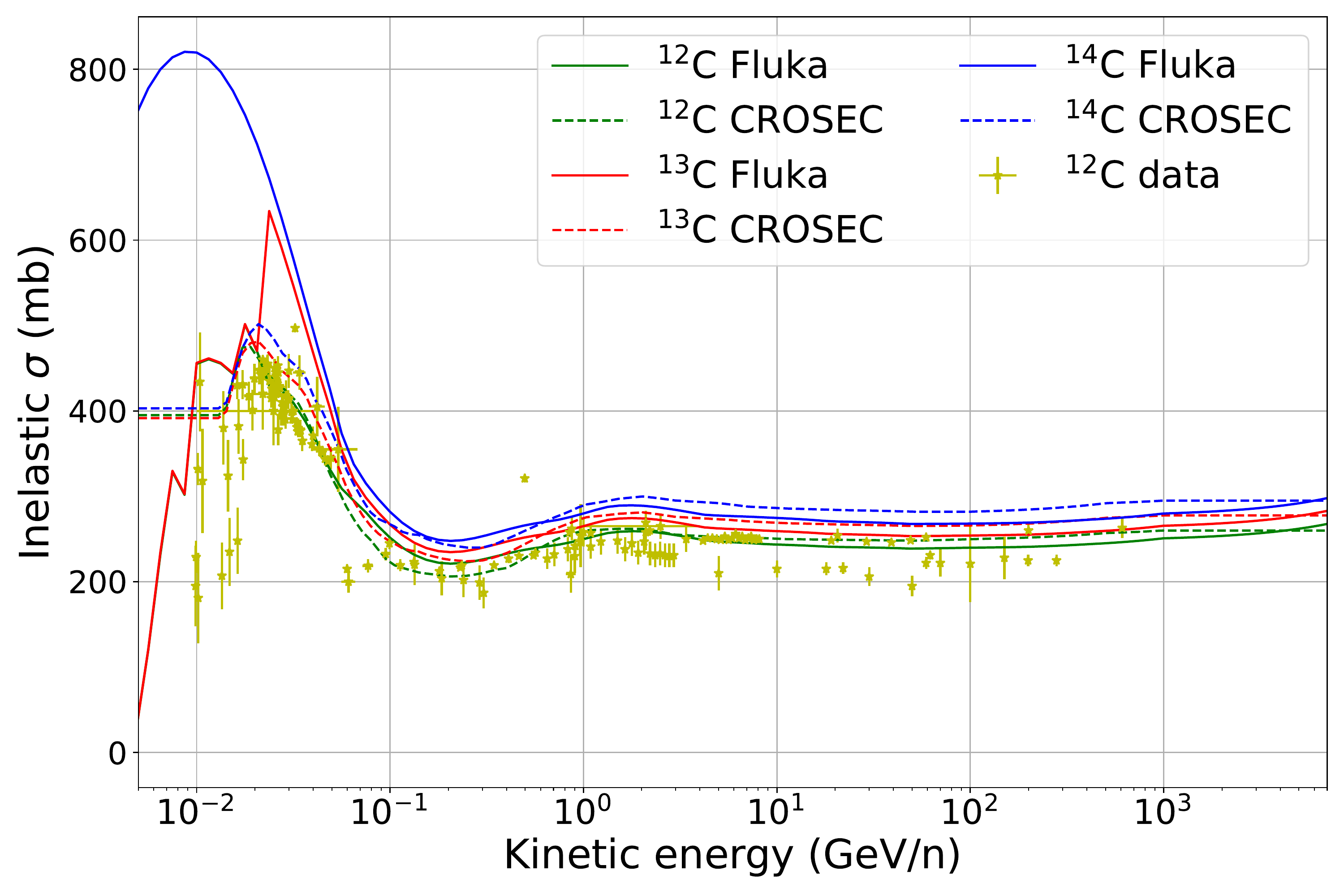}

\footnotesize{e)}\includegraphics[width=0.45\textwidth,height=0.24\textheight,clip] {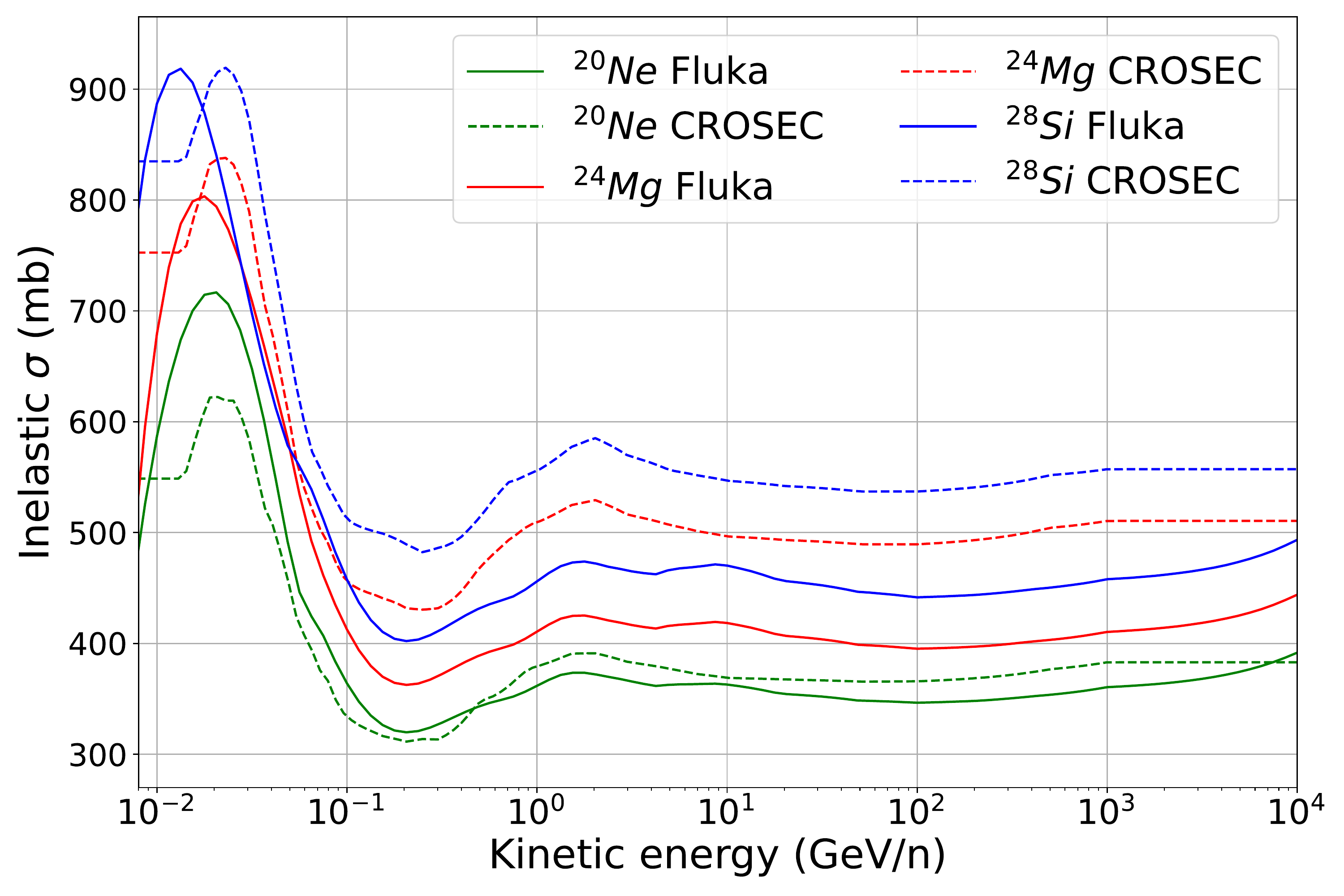} \hspace{0.3cm}
\footnotesize{f)}\includegraphics[width=0.45\textwidth,height=0.24\textheight,clip] {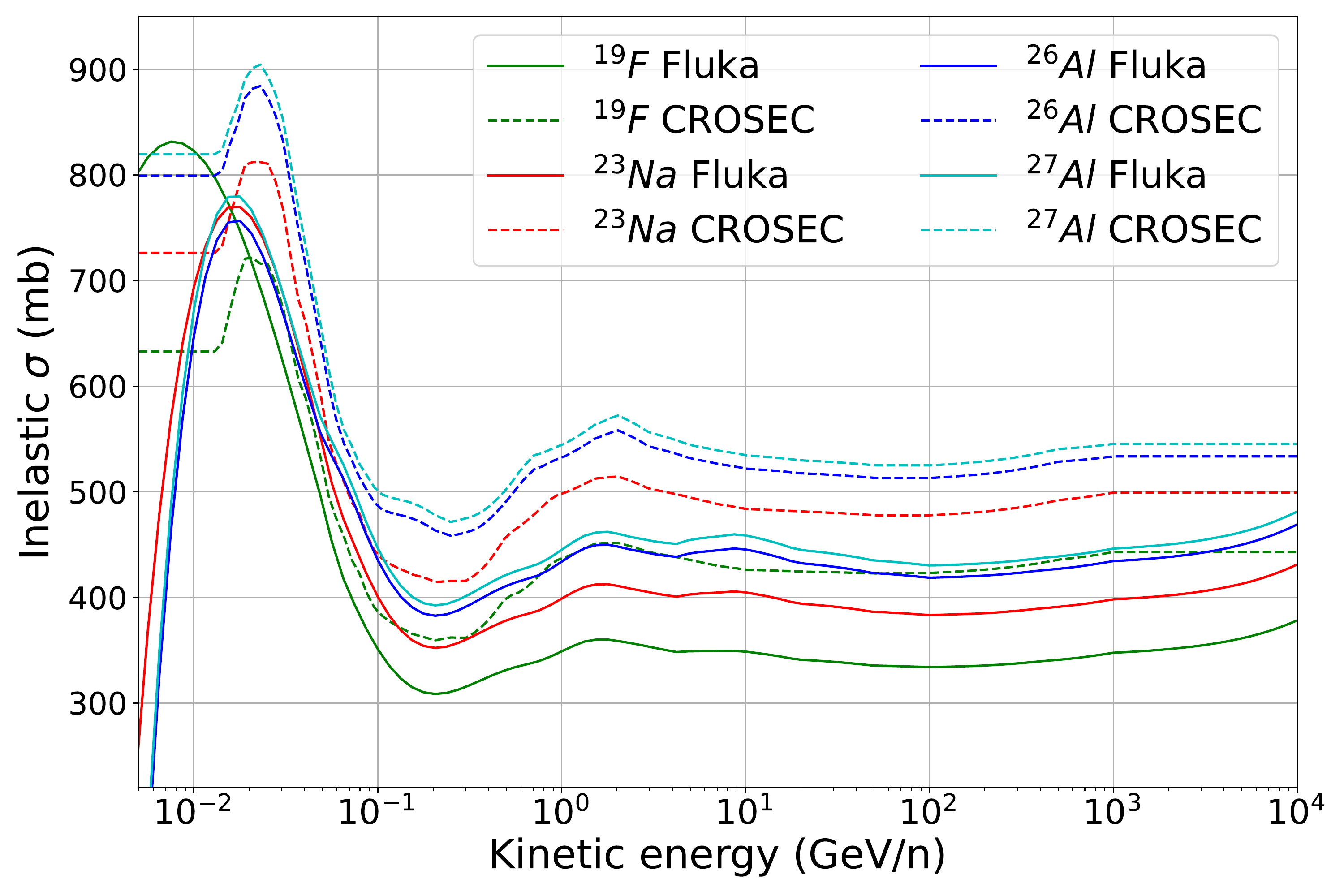}
\end{center}
\caption{Comparison of the CROSEC and FLUKA inelastic cross sections on hydrogen target for diverse species of projectiles relevant for CR physics: a) protons; b) helium; c) boron; d) carbon; e) other heavier primary CRs (Ne, Mg and Si); f) other heavier secondary CRs ($^{19}F$, $^{23}Na$, $^{26}Al$, $^{27}Al$). Experimental data are also shown when they are available. Proton-induced reactions data are taken from~\cite{Bobchenko:1979hp} and are assembled in tables in \url{http://www.oecd-nea.org/dbdata/bara.html}. Data for heavier nuclei come from a compilation using the EXFOR database \cite{OTUKA2014272, Zerkin_2018}. Full references can be found in \cite{GenoliniRanking}.}
\label{fig:ineplots}
\end{figure*} 

Here, we compute the inelastic and inclusive cross sections of all stable isotopes from protons to iron impinging in helium and hydrogen as targets, following the strategy of Ref.~\cite{Mazziot}, from $1 \units{MeV/n}$ to $35 \units{TeV/n}$, using 176 bins equally spaced in a logarithmic scale. Ghost nuclei (i.e. those unstable nuclei, generated via spallation reactions, which have a negligible lifetime compared to typical CR propagation times, and whose contribution is directly added up to the cross sections of formation of their daughter nuclei) are carefully taken into account by making them decay before calculating the yield of secondary particles produced after the simulated interaction. The results can be summarized in a set of spallation and inelastic cross section tables, with the spallation ones calculated for both the direct and cumulative channels, which will be soon publicly available. 

Some of the most important inelastic cross sections, for hydrogen as target, computed with {\tt FLUKA} are shown in Figure~\ref{fig:ineplots}. It shows the proton, helium (as examples of light primary CR), carbon (as an example of heavy primary CR), boron (as an example of secondary CR) and other heavy elements (as a sample of other CR nuclei with very few measurements available) inelastic cross sections calculated with {\tt FLUKA} and with the CROSEC code\footnote{This is the default option for inelastic cross sections in the {\tt DRAGON2} code.} (also called CRN6)~\cite{Barashenkov:1994cp} as a function of energy per nucleon compared to experimental data. For protons, {\tt FLUKA} adopts the parameterisation found in~\cite{Kafexhiu:2014cua}. Notice that in the case of boron (panel c), the experimental data correspond to the isotopic composition (with $A\sim 10.8$). Comparisons for the primary CRs $^{20}$Ne, $^{24}$Mg and $^{28}$Si are shown to emphasize the differences found between both computations in channels where the amount of experimental data is very scarce. In fact, as in the case of the spallation inclusive cross sections, parameterisations for inelastic (total) cross sections with few or no experimental data (as in Figure~\ref{fig:ineplots}, f) are just extrapolations of the better measured channels, which may significantly bias the final results. Channels above silicon turn out to be not very important in CR studies of light secondary CRs: they have an impact on the production of light secondary CRs $< 4\%$ for Be and Li, and $<1\%$ for B (at $\sim$ 10~GeV). In the case of heavier species (Figure~\ref{fig:ineplots}, e, f), discrepancies around $10$-$20\%$ are found between the two computations at atomic numbers above that of Ne ($Z \geq 10$). For all these heavy nuclei, the energy dependence of their cross sections is very similar. The lack of data in these channels makes difficult to decide which computation is more suitable, but, in general, it seems that the inelastic cross sections calculated with {\tt FLUKA} show a good agreement to one of the most updated semi-analytical parameterisations used in the CR community, as the CROSEC ones. 

Besides, the inclusion of a rise cross sections at high energies is evident in the {\tt FLUKA} cross sections, while this is not incorporated neither in the inelastic nor in the inclusive cross sections parameterisations, although it has been already experimentally observed~\cite{RiseofXS, Block_2011, PhysRevD_rise}. This fact can have important implications on the interpretation of the break at high energies in CR spectra, as we comment in Appendix~\ref{sec:appendixA}.

\begin{figure*}[ptb]
\begin{center}
\hskip -0.4 cm
\includegraphics[width=0.53\textwidth,height=0.26\textheight,clip] {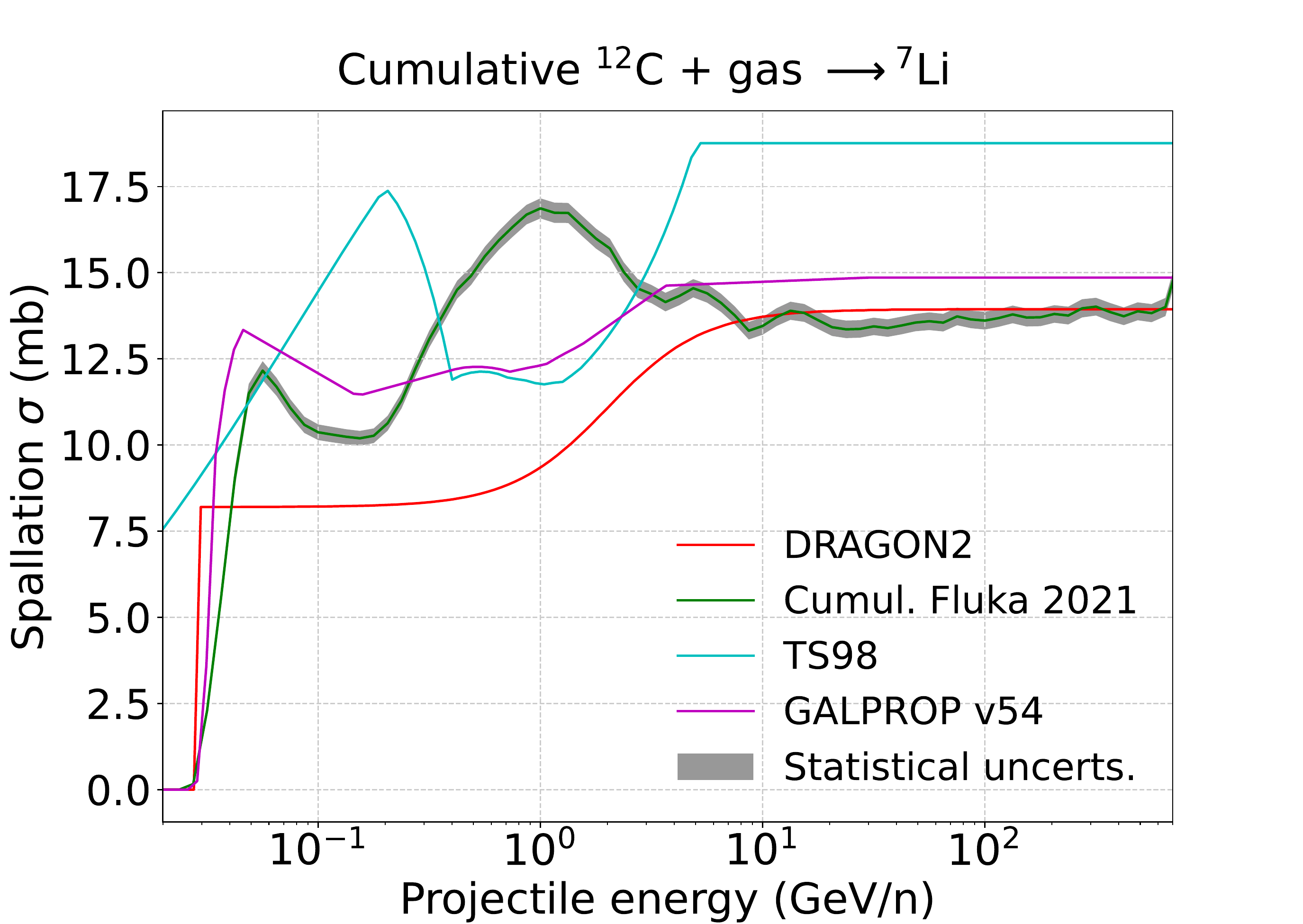}
\hspace{-0.65cm}
\includegraphics[width=0.53\textwidth,height=0.26\textheight,clip] {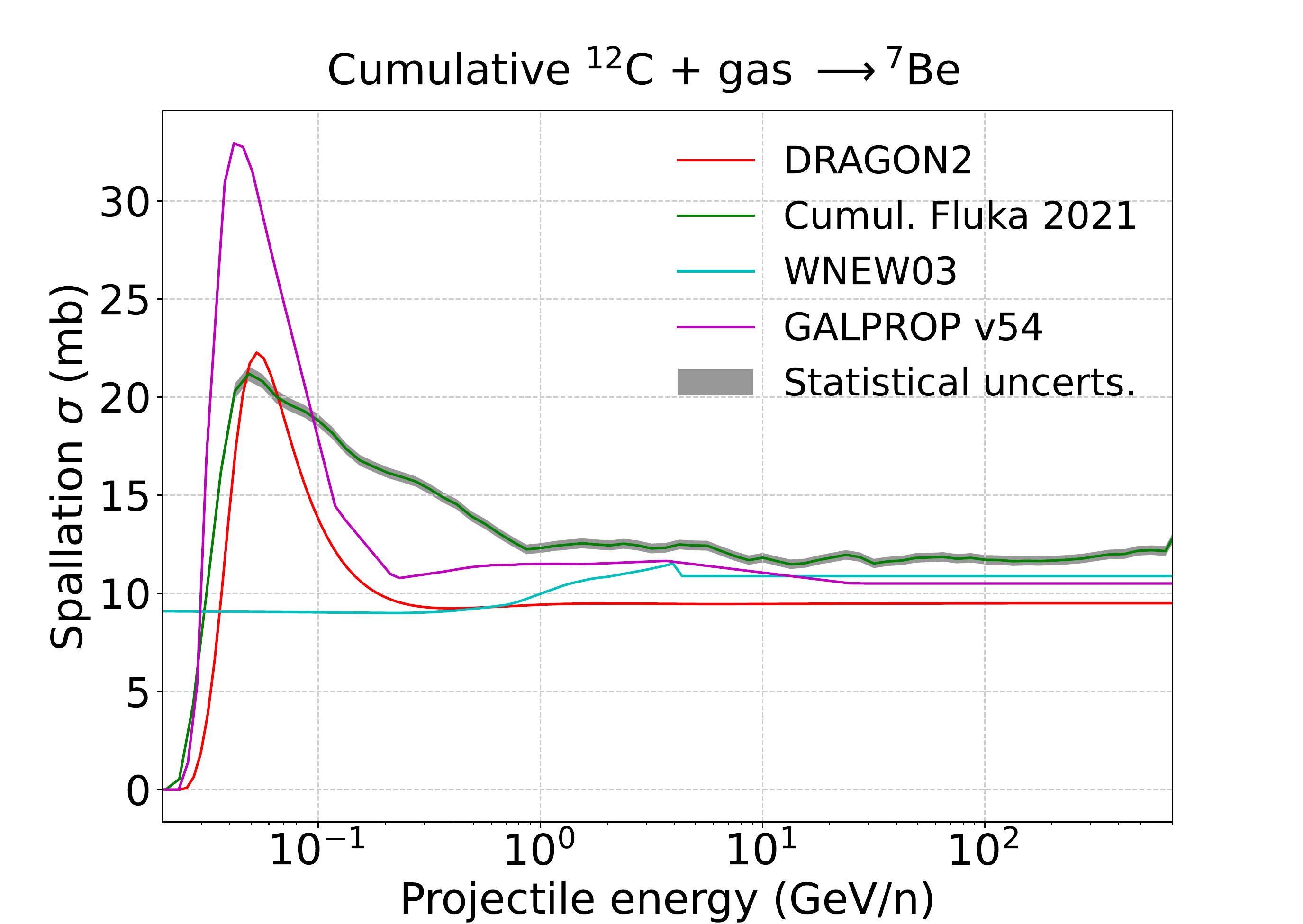}
\rightskip -0.4cm

\hskip -0.4 cm
\includegraphics[width=0.53\textwidth,height=0.26\textheight,clip] {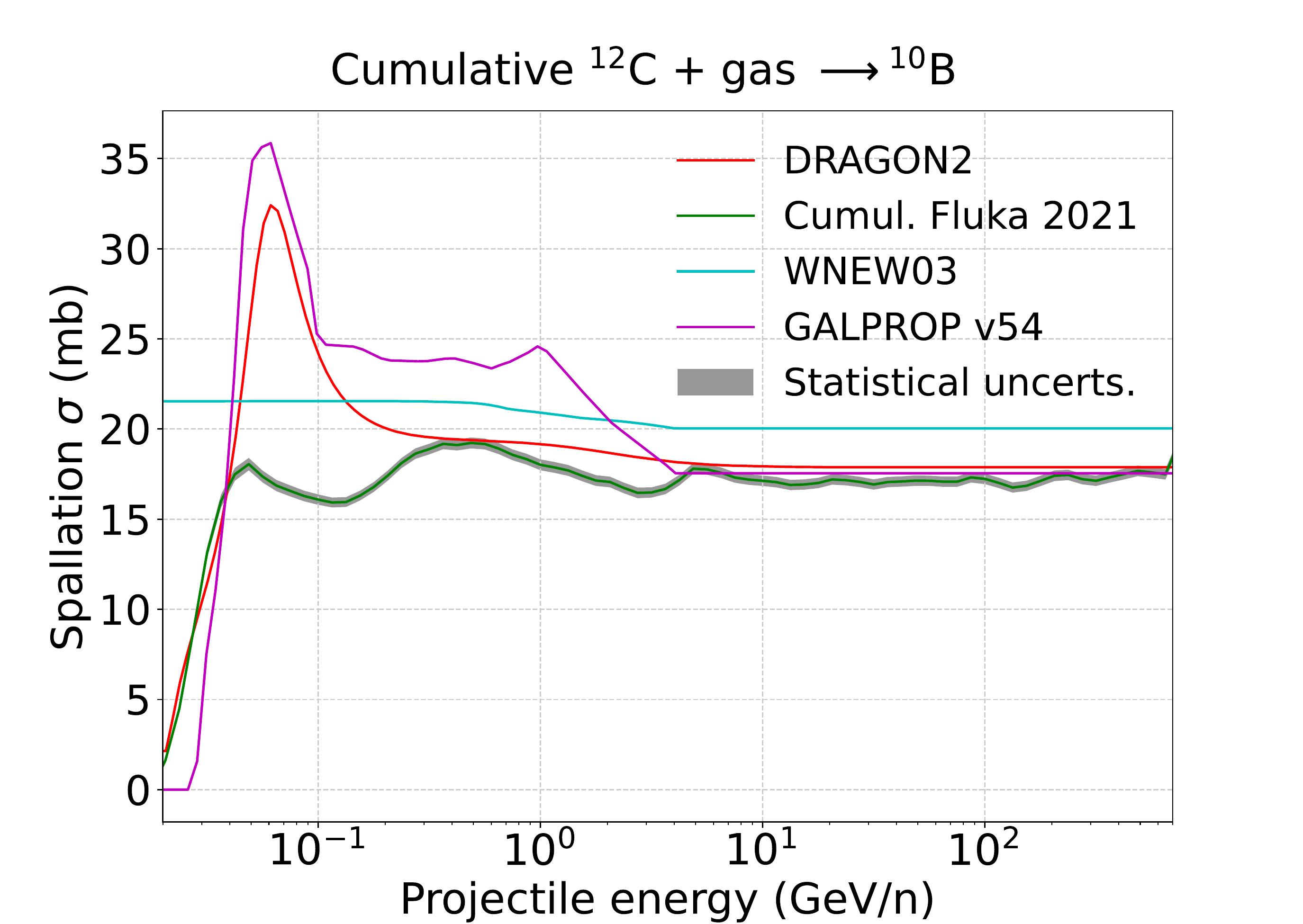}
\hspace{-0.65cm}
\includegraphics[width=0.53\textwidth,height=0.26\textheight,clip] {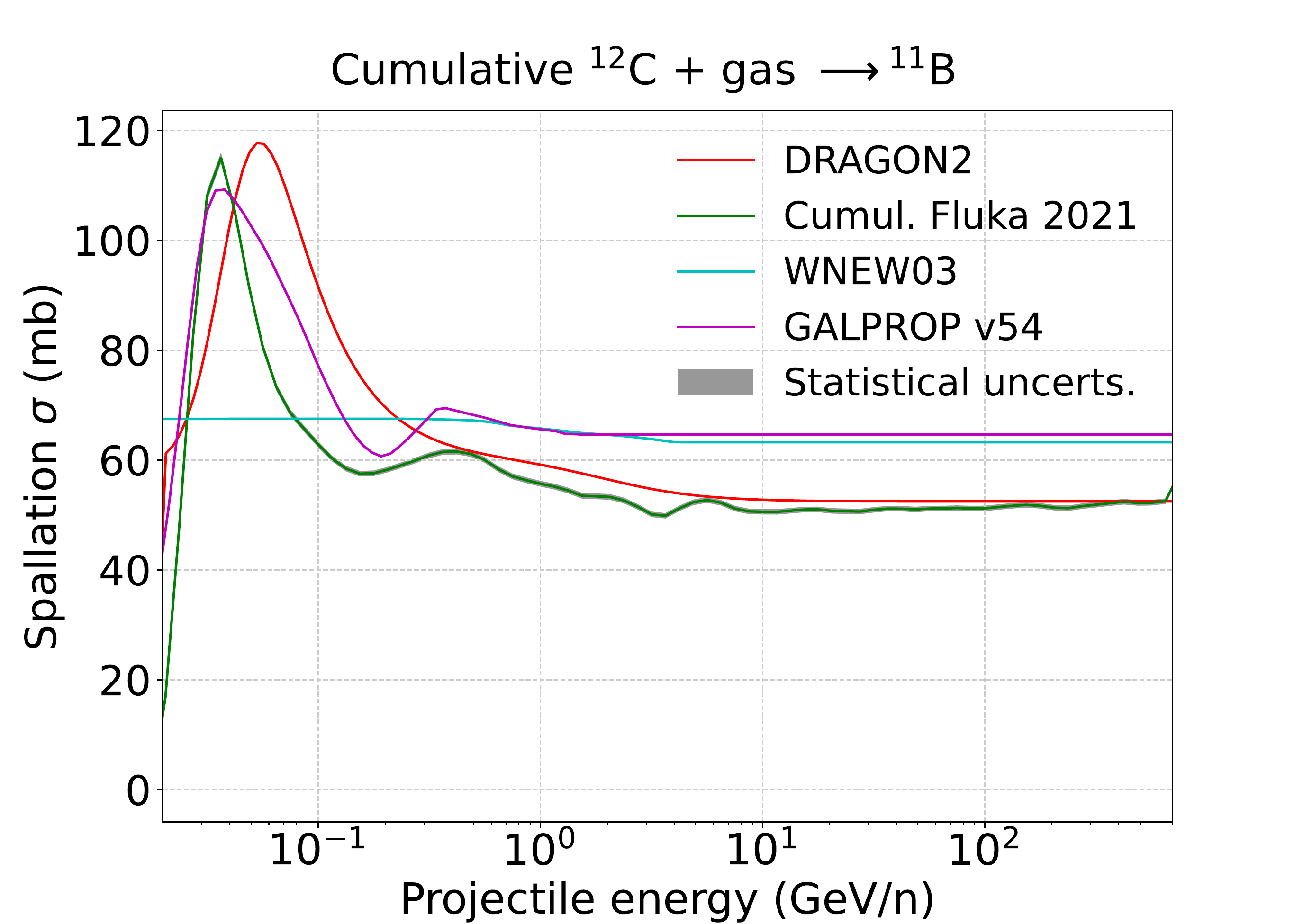}
\rightskip -0.4cm
\end{center}
\caption{Comparison of the spallation cross sections of CR interactions with ISM gas computed with {\tt FLUKA} and the most widespread parameterisations, for the production of various isotopes of B, Be and Li from $^{12}$C as projectile.}
\label{fig:Tot_XS_C}
\end{figure*}

\begin{figure*}[!tb]
\begin{center}
\hskip -0.4 cm
\includegraphics[width=0.53\textwidth,height=0.26\textheight,clip] {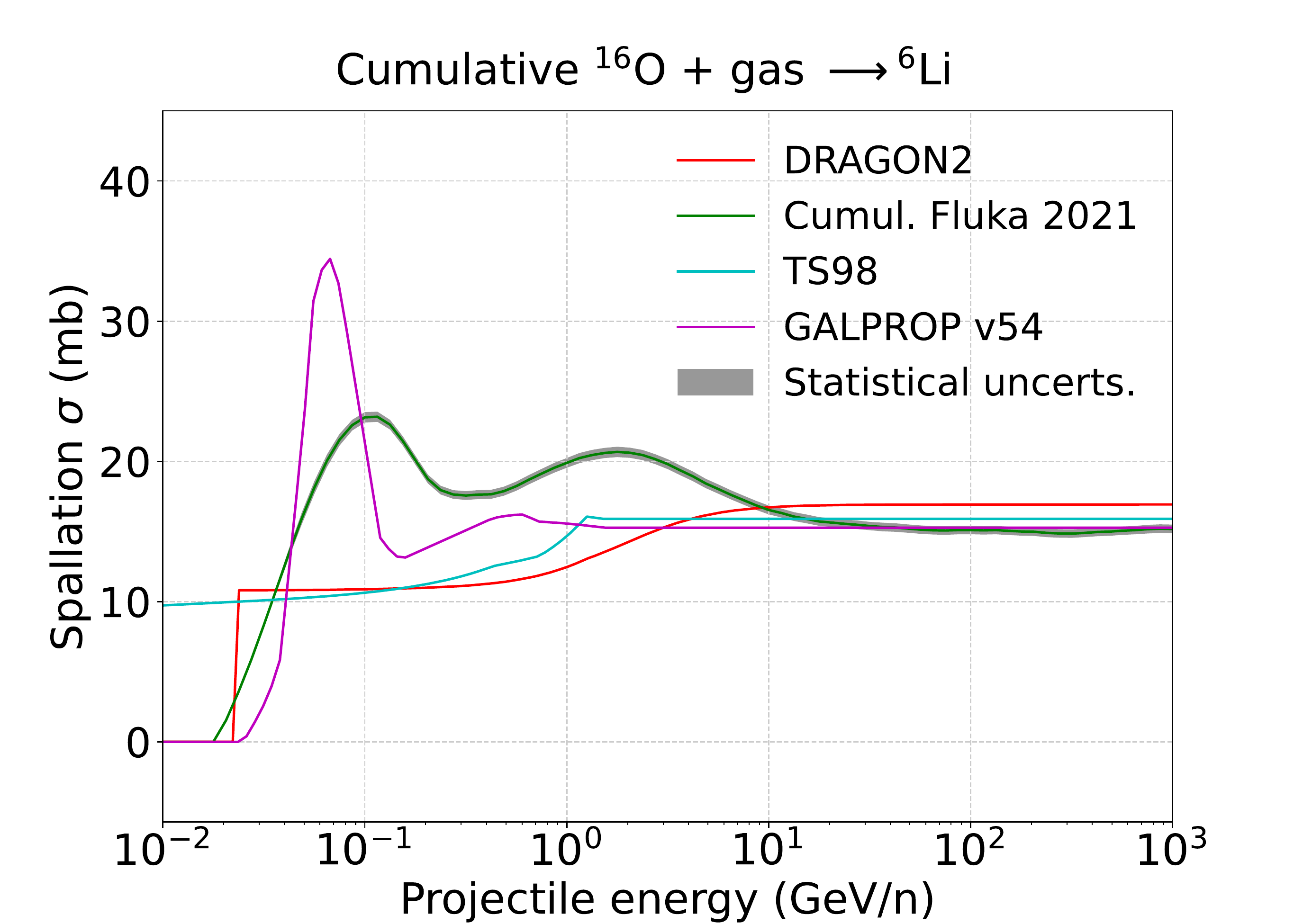}
\hspace{-0.65cm}
\includegraphics[width=0.53\textwidth,height=0.26\textheight,clip] {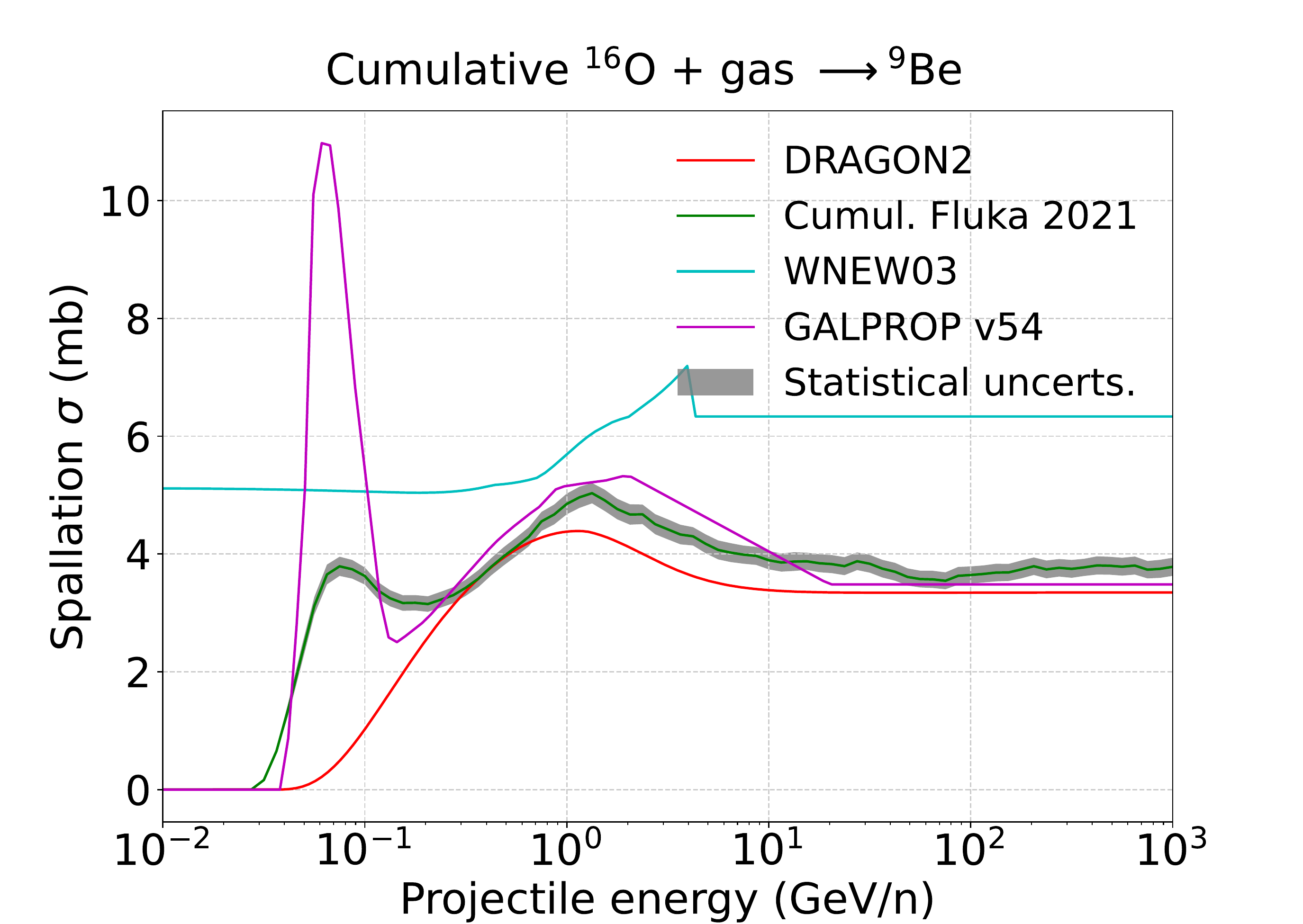}
\rightskip -0.4cm

\hskip -0.4 cm
\includegraphics[width=0.53\textwidth,height=0.26\textheight,clip] {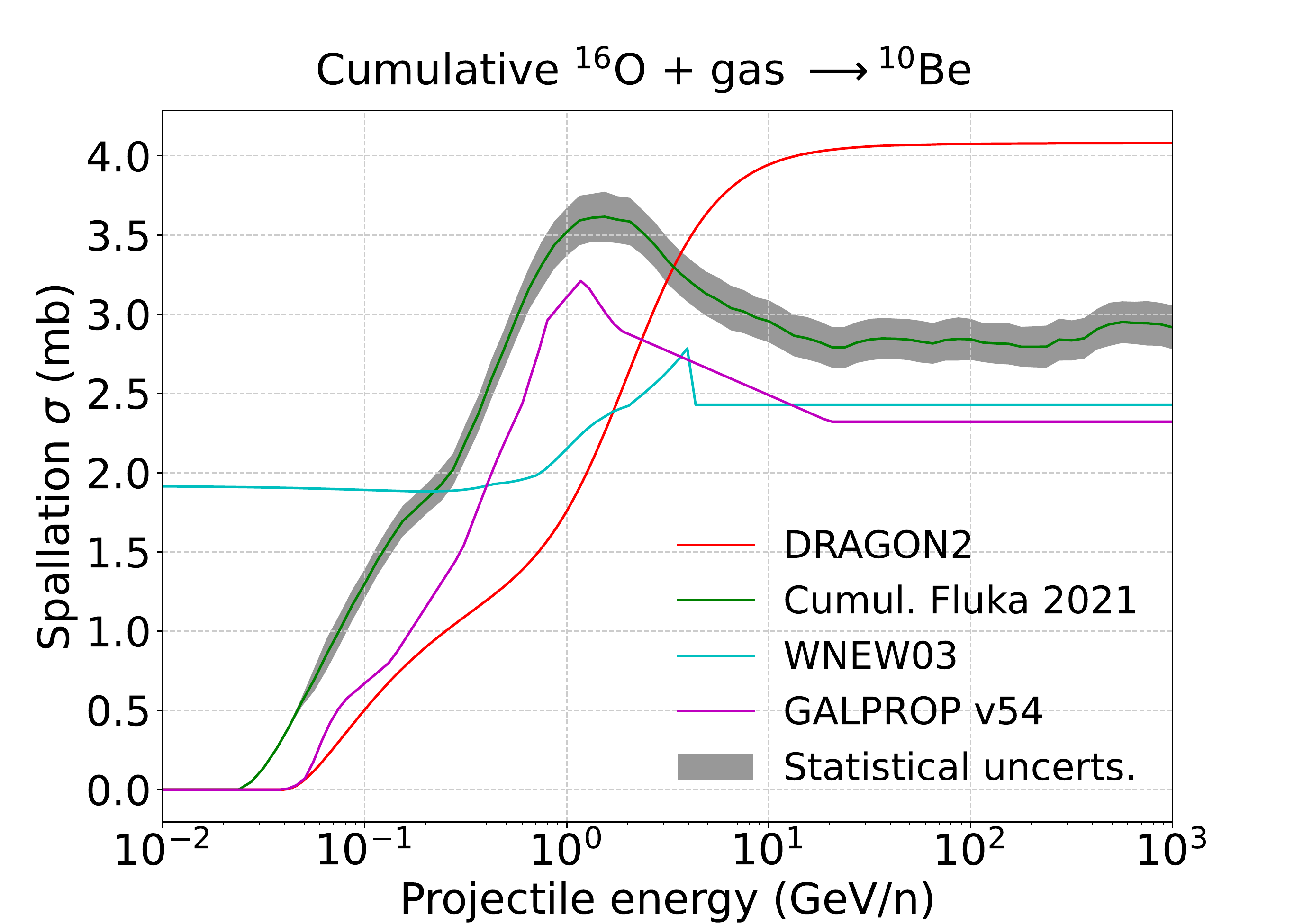}
\hspace{-0.65cm}
\includegraphics[width=0.53\textwidth,height=0.26\textheight,clip] {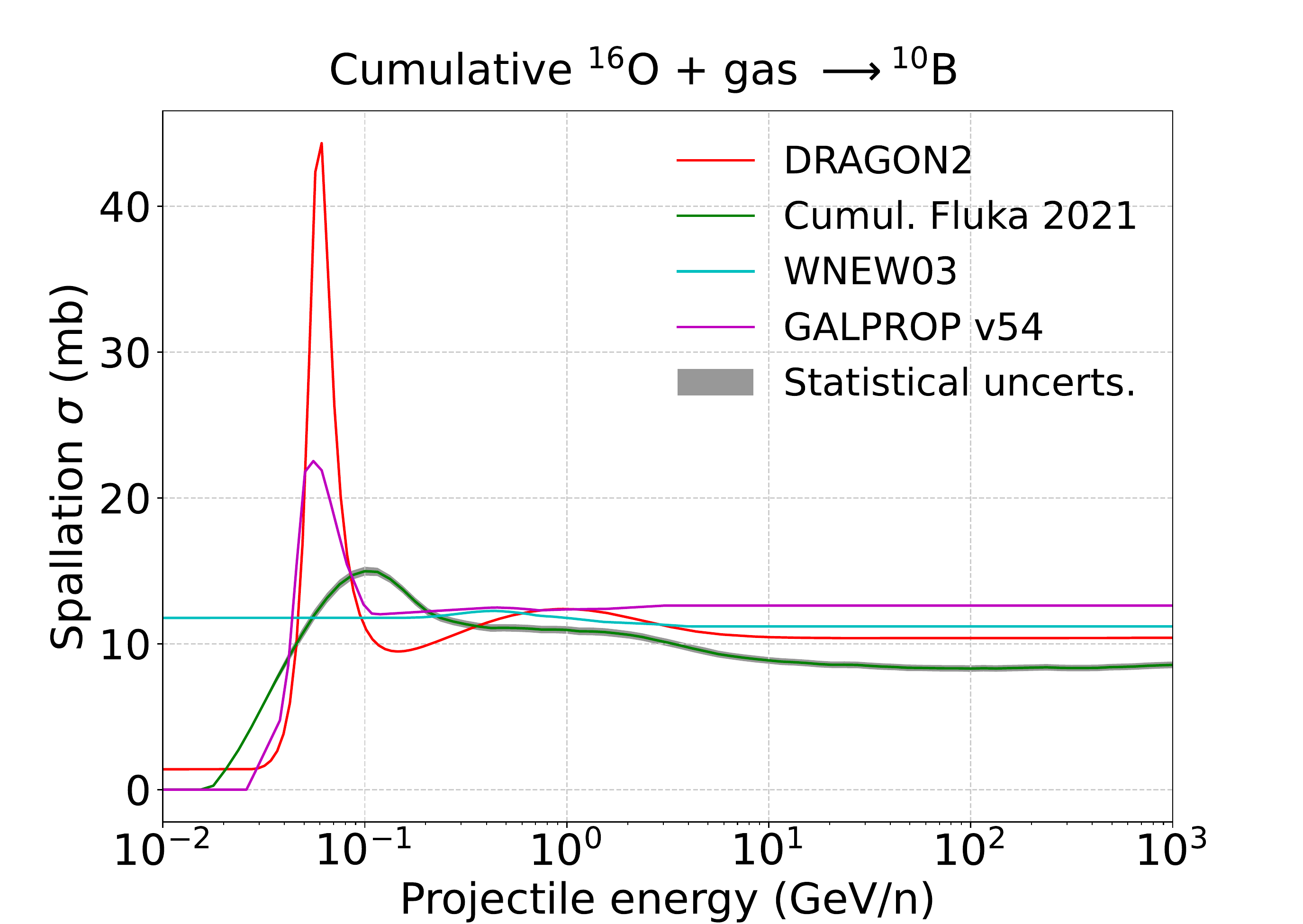}
\rightskip -0.4cm
\end{center}
\caption{As in Fig.\ref{fig:Tot_XS_C} but for $^{16}$O as projectile.}
\label{fig:Tot_XS_O}
\end{figure*}

In turn, the case of inclusive cross sections is more delicate, due to the amount of isotopes involved in the CR network. Figures~\ref{fig:Tot_XS_C} and~\ref{fig:Tot_XS_O} show the total (i.e. cross sections of interactions with a gas with ISM composition) cumulative (i.e. including the contribution of ghost nuclei) cross sections of production of the main secondary CR species. The cross sections calculated from {\tt FLUKA} have associated a small statistical uncertainty, since it is a Monte Carlo-based computation, that we highlight as a grey band. We compare the computed {\tt FLUKA} cross sections with the most widespread cross sections parameterisations used in CR codes: the GALPROP and DRAGON2 parameterisations and the cross sections calculated with the {\tt WNEW03}~\cite{webber2003updated, webber1990formula} and {\tt YIELDX}~\cite{tsao1998partial,silberberg1998updated} (labelled in the legends as TS98) codes. These four cross section models give similar predictions even though they follow different strategies to fit data: whereas the Webber (WNEW03) and DRAGON2 parameterisations followed different phenomenological formulations (and even different experimental data sets to adjust them), the GALPROP cross sections make use of a semi-analytical approach, using Monte Carlo codes in combination with their phenomenological parameterisations and giving special attention to the cross sections measurements more reliable in case there are sizeable discrepancies between experiments in a certain energy region. On the contrary the FLUKA cross sections derived in this work completely relies in theory-based interaction models.

In general, they seem to be consistent with each other, sharing some common trends that should be remarked: 
\begin{itemize}
    \item We find that the FLUKA cumulative cross sections are compatible within $30\%$ with the other cross sections parameterisations. In addition, not only the normalization of these cross sections is compatible with the most updated parameterisations, but their energy dependence seem to be in great agreement. 

    \item The position of the predicted resonances are, in general, in good agreement with those parameterised from data. The main discrepancy is found in the production of $^{10}$B, which could also be related to the cross sections from other resonances decaying to $^{10}$B, as, e.g., $^{10}$C. We highlight here that the different parameterizations also differ in the predicted intensity of resonances, which can be seen from the $^{16}$O $\longrightarrow$ $^{9}$Be reaction (upper right panel of Fig.~\ref{fig:Tot_XS_O}). This would affect our predictions in the energy range of the Voyager-1 data. 
    However, most of the resonances are well reproduced, as it can be seen in the $^{12}$C $\longrightarrow$ $^{11}$B reaction (lower right panel of Fig.~\ref{fig:Tot_XS_C}).

    \item One of the main features expected in the FLUKA cross sections is their smooth rise with energy, that is reliably observed in particle physics experiments~\cite{PhysRevD_rise, Block_2011, RiseofXS}, but which is not incorporated in any of the current relevant parameterisations. This logarithmic rise in the cross sections spectra is present in the FLUKA inelastic and inclusive cross sections, clearer above $100 \units{GeV/n}$, and can affect our predictions of the spectral index of the diffusion coefficient, $\delta$.

\end{itemize}

As demonstrated, the {\tt FLUKA} code is able to make inelastic and spallation cross sections calculations which are, in general, compatible with experimental data and also with the most used and updated parameterisations. Given the scarcity of cross sections measurements (especially for spallation reactions), these nuclear codes, in combination with existent data, are expected to provide a meaningful improvement in the cross sections sets employed for CR studies. In the next section, the capability of the FLUKA cross sections data sets to be consistent with CR data and to provide predictions on CR propagation is broadly studied.

\section{Implementation of the FLUKA cross sections in CR propagation codes and compatibility with CR data}
\label{sec:FDRAGON}

The implementation of these new sets of cross sections into the {\tt DRAGON2} code has been performed in order to test the reproducibility of the current CR data for all species up to $Z=26$. The general formula describing the propagation of CRs solved by {\tt DRAGON2} is given by the following set of equations:
\begin{equation}
\label{eq:caprate}
\begin{split}
 \vec{\nabla}\cdot(\vec{J}_i - \vec{v}_{\omega}N_i) + \frac{\partial}{\partial p} \left[p^2D_{pp}\frac{\partial}{\partial p} \left( \frac{N_i}{p^2}\right) \right] = Q_i + \frac{\partial}{\partial p}  \left[\dot{p} N_i - \frac{p}{3}\left( \vec{\nabla}\cdot \vec{v}_{\omega} N_i \right)\right] \\
  - \frac{N_i}{\tau^f_i} + \sum \Gamma^s_{j\rightarrow i}(N_j) - \frac{N_i}{\tau^r_i} + \sum \frac{N_j}{\tau^r_{j\rightarrow i}} \hspace{1.2cm}
\end{split}
\end{equation}
Here, $\vec{J_i}$ indicates the CR diffusive flux of the $i$-th species and $N_i$ is the density per unit momentum. The term $\vec{v}_{\omega}$ represents the advection speed, which we assume to be null. The second term in the left-hand side accounts for the diffusion in momentum space. The first term in the right-hand side, $Q_i$, represents the distribution and energy spectra of particle sources (injection spectra); the second term describes the momentum losses; finally, the remaining four terms describe losses and gains due to decay and fragmentation. 

\begin{figure*}[t]
\begin{center}
\includegraphics[width=0.76\textwidth] {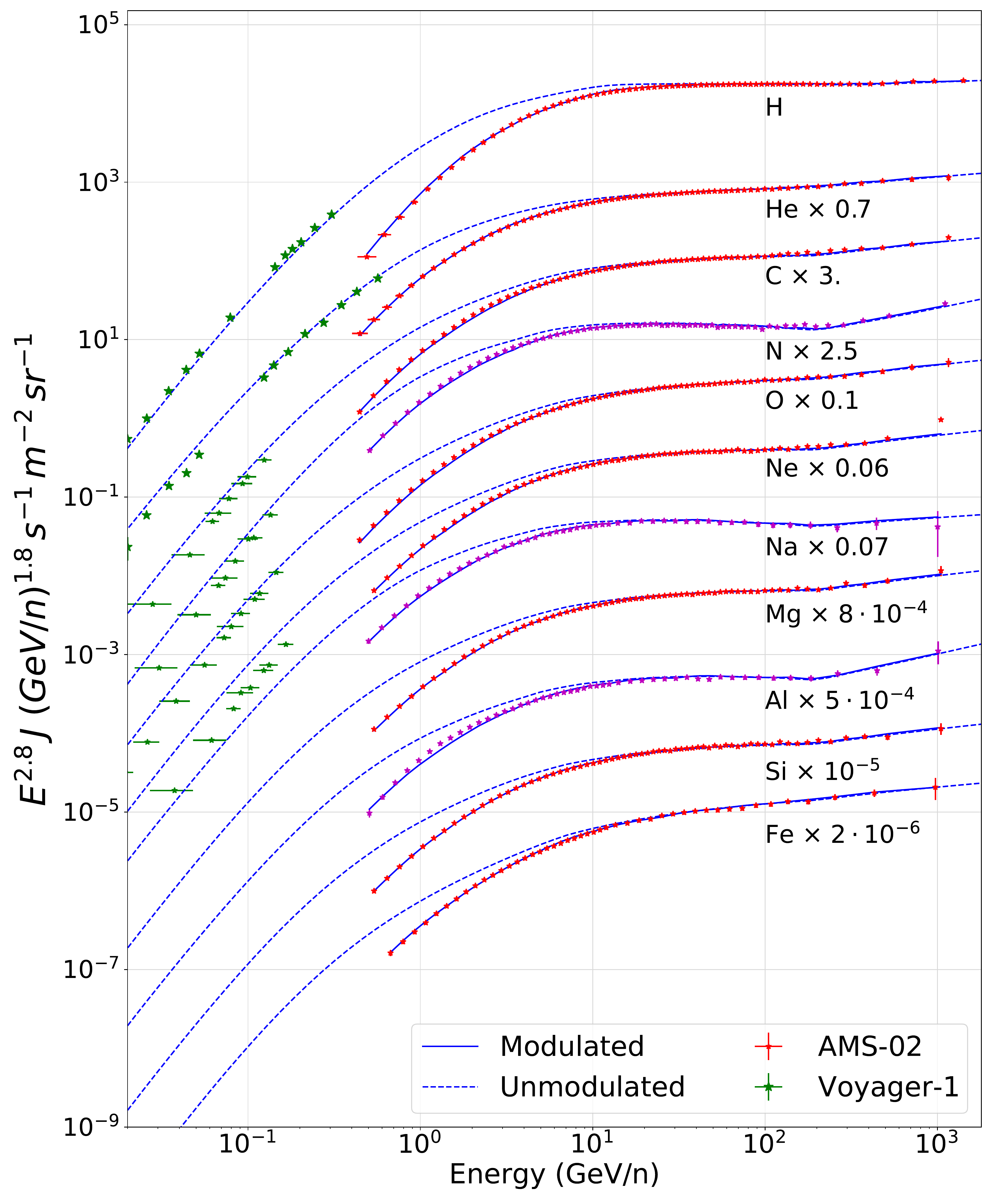}
\end{center}
\caption{Spectra of the CRs injected in the simulations, compared to AMS-02 and Voyager-1 data (available up to Ne). Modulated spectra (see the text for the details in the Fisk potential used for the data collected in different periods) are shown as a solid line and the unmodulated spectra as a dashed line. Voyager data is shown in green, AMS-02 data corresponding to CR species with a significant secondary contribution (namely N, Na and Al) is shown in magenta and AMS-02 for the main primary CR nuclei is shown in red.}
\label{fig:primFluk}
\end{figure*} 

The same injection and diffusion set-up as presented in Refs.~\cite{Luque:2021joz, Luque_MCMC} is used for the simulations performed with the derived FLUKA cross sections to facilitate comparisons. We have employed two diffusion coefficient parameterisations commonly used in previous works~\cite{genolini2017indications, Weinrich:2020cmw, Mazziot}, which differ in the interpretation of the high energy break observed in CR data~\cite{Adriani69, Ahn_2010, AMS_hardening, Panov:2006kf}. We refer to them as ``diffusion'' (Eq.~\ref{eq:breakhyp}) and ``source'' (Eq.~\ref{eq:sourcehyp}) hypotheses, as discussed in Ref.~\cite{Luque_MCMC}.
\begin{equation}
D = D_0 \beta^{\eta}\left(\frac{R}{R_0} \right)^{\delta} \text{ \hspace{1.8 cm} \textbf{Source hypothesis}}
\label{eq:sourcehyp}
\end{equation}

\begin{equation}
 D = D_0 \beta^{\eta}\frac{\left(R/R_0 \right)^{\delta}}{\left[1 + \left(R/R_b\right)^{\Delta \delta / s}\right]^s}
 \text{  \hspace{0.7 cm} \textbf{Diffusion hypothesis}} \,\,\, ,
\label{eq:breakhyp}
\end{equation}
where $R_0$, the rigidity at which the diffusion coefficient is normalized, is set to $4\units{GV}$.
For the diffusion coefficient of Equation~\ref{eq:breakhyp}, we use the values $\Delta\delta = 0.14$, $R_b = 312 \units{GV}$ and $s = 0.040$, as explained in~\cite{Luque_MCMC} and derived in Ref.~\cite{genolini2017indications}. The use of the diffusion coefficient of Eq.~\ref{eq:breakhyp} implies using as injection a broken power-law (with break at $8$~GV), while the use of that defined in Eq.~\ref{eq:sourcehyp} implies using a doubly broken power-law (with breaks at $8$~GV and  $320$~GV) for the injection.

We show in Figure~\ref{fig:primFluk} the spectra of the main primary CRs (mainly dependent on the injection parameters employed) compared to AMS-02. The injection spectra used are freely adjusted to reproduced AMS-02 data. All the experimental data from CR experiments are taken from \url{//https://lpsc.in2p3.fr/crdb/}~\cite{Maurin_db1, Maurin_db2} and \url{https://tools.ssdc.asi.it/CosmicRays/} \cite{ssdc}. We find here that, in order to reproduce those spectra, we need to employ different injection parameters for $^{1}$H, $^{4}$He, C-O group (i.e. $^{12}$C and $^{16}$O), Ne-Mg-Si group ($^{20}$Ne, $^{24}$Mg and $^{28}$Si), N-Na-Al ($^{14}$N, $^{23}$Na and $^{27}$Al) group and Fe ($^{56}$Fe, $^{57}$Fe and $^{58}$Fe). These discrepancies may be due to the fact that sources with different properties (as mass or metallicity) may accelerate particles with slightly different power-laws, but the actual reason is still a mystery (see also Ref.~\cite{Korsmeier:2021bkw} for a discussion).

\begin{figure*}[!t]
\begin{center}
\hskip -0.25 cm
\includegraphics[width=0.53\textwidth,height=0.25\textheight,clip] {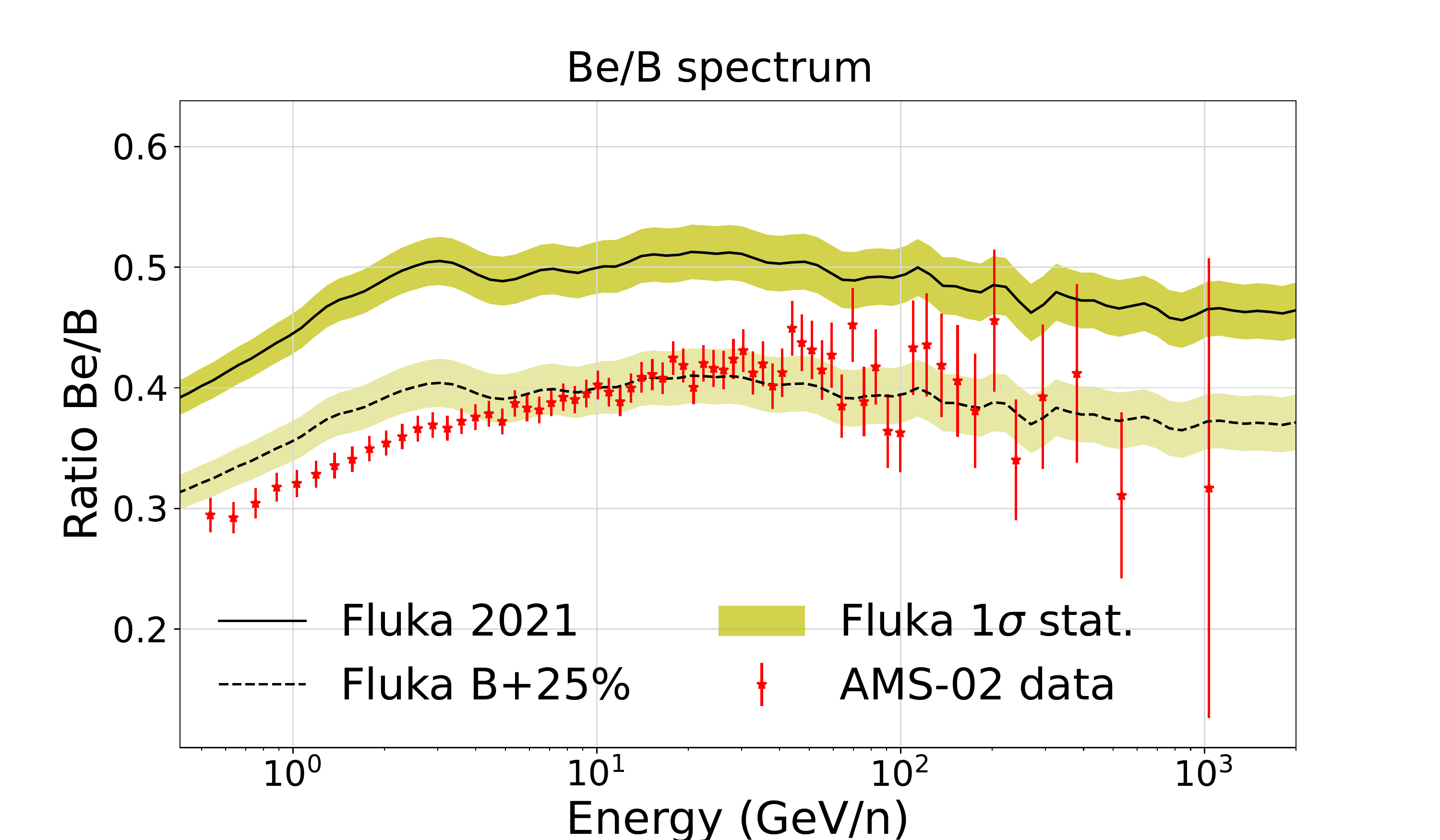} \hspace{-0.5cm}
\includegraphics[width=0.53\textwidth,height=0.25\textheight,clip] {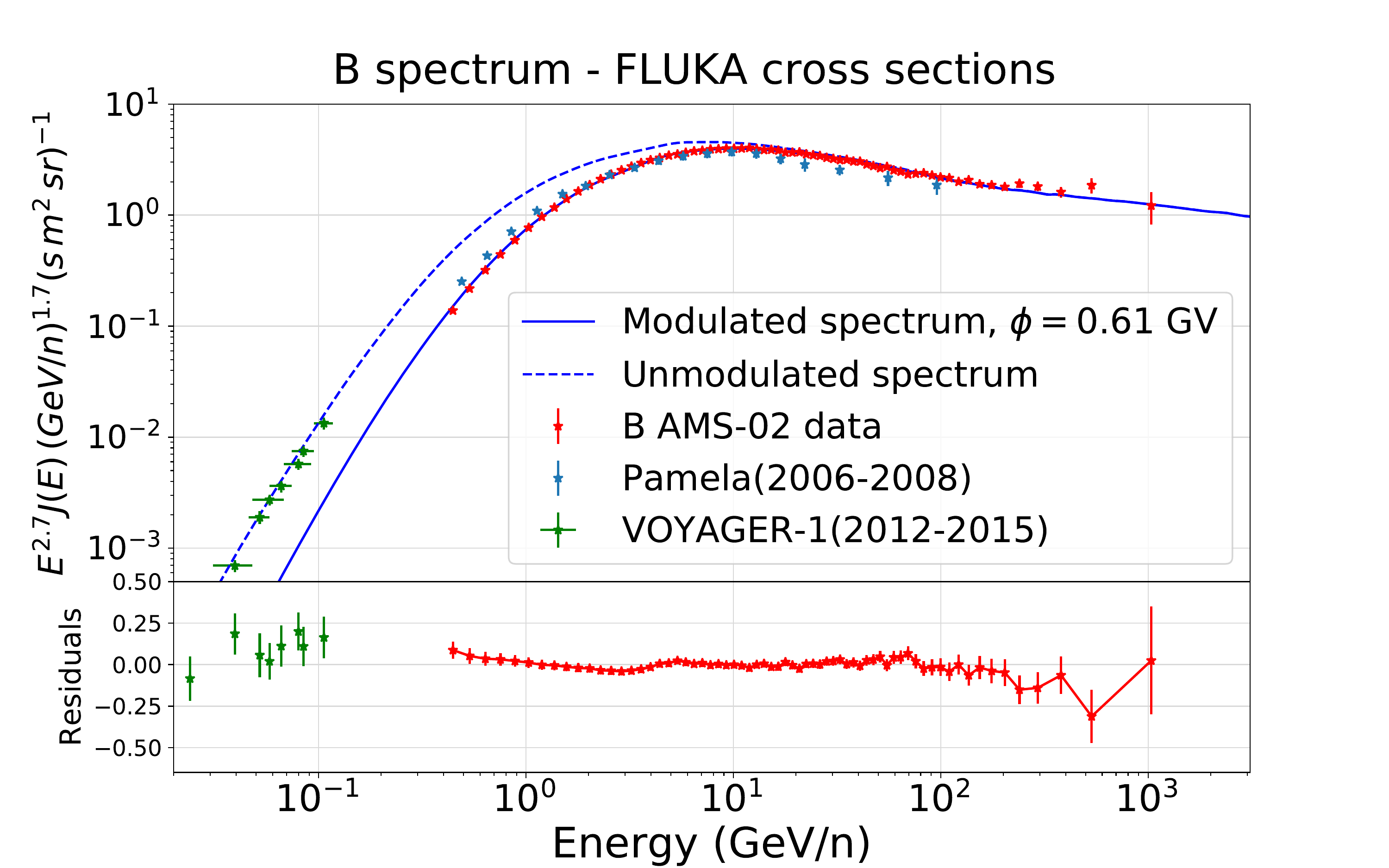}
\rightskip -0.45cm

\hskip -0.25 cm
\includegraphics[width=0.53\textwidth,height=0.25\textheight,clip] {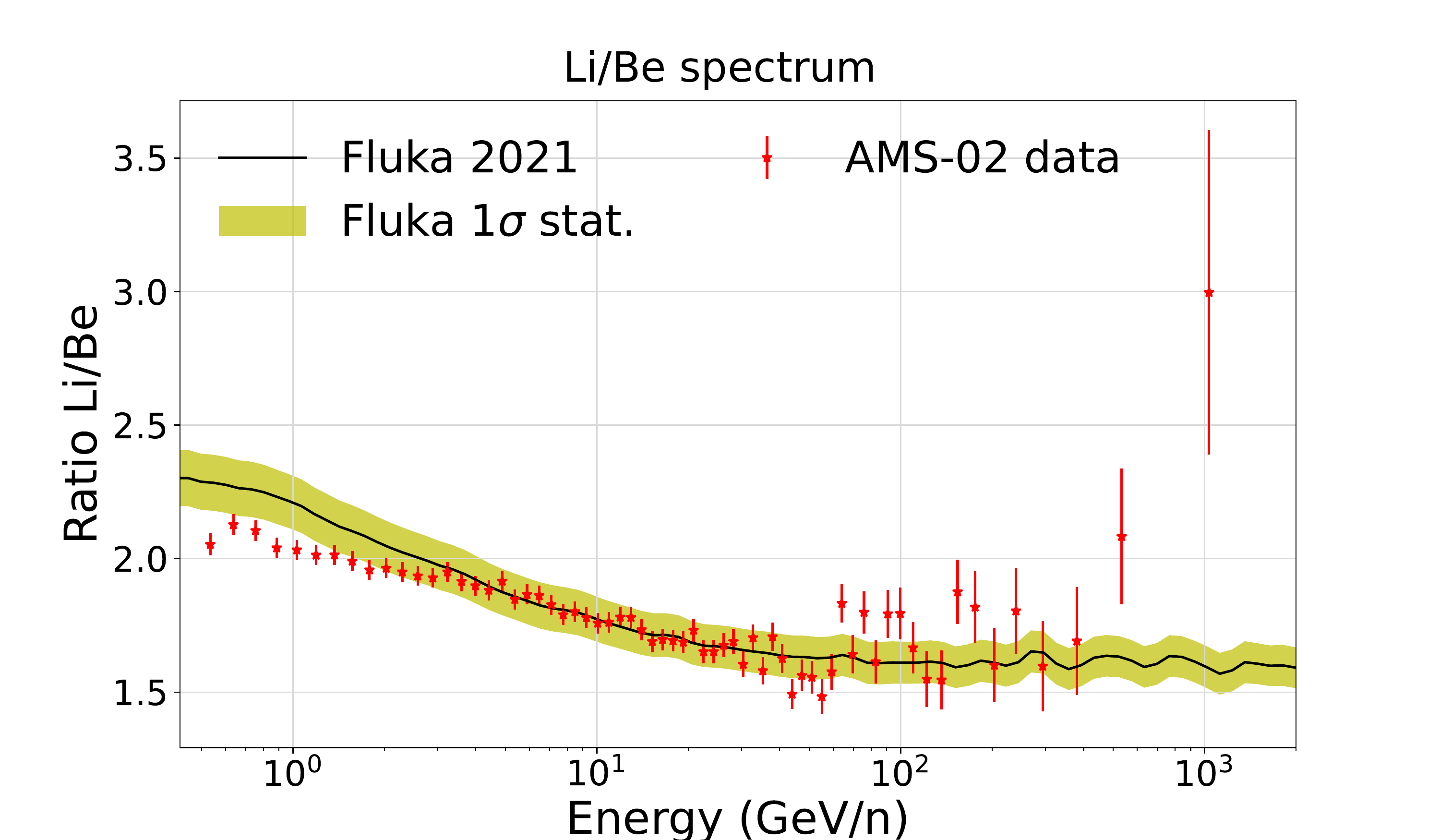}
\hspace{-0.5cm}
\includegraphics[width=0.53\textwidth,height=0.25\textheight,clip] {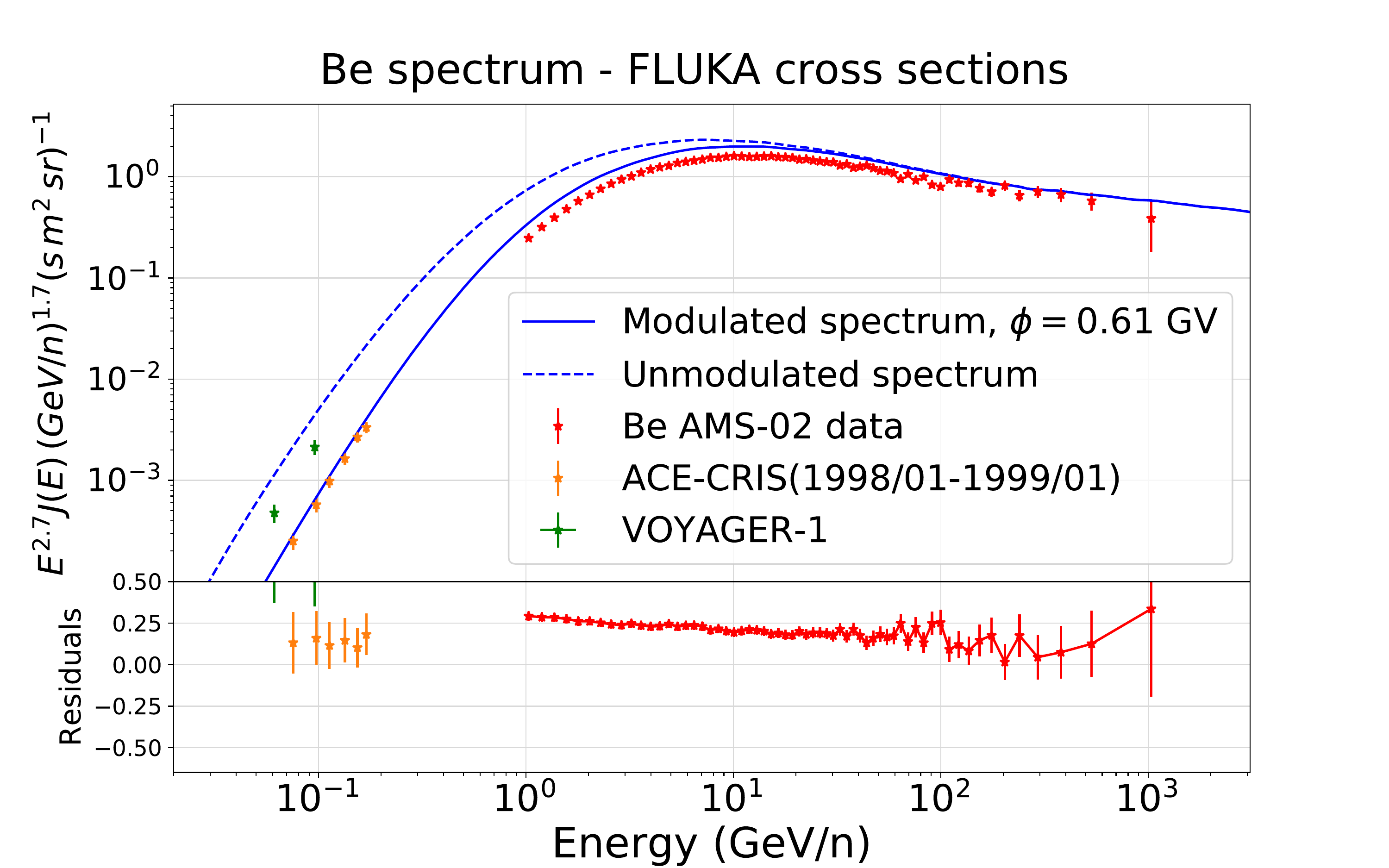}
\rightskip -0.45cm

\hskip -0.25 cm
\includegraphics[width=0.53\textwidth,height=0.25\textheight,clip] {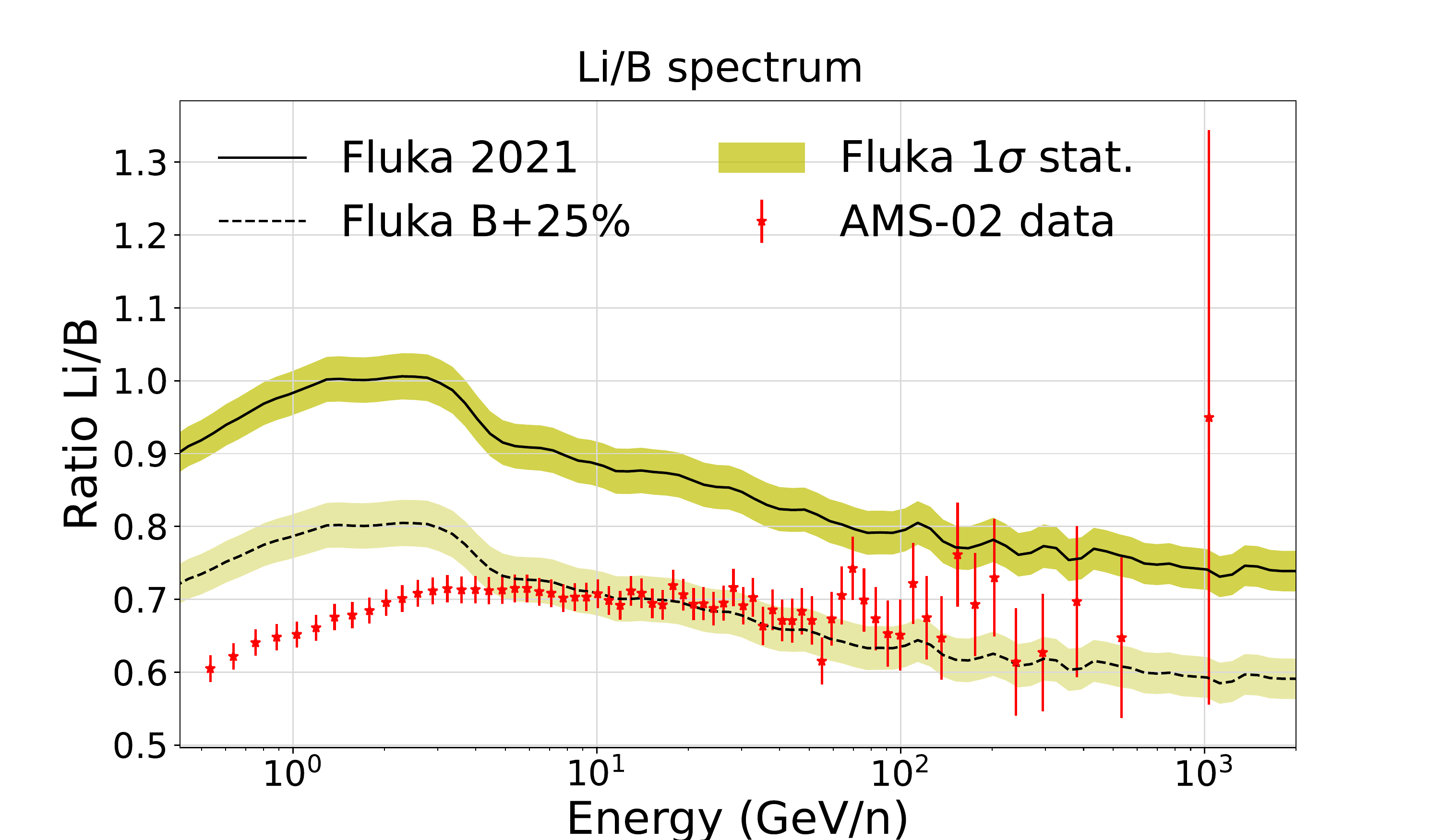} 
\hspace{-0.5cm}
\includegraphics[width=0.53\textwidth,height=0.25\textheight,clip] {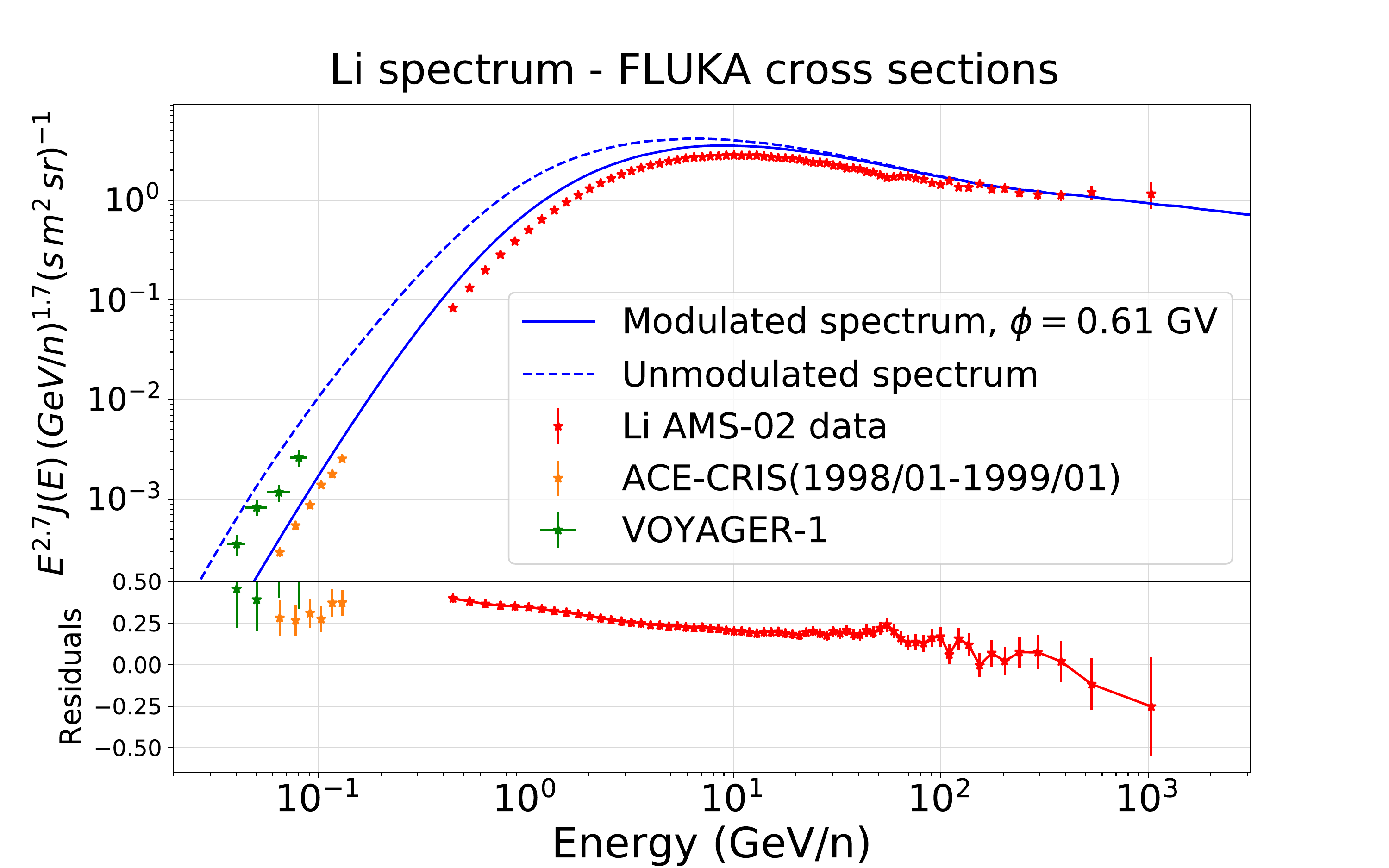}
\rightskip -0.45cm
\end{center}
\caption{Left column: secondary-over-secondary flux ratios for Li, Be and B obtained with the FLUKA computations, together with the band of statistical uncertainties due to the spallation cross sections calculation (since this is a Monte Carlo computation). Right column: B, Be and Li fluxes obtained from the simulation compared with experimental data, using the diffusion parameters obtained from the fit of the B/C flux ratio (see section~\ref{sec:MCMC}).}
\label{fig:secsec_Fluka}
\end{figure*}

Given their importance to study the transport of CRs, we specially focus on the spectra of B, Be and Li and the flux ratios among them, which are depicted in Fig.~\ref{fig:secsec_Fluka}. These spectra are derived from a Markov chain Monte Carlo analysis (section~\ref{sec:MCMC}) of the AMS-02 boron-over-carbon flux ratio (see Fig.~\ref{fig:Box_plot} and upper table~\ref{tab:MCMC_table}) for the diffusion coefficient of Eq.~\ref{eq:sourcehyp}. 
As it is observed in the figure, there is a $\sim 20-25$\% discrepancy in the Li and Be spectra (right panels of Fig.~\ref{fig:secsec_Fluka}) with these diffusion parameters, which is well within the uncertainties usually reported in cross sections data~\cite{Luque:2021joz}. We remark that this discrepancy can be resolved with a $\sim12\%$ overall increase of the B production cross sections and a $\sim12\%$ reduction of the Be and Li production cross sections, which predicts the same ratios as for the case with a $25\%$ increase of the B cross sections. 
Notably, the energy dependence of the derived spectra is well compatible with that of the AMS-02 data for Be, while the Li spectrum seems to be slightly more deviated, as easily seen from their residuals. We notice that the use of the straight-ahead approximation (where the kinetic energy per nucleon of the primary nucleus is conserved to the secondary nucleus) in the {\tt DRAGON2} code can affect the predictions of light nuclei at low energy, which would, mainly, add more uncertainty to the low energy part of the Li spectrum. The energy dependence of their respective flux ratios (left column), whose main dependence above $10 \units{GeV/n}$ is the inclusive cross sections, shows this as well. Specially, the predicted Li/Be flux ratio is compatible with AMS-02 data without any scaling of the predicted cross sections above a few GeV/n, while the other ratios need a scale to be reproduced. In addition, the energy dependence of the predicted Li/B spectrum seems to be slightly discrepant, although still in agreement within the $1\sigma$ uncertainties. We have renormalised the B cross sections by $25\%$ to show this more clearly (dashed lines).  
In Appendix~\ref{sec:appendixB} we show the predicted flux of another important light secondary CR, the $^3$He isotope, compared to data.

The high importance of having good estimations of the cross sections of production of B, Be and Li stands for the fact that they allow us to constrain the magnetized halo size, the diffusion parameters, the Alfv\'en velocity and other parameters important for understanding the transport of CRs. These parameters are carefully studied in sections~\ref{sec:halo_Fluka} and~\ref{sec:MCMC}.

\subsection{Halo size predictions with the FLUKA cross sections}
\label{sec:halo_Fluka}

\begin{figure*}[t]
\begin{center}
\includegraphics[width=0.52\textwidth,height=0.24\textheight,clip] {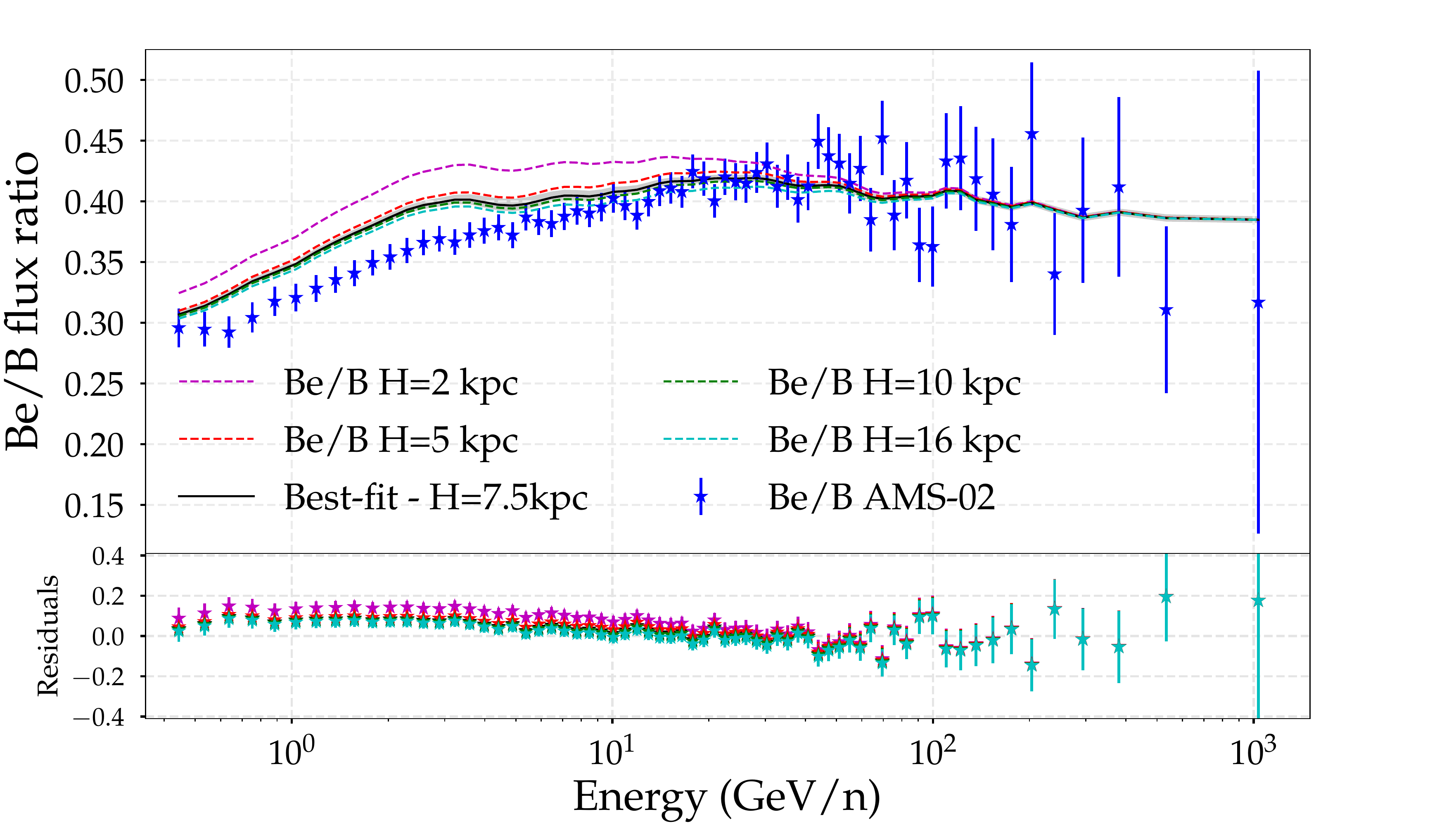} 

\includegraphics[width=0.49\textwidth,height=0.24\textheight,clip] {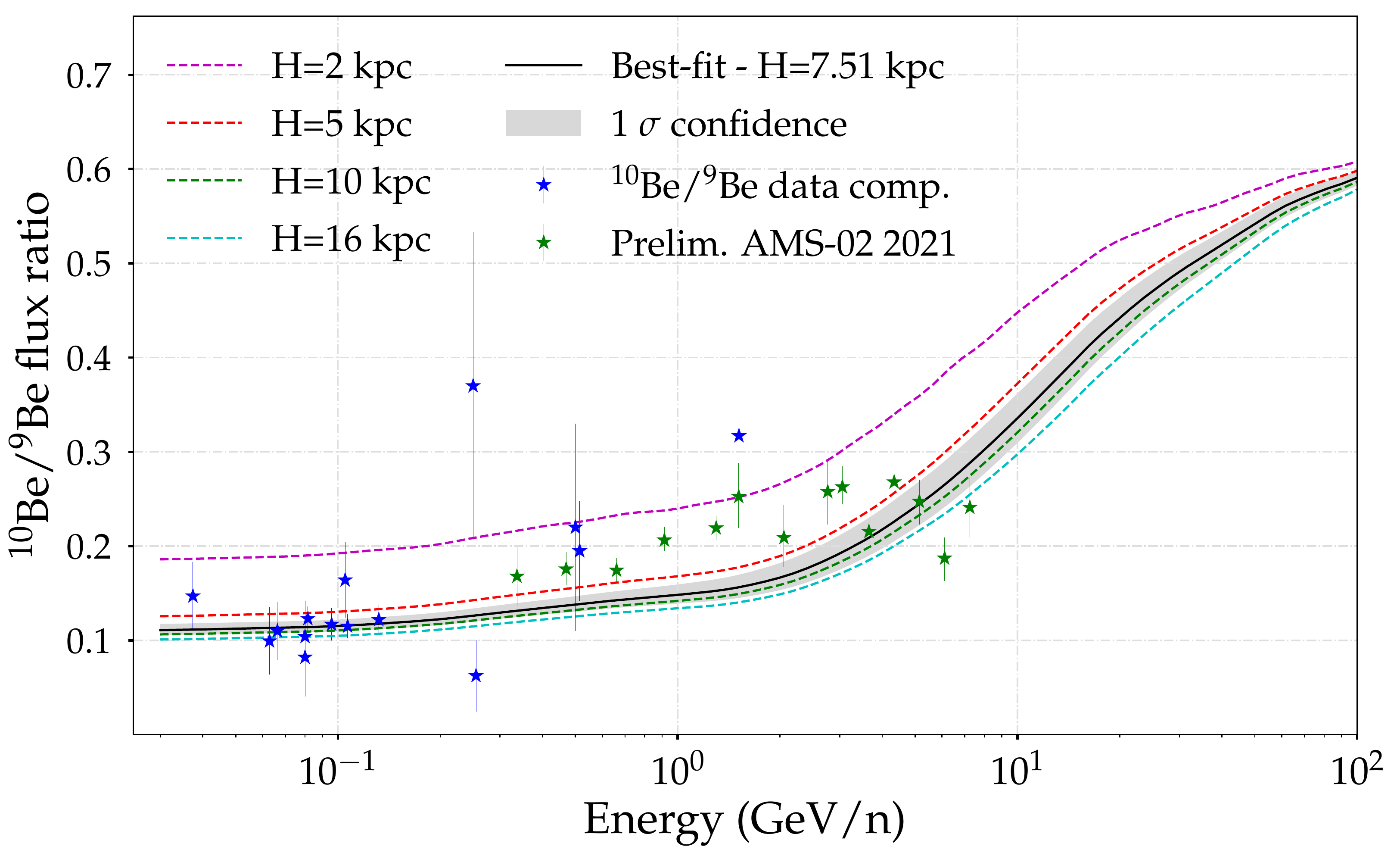}
\includegraphics[width=0.49\textwidth,height=0.24\textheight,clip] {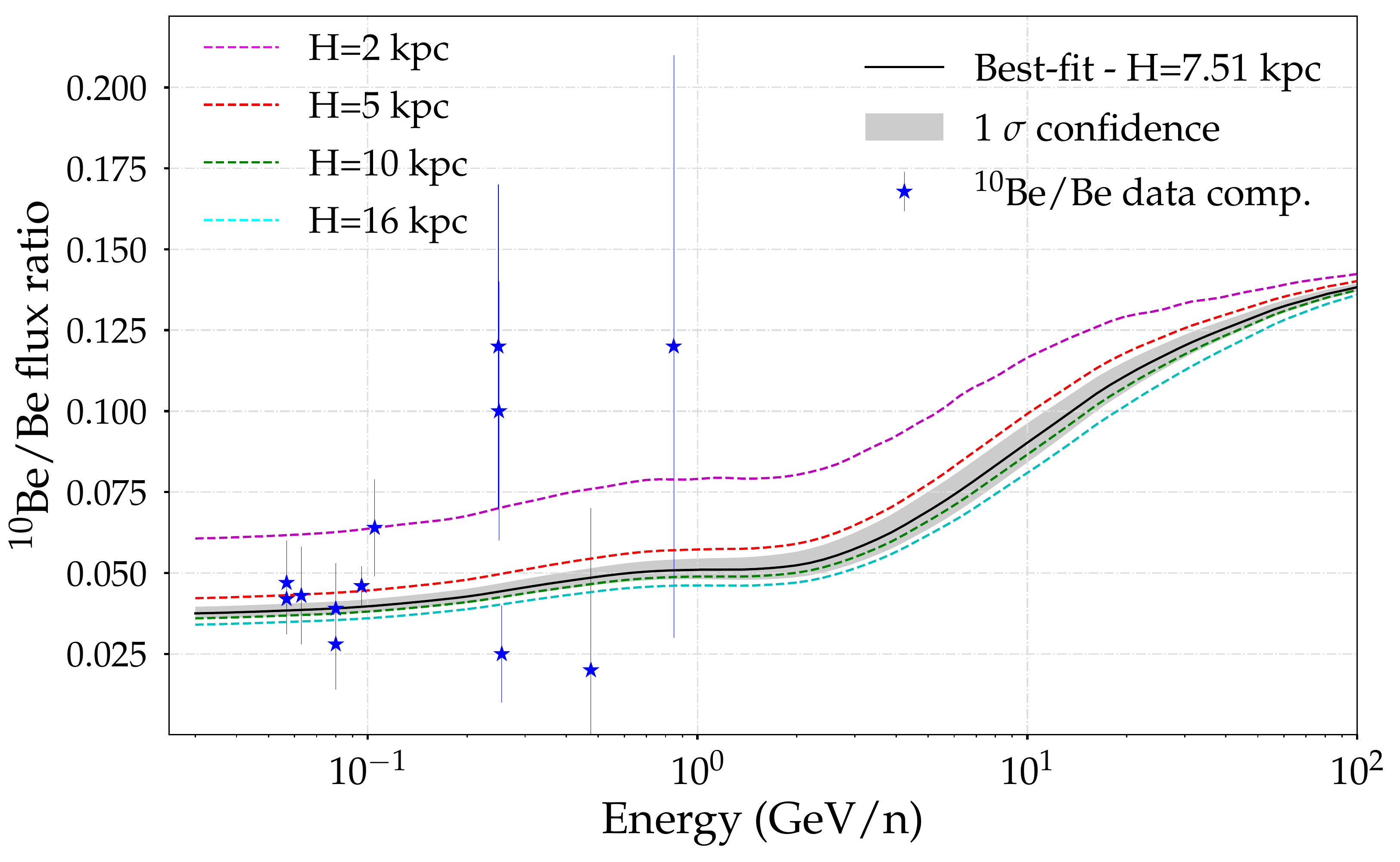}
\end{center}
\caption{\textbf{Top panel:} secondary-over-secondary flux ratios of the light secondary CRs (Li, Be, B). In these plots, the spectrum for the best-fit halo size value found from the $^{10}$Be/$^{9}$Be ratios is also shown. \textbf{Bottom panels:} $^{10}$Be/$^{9}$Be and $^{10}$Be/Be flux ratios. The predictions are compared to experimental data for simulations performed with various halo sizes. }
\label{fig:Fhalos}
\end{figure*} 

The evaluation of the magnetized halo size, i.e. the region in the Galaxy where CRs diffuse, can be constrained from several observables, as radio~\cite{bringmann2012radio}, X-ray and gamma-ray emission~\cite{biswas2018constraining} from lepton synchrotron and inverse-Compton processes or from other CR observables like unstable heavy nuclei~\cite{moskalenko2000diffuse}, whose mean lifetime is of the order of the propagation time.

Here, we estimate the magnetized halo size, derived with the FLUKA cross section, from a combined analysis of the ratios of $^{10}$Be (mean life time $\sim 1.4 \units{My}$~\cite{chmeleff2010determination})), namely $^{10}$Be over $^{9}$Be and $^{10}$Be over the total Be flux. To do so, the halo size value is fitted to experimental data on the $^{10}$Be flux ratios, from the ACE~\cite{ACEBe}, IMP~\cite{IMP1, IMP2}, ISEE~\cite{ISEE}, ISOMAX~\cite{Hams_2004}, Ulysses~\cite{Ulysses} and Voyager~\cite{VoyagerMO} missions, using the {\tt curve\_fit} package from the {\tt scipy.optimize} library, as it is done in~\cite{Luque:2021joz}, where we refer the reader for more details about the analysis. This fit yields a value of the halo size of $\sim7.5^{+1.13}_{-0.95} \units{kpc}$, similar to estimations with other cross sections, usually quoted to be around $6 \units{kpc}$~\cite{Luque:2021joz, CarmeloBeB, Weinrich_halo} and also favored by gamma-ray~\cite{zaharijas2012fermi} and radio observations~\cite{di2013cosmic,beuermann1985radio,orlando2013galactic}. 

In the top panel of Figure~\ref{fig:Fhalos}, we display the Be/B flux ratio, with B cross sections scaled by $25\%$, for various values of the halo size, to show the variations of the predicted ratio for different values of $H$. In the bottom panel, we show the predicted $^{10}$Be ratios for different halo size values. We have also included the $^{10}$Be/$^9$Be AMS-02 preliminary data in the figure (lower left panel of Fig.~\ref{fig:Fhalos}) to show that these predictions are compatible with these data as well, but we did not include them in the fit procedure since they are yet preliminary. From these figures one can conclude that a halo size value between $5$ and $10$~kpc would be well compatible with the data. However, we employ the best-fit halo size value in the determination of the rest of propagation parameters from the secondary-to-primary and secondary-to-primary flux ratios, as explained in the next section.

\subsection{Markov chain Monte Carlo analysis of the diffusion coefficients parameters with the FLUKA cross sections}
\label{sec:MCMC}
In this section, we show and discuss the results obtained from the analyses carried out for the flux ratios of B, Be and Li to C and O (B/C, B/O, Be/C, Be/O, Li/C and Li/O) and its respective secondary-to-secondary flux ratios (Be/B, Li/B, Li/Be), using the Markov chain Monte Carlo algorithm (MCMC) presented in Ref.~\cite{Luque_MCMC}. We also performed the combined analysis of all these spectra presented in that work, which adds a scale factor to renormalise the cross sections of production of B, Be and Li. The best-fit propagation parameters and their confidence intervals are depicted for both the parameterisations of the diffusion coefficient used, in the form of a box-plot in Figure~\ref{fig:Box_plot} and explicitly given in tables in appendix~\ref{sec:appendixA}. On one hand, it is convenient to remark some of the implications of the values found in these analyses of independent ratios:
\begin{itemize}
\item Interestingly, an average value of $\delta\sim0.42$ is favored by the B and Be ratios, whereas the Li ratios predict a $\delta$ value below $0.4$. Nevertheless, these values seem to be generally compatible within the $2\sigma$ uncertainties in their determination.

\item The analyses of the Li and B ratios favor an Alfv\'en velocity value around $30\units{km/s}$, not too far from theoretical estimations~\cite{Spanier_2005, Spangler:2010nu, Lerche_2001} and with predictions using other cross sections parameterisations~\cite{Luque_MCMC}. Nevertheless, these values are in tension with energetic considerations, as discussed by Ref.~\cite{Drudy_VA_Energetics}. The lower values obtained for $V_A$ in the analyses of the Be ratios were also found for the GALPROP cross sections by Ref.~\cite{Luque_MCMC}. 

\item Notably, $\eta$ is negative for all the analyses, as revealed by other recent analyses using different cross sections (see, for example, Ref.~\cite{Weinrich_combined}), which reflects a change in the trend of the diffusion coefficient at low energies, towards lower confinement of low-energy CRs. This seems to be due to the dissipation or damping of MHD waves at the scales (Larmor radius of the particles) associated to these energies~\cite{Ptuskin_2006, Fornieri:2020wrr}. 

\item Finally, it can be observed that the $D_0/H$ values obtained are pretty coincident for pairs of ratios with same secondary CR, while the ratios of different secondary CRs significantly deviate, which is due to the spallation cross sections related to the production of each of these secondaries. This highlights the need to add a scale factor for each of these CRs in order to renormalise their cross sections of production.    

\end{itemize}

\begin{figure}[!t]
\centering
\includegraphics[width=\textwidth]{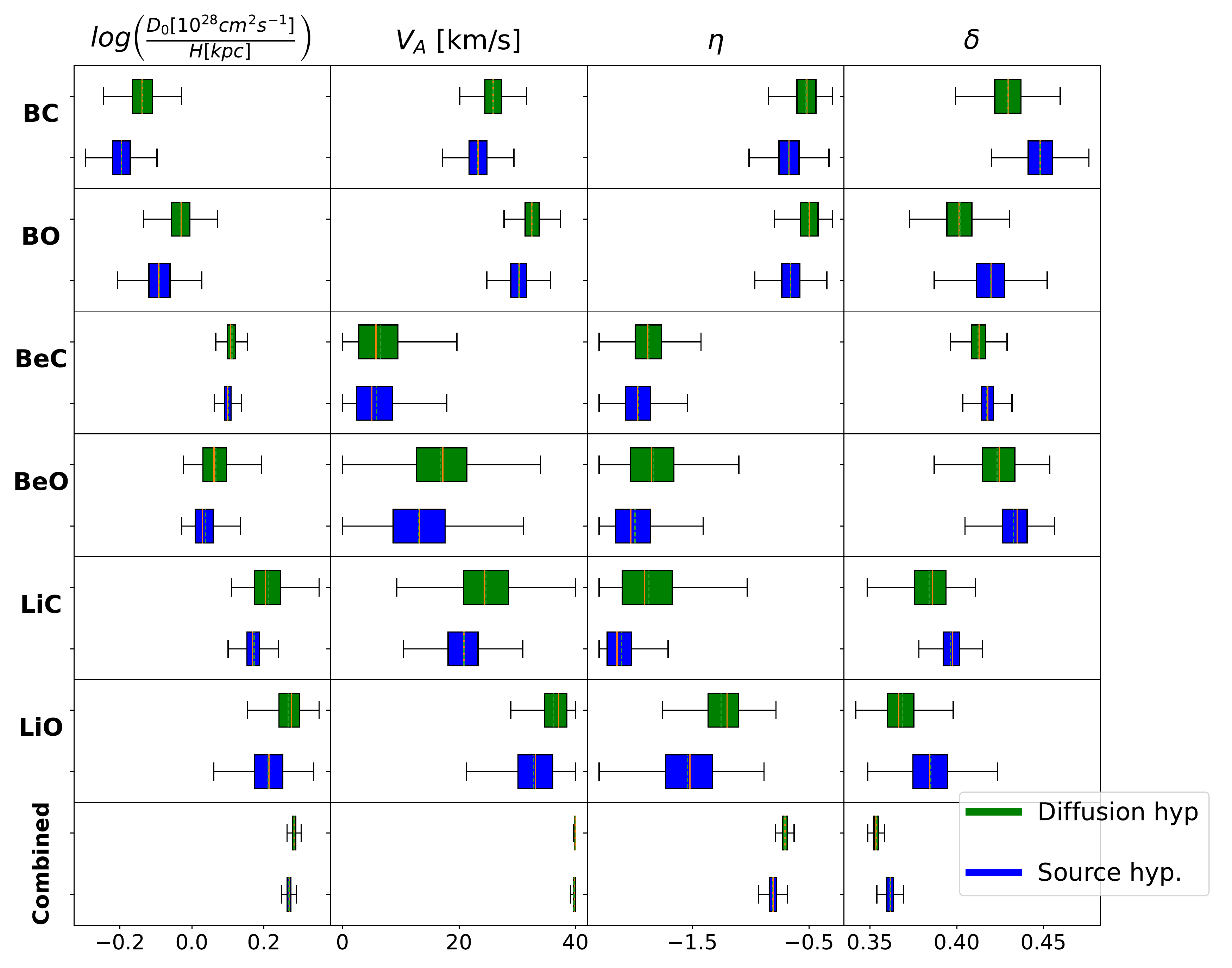} 
\caption{Box-plot summarising the information from the Probability Distribution Function (PDF) of every parameter included in the MCMC analyses performed with the FLUKA cross sections. Different colors represent the different diffusion coefficients employed (Eq.~\ref{eq:sourcehyp} and Eq.~\ref{eq:breakhyp}). This figure is shown in the same format as in Ref~\cite{Luque_MCMC} to facilitate the comparison with the results obtained from the GALPROP and DRAGON2 cross sections.}
	\label{fig:Box_plot}
\end{figure}

\begin{figure}[!t]
\centering
\includegraphics[width=0.55\textwidth, height=0.27\textheight]{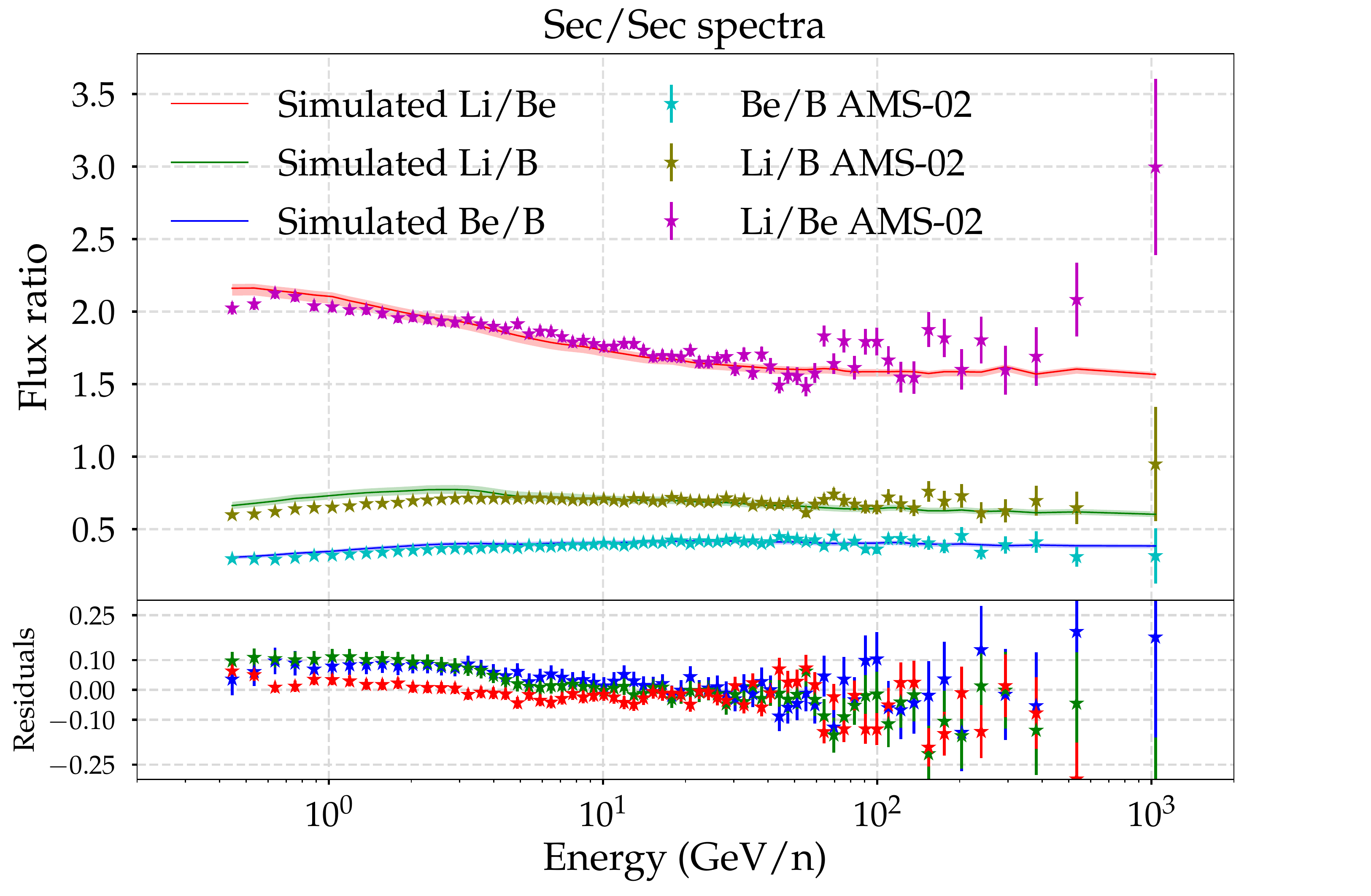} 

\includegraphics[width=0.5\textwidth, height=0.24\textheight]{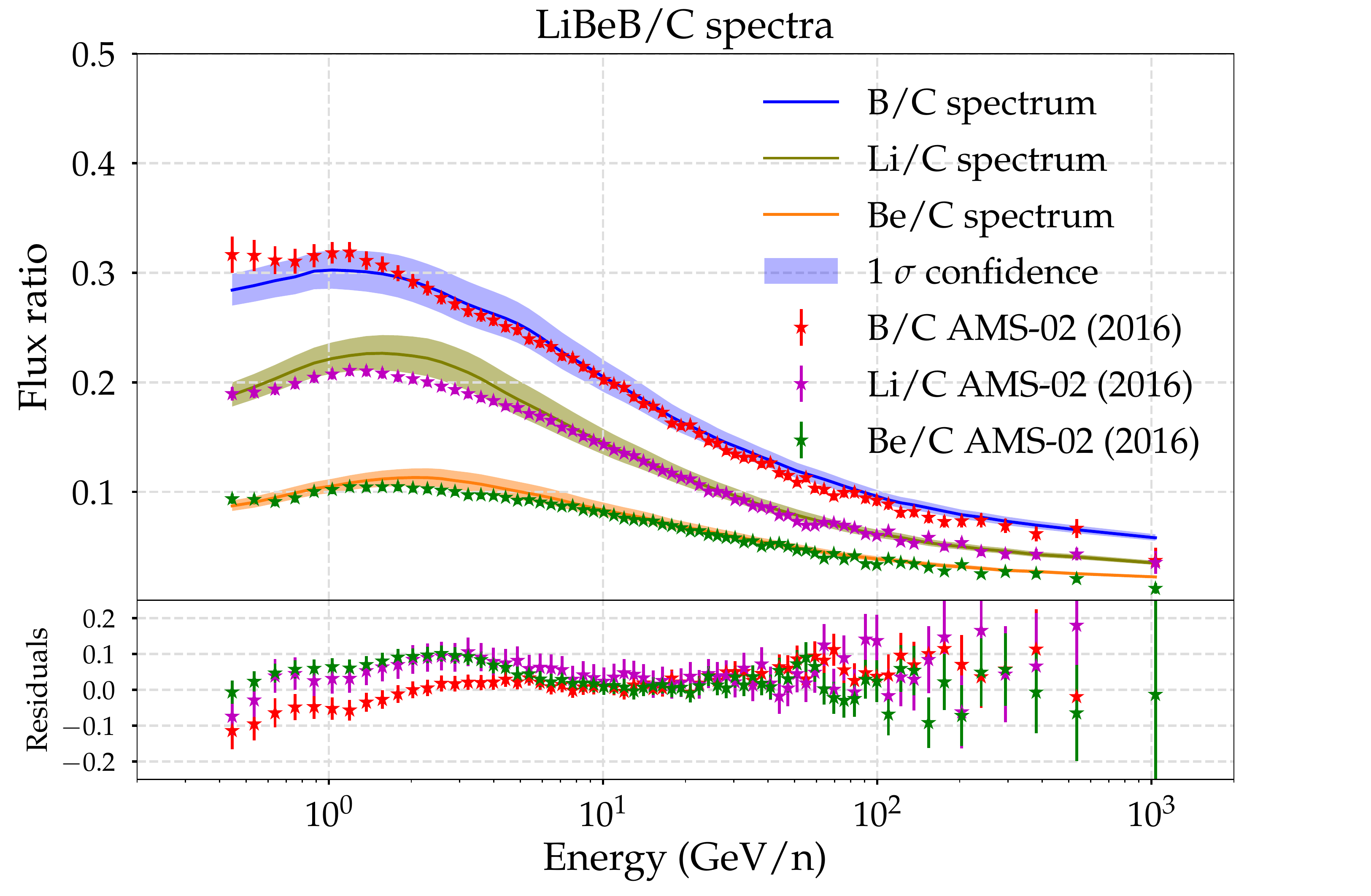}
\hspace{-0.4cm}
\includegraphics[width=0.5\textwidth, height=0.24\textheight]{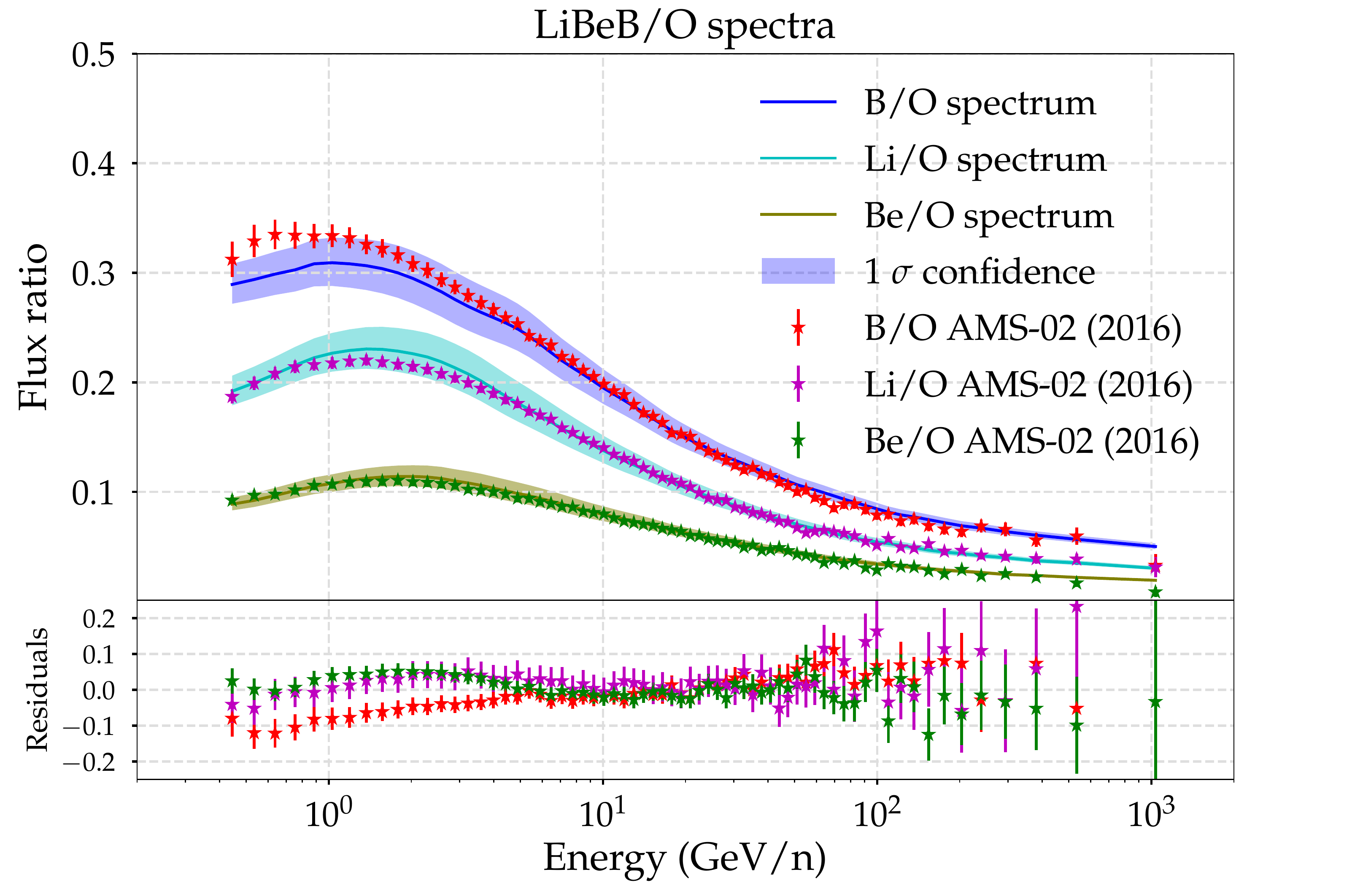} 
\caption{\textbf{Top panel:} Secondary-over-secondary flux ratios of B, Be and Li computed using the propagation parameters and scaling factors found in the combined analysis, compared to the AMS-02 experimental data. \textbf{Lower panels:} Secondary-over-primary spectra (B/C, B/O, Be/C, Be/O, Li/C, Li/O) predicted with the parameters determined by the MCMC combined analysis for the DRAGON2 cross sections. The grid lines at the level of 5\% residuals are highlighted in the lower panel in a different color for clarity. In addition, the Voyager-1 data are also included for completeness.}
\label{fig:Sec_combined}
\end{figure}

On the other hand, the combined analyses converge to propagation parameters closer to those predicted by Li ratios. These analyses predict a spectral index of the diffusion coefficient $\sim 0.36$, which is significantly lower than the typical values found from B ratios, but it is compatible with the values found for the GALPROP cross sections in Ref.\cite{Luque_MCMC}.
A crucial point of this analysis is the cross sections scale factors obtained. These are the same for both diffusion coefficient used 
 and are $1.18$, $0.94$ and $0.93$, for B, Be and Li, respectively, with $1\sigma$ statistical uncertainties of $\sim \pm 0.01$. However, we highlight that there are other systematic uncertainties related to the determination of these scale factors, like those related to the inelastic cross sections used, and could make the total systematic uncertainties in the determination of these scale factors to larger than $5\%$~\cite{Luque_MCMC}. In addition, we highlight that the diffusion coefficient obtained from these combined analyses allows us to reproduce, at the same time, the $^3$He/$^4$He flux ratio (see Fig.~\ref{fig:He3_Fluka}), which evidences that the diffusion coefficient predicted by the light secondary CRs is compatible for the different nuclei within uncertainties (mainly cross sections uncertainties).

The spectra of secondary-to-secondary and secondary-to-primary flux ratios are shown in Figure~\ref{fig:Sec_combined} for the best-fit parameters obtained in the combined analysis for the diffusion coefficient of Eq.~\ref{eq:breakhyp} (diffusion hypothesis for the origin of the break). These panels demonstrate that the ratios of secondary CRs can be successfully reproduced at the same time with the FLUKA cross sections within $1\sigma$ uncertainties, after scaling the overall cross sections of production of B, Be and Li which is well below the typical average uncertainties in the experimental measurements of these cross sections for the best-known channels: $\sim 20-30$\%, which are those from $^{12}$C and $^{16}$O as projectiles. We observe small discrepancies at low energies (below $5\%$) that could be related with these cross sections but could also be related to the parameterisation of the diffusion coefficient used.

It must be highlighted that very few nuclear codes have been able to produce suitable cross sections sets intended to be used in CR propagation codes and none has demonstrated to be able to reproduce the secondary-to-primary flux ratios just with the use of renormalisation (scale) factors. This, in fact, means that the FLUKA cross sections are at the level of precision of the most updated dedicated cross sections parameterisations of CR spallation reactions.

\section{Study of the gamma-ray diffuse emission with the FLUKA cross sections}
\label{sec:FGamma}

In addition to the information about the transport of CRs from secondary CRs, other secondary particles - namely leptons, antinuclei, gamma rays and neutrinos - are formed and provide essential insight about acceleration mechanisms, magnetic field configurations, environment of sources, etc. An important fraction of the particles produced from spallation reactions of CR nuclei with the ISM gas are short-living mesons (mainly pions), which are the responsible of the formation of leptons and gamma rays from their decays. Secondary electrons and positrons are mainly produced via the decay of charged pions (accompanied by neutrinos). Gamma rays are mainly produced by the decays of neutral mesons and by the interactions of CR leptons with magnetic and radiation fields in the ISM, carrying important information about the environmental conditions. The study of the galactic gamma-ray diffuse emission is important to constrain our models and opens a new window to unveil the mechanisms behind CR interactions and acceleration~\cite{Ackermann_2012, Acero_2016}.

In this section we demonstrate that the gamma-ray sky maps and the local HI emissivity spectrum, produced using the FLUKA cross sections for the production of leptons and gamma rays from CR interactions with the interstellar gas, are in good agreement with the current experimental data. We adjust the spectra of protons and leptons with the injection parameters tuned to reproduce AMS-02 data and using the diffusion coefficient given in Eq.~\ref{eq:breakhyp}, with the parameters obtained in the fit of the B/C spectrum (see Fig~\ref{fig:Electrons_posit} in appendix~\ref{sec:appendixD}). Here, the CR e$^{+}$ and e$^{-}$ spectra require further considerations: while the vast majority of CR electrons are supposed to be accelerated in supernova remnants (SNRs), a significant amount of positrons seem to be emitted from pulsar wind nebulae (PWN), expected to be symmetric $e^+$ and $e^{-}$ pairs emitters~\cite{Amato:2013fua, Serpico_efraction}. Therefore, to reproduce this component of primary positrons and electrons we inject, in the DRAGON2 simulation, an extra source of leptons (using a single power-law) emitting them evenly. We have adjusted this extra component by reproducing the high energy points of the AMS-02 positrons data. In this way, we reproduce the AMS-02 $e^+$ spectrum as a sum of secondary positrons + primary positrons from extra sources (likely PWN) and the AMS-02 $e^-$ spectrum as primary electrons injected by SNRs + secondary electrons from CR interactions + extra sources (right panel of Figure~\ref{fig:Electrons_posit}). We leave a detailed study of secondary leptons from {\tt FLUKA} for a separated paper.

\begin{figure}[!t]
	\centering
	\includegraphics[width=0.495\textwidth, height=0.22\textheight]{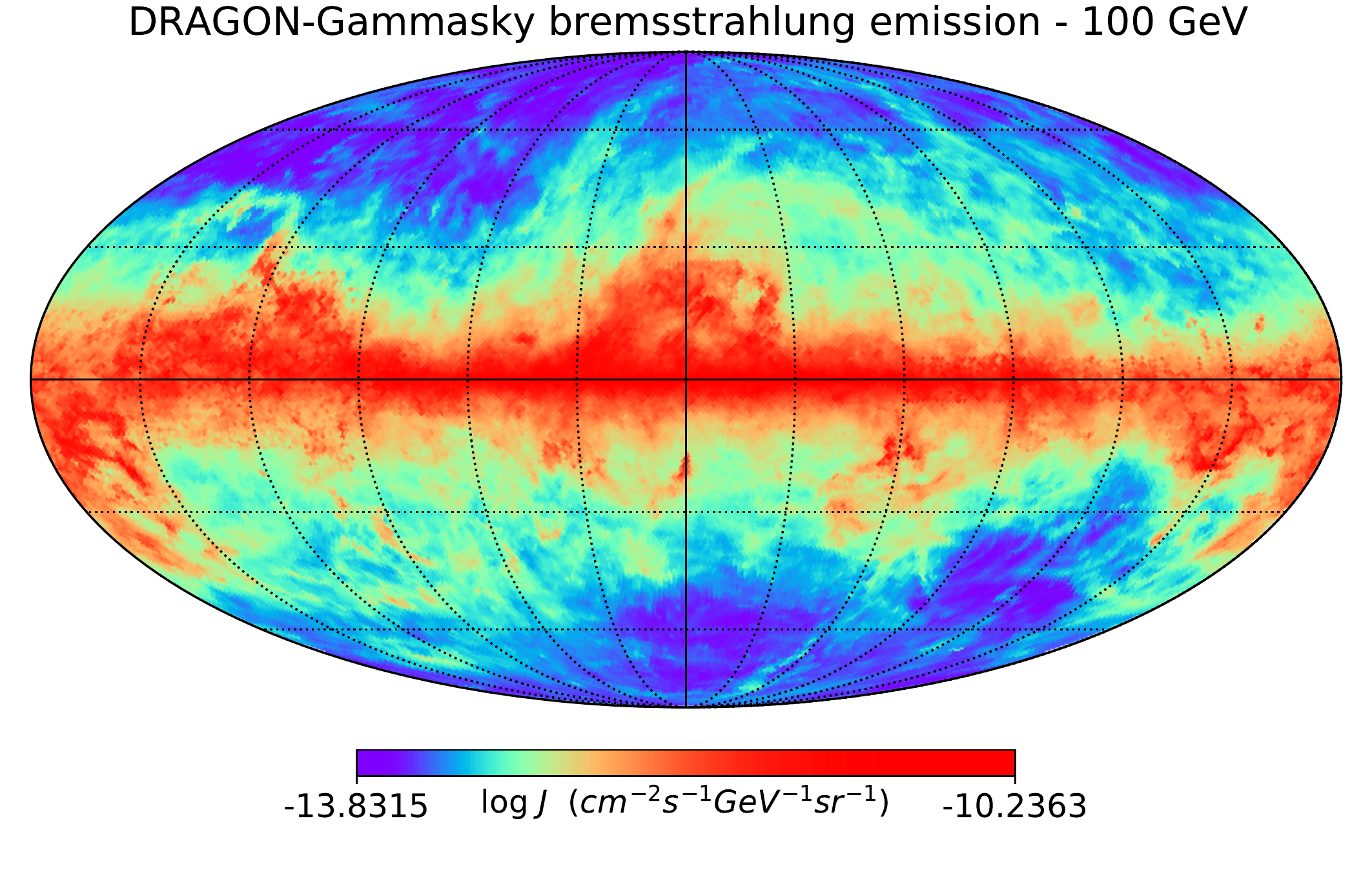}
	\includegraphics[width=0.495\textwidth, height=0.22\textheight]{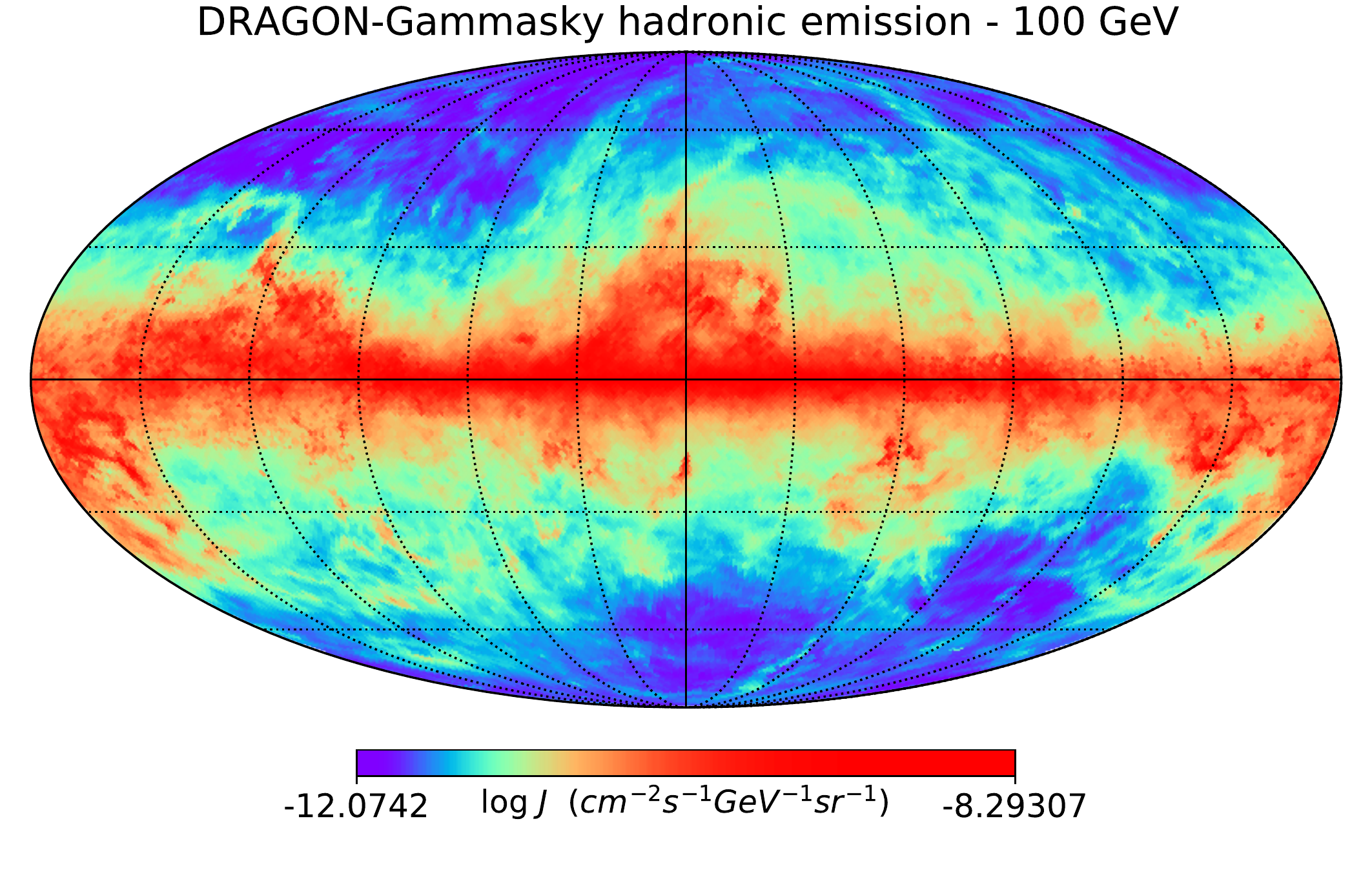}
	
	\vspace{0.35cm}
	
	\includegraphics[width=0.495\textwidth, height=0.22\textheight]{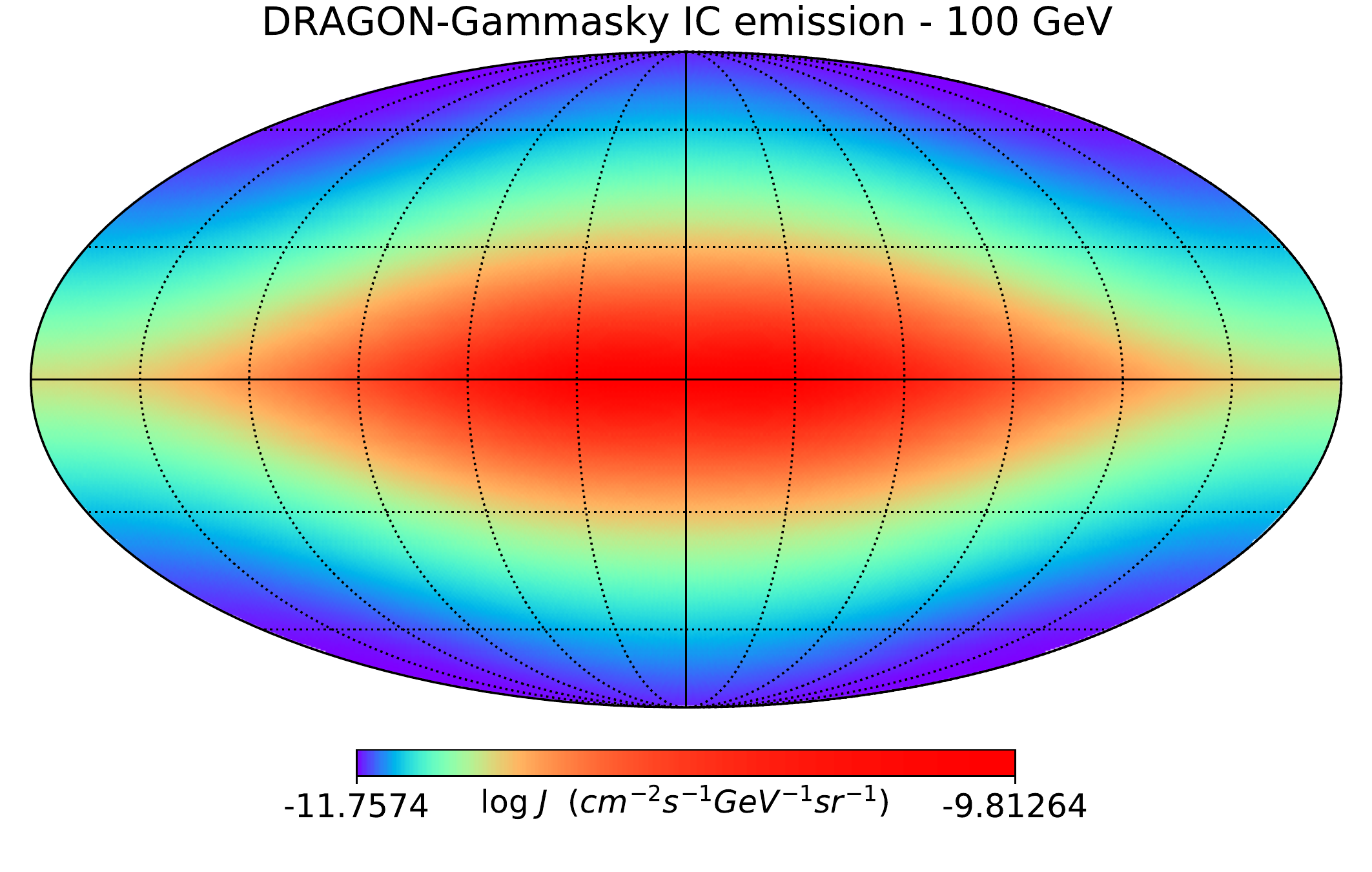}
	\includegraphics[width=0.495\textwidth, height=0.22\textheight]{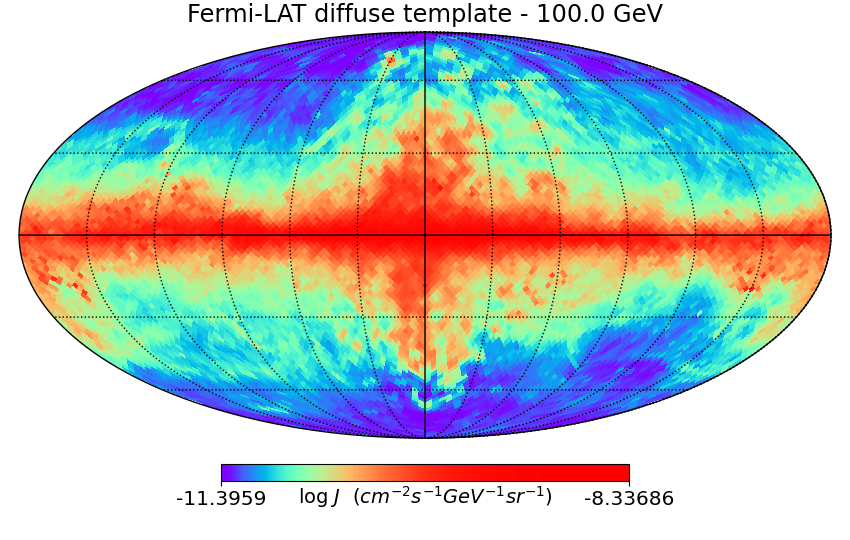}
	\caption{Gamma-ray intensity maps of the Galaxy, in Mollweide projection, for the propagation parameters inferred from the B/C analysis with the FLUKA cross sections. The upper-left map displays the bremsstrahlung emission, the upper-right map the gamma-ray emission from $\pi^0$ decays, and the lower-left map the IC emission. The lower-right map is the Fermi-LAT template for the diffuse emission, derived from the 8 first years of operation (PASS 8 data; $P8\_R3$, taken from {\tt gll\_iem\_v07.fits} \url{https://fermi.gsfc.nasa.gov/ssc/data/access/lat/BackgroundModels.html}).}
	\label{fig:sky_maps}
\end{figure}

Once the CR spectra of these species match local data, we compute the gamma-ray diffuse emission by using the {\tt Gammasky} code, a dedicated code able to calculate sky maps of various radiative processes in our Galaxy~\cite{Cirelli:2014lwa, Gaggero:2015xza, Mazziot}. Our computation is performed using the gas maps developed by the GALPROP team~\cite{Moskalenko:2001ya, Ackermann:2012pya}. Each of its main components and the total gamma-ray intensity predicted at $100\units{GeV}$ with the FLUKA cross sections are shown in Figure~\ref{fig:sky_maps}, together with the Fermi-LAT template of the gamma-ray diffuse emission (lower ight panel). Following the calculations performed in Ref.~\cite{Mazziot}, the IC, bremsstrahlung and hadronic cross sections of gamma-ray production are calculated with {\tt FLUKA} assuming an ISM composition with relative abundance of H : He : C : N : O : Ne : Mg : Si = $1:0.096:4.65\times 10^{-4}:8.3\times 10^{-5}:8.3\times 10^{-4}:1.3\times 10^{-4}:3.9\times 10^{-5}:3.69\times 10^{-5}$.

As we can see from Figure~\ref{fig:sky_maps}, the hadronic emission and bremsstrahlung intensity maps mainly track the gas composition, with the IC component principally probing the ISRF, which is more intense at the Galactic Center. The most recent Fermi-LAT template for the total diffuse emission {\tt gll\_iem\_v07.fits} (\url{https://fermi.gsfc.nasa.gov/ssc/data/access/lat/BackgroundModels.html}) is also displayed for comparison. In this template, the contributions from large scale structures, such as the gamma-ray Loop-I~\cite{Casandjian:2009wq} and the Fermi Bubbles~\cite{Fermi-LAT:2014sfa} are also added, which explains the differences observed at high latitudes around the Galactic Center.

Then, in order to avoid the uncertainties related to the gas and radiation fields, we turn our attention to the gamma-ray emissivity produced in the local ISM (LISM). In this context, the local HI emissivity is defined as the diffuse gamma-ray flux, coming from CR interactions with the local gas per unit of gas atom (HI). This quantity is inferred from the Fermi-LAT gamma-ray data~\cite{Casandjian:2015hja} and its determination is useful to help constraining different propagation models. The main contribution to the local HI emissivity at low energy is originated from Coulomb scattering of CRs with gas and electron bremsstrahlung; a further contribution, dominant above $\sim 100 \units{MeV}$, comes from the gamma-ray emissions of unstable particles formed via nuclear reactions (hadronic emissions). Formally, emissivity can be calculated by means of the following equation:
\begin{equation}
\frac{Q_S (E_S)}{n_{gas}} = 4\pi \int J(E) \frac{d\sigma(E_S | E)}{dE_S} dE , 
    \label{eq:emissivity}
\end{equation}
where $n_{gas}$ is the number of target gas atoms per unit volume and $\sigma (E_S | E)$ the cross section of production of any secondary particle, S, as for example neutrinos, gamma rays, positrons, etc. The ratio $Q_S ( E_S )/ n_{gas}$ is expressed in units of $\units{GeV} s^{-1}$. 

\begin{figure}[!t]
	\centering
	\includegraphics[width=0.75\textwidth, height=0.3\textheight]{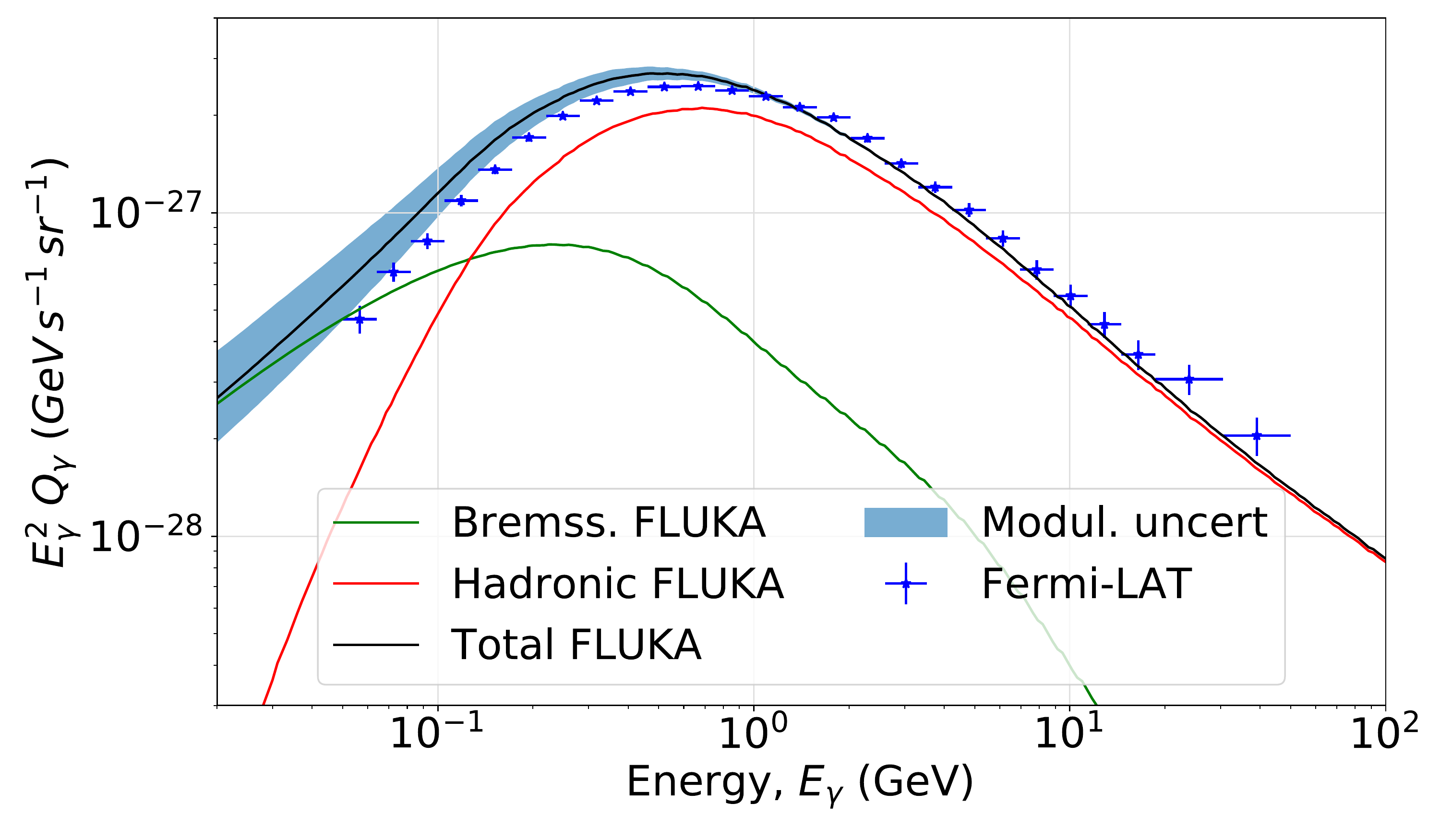}
	\caption{Local HI gamma-ray emissivity spectrum for the propagation parameters derived from the B/C analysis for the FLUKA cross sections. An uncertainty band, for the adjustment of the spectra of p, He and electrons, related to solar modulation uncertainty (variation of $\pm0.1$ in the Fisk potential) is also shown. Bremsstrahlung and hadronic emissions are also displayed to visualize the region of dominance of each component of the total emissivity. The experimental data is taken from Ref.~\cite{Casandjian:2015hja}.}
	\label{fig:Emiss}
\end{figure}

The computed emissivity spectra, calculated for the propagation parameters derived from the B/C analysis of the FLUKA cross sections, are shown for the diffusion coefficient given in Equation~\ref{eq:breakhyp} (diffusion hypothesis) in Figure~\ref{fig:Emiss}. They are compared to the emissivity observed by the Fermi-LAT, inferred from the recent PASS8 data set\footnote{These are public data taken from the  \href{https://fermi.gsfc.nasa.gov/ssc/data/access/lat/BackgroundModels.html}{Fermi background models website.}}.

In order to reproduce the local emissivity spectrum at low energies, we are forced to include a low energy break in the injection spectrum of electrons to prevent the bremsstrahlung emission to overpredict the data (or a too small solar modulation potential, which is not in agreement with the rest of observations from neutron monitors and other cosmic rays). This makes us use a doubly broken power-law injection spectrum for CR electrons, with a low energy break at $8$~GeV (note that this is the same break position as for protons), a high energy break at $65$~GeV and an exponential cut-off at around $1$~TeV (although the AMS-02 spectrum can be perfectly reproduced only with the high energy break and the cut-off). We remark that the high-energy break and cut-off used are in agreement with what other works have recently found~\cite{Fornieri:2019ddi}, however the low energy break is added ad hoc, without a clear theoretical motivation.

Finally, it is worth mentioning that, thanks to the use of the {\tt FLUKA} nuclear code to calculate the gamma-ray production cross sections, we are able to study the gamma-ray emission lines coming from the hadronic interactions of CR protons and helium with ISM nuclei (even being found as traces) such as $p + p \longrightarrow \gamma + X$ or $p + ^{14}$N$ \longrightarrow \gamma + X$. These lines could be detected in the MeV region by the future space experiments like e-Astrogam~\cite{DEANGELIS20181} and have been explored in very few works~\cite{gamma_lines}. 

\section{Conclusions}
\label{sec:Fconc}

In this work, the {\tt FLUKA} calculations for the inelastic and inclusive cross sections important for CR spallation reactions have been presented and widely discussed. We found that the inelastic cross sections for H, He, B and C are in good agreement with data and are quite similar to the CROSEC predictions at the level of a few percent. Nevertheless, elements with atomic number $Z\geq10$ (from Ne) start to differ significantly. 
Then, the total cumulative inclusive cross sections involving the production of Be, Li and B isotopes have been tested in comparison to the most recent dedicated parameterisations, showing a good general agreement, within $30\%$ discrepancies. 

Then, we evaluate the spectra of secondary CR species after implementing the FLUKA inelastic and inclusive cross sections in the {\tt DRAGON2} propagation code, in order to demonstrate that every CR observable can be reproduced with accuracy and at the level of precision of current CR parameterisations.  Notably, we have studied the B, Be and Li over C and O flux ratios under two different parameterisations of the diffusion coefficient, and the results of these fits and their confidence intervals have been reported. We obtain a best-fit value of the halo size, from the $^{10}$Be ratios of $H=7.5^{+1.13}_{-0.95} \units{kpc}$. All these ratios favor a change in the trend of the diffusion coefficient at sub-GeV energies and the predicted slope of the diffusion coefficient, $\delta$, varies from $\sim0.42$ (for B and Be ratios) to $\sim0.39$ (for Li ratios), which is very similar to what is found from the GALPROP and DRAGON2 parameterisations. We have also performed a combined analysis of these ratios, which includes also scale factors intended to renormalise the overall FLUKA cross sections of production of B, Be and Li. We have demonstrated that, with the use of these scale factors, we are able to match AMS-02 observations of these secondary-to-primary species within $5$\% errors above $5$~GeV. The scaled factors obtained are of $\sim 1.18$, $\sim0.94$ and $\sim0.93$ for B, Be and Li, respectively, which are below the typical average experimental uncertainties of cross sections measurements for the best known channels (i.e. those from $^{12}$C and $^{16}$O).

Finally, we have shown the derived diffuse gamma-ray sky maps and the local HI emissivity spectrum using the FLUKA cross sections for production of gamma rays, implemented in the {\tt Gammasky} code. The derived maps are shown to be coherent with the standard picture of the different mechanisms of gamma-ray emission (Inverse-Compton, bremsstrahlung and hadronic decay into photons) in the Galaxy. In addition, we demonstrate that the local HI emissivity spectrum can be perfectly reproduced from the models derived with the FLUKA cross sections, although with the need of inserting a low energy break in the injection spectrum of electrons.

In conclusion, the cross sections obtained with the {\tt FLUKA} code seem to be consistent with the current CR parameterisations and with experimental data, and are able to reproduce every CR observable with a low level of uncertainty. They represent a promising tool for future more precise CR studies.

\acknowledgments

P. De la Torre is supported by the Swedish National Space Agency under contract 117/19.
We acknowledge the {\tt FLUKA} collaboration for providing and supporting the code.
This work has been partly carried out using the RECAS computing infrastructure in Bari (\url{https://www.recas-bari.it/index.php/en/}). A particular acknowledgment goes to G. Donvito and A. Italiano for their valuable support. This project used computing resources from the Swedish National Infrastructure for Computing (SNIC) under project Nos. 2021/3-42 and 2021/6-326 partially funded by the Swedish Research Council through grant no. 2018-05973. Some of the discussions included in the text were triggered thanks to the Paris-Saclay Astroparticle Symposium 2021, with the support of the P2IO Laboratory of Excellence (programme “Investissements d’avenir” ANR-11-IDEX-0003-01 Paris-Saclay and ANR-10-LABX-0038), the P2I research departments of the Paris-Saclay university, as well as IJCLab, CEA, IPhT, APPEC, the IN2P3 master projet UCMN and EuCAPT.

\bibliographystyle{apsrev4-1}
\bibliography{biblio}

\begin{thebibliography}{123}%
\makeatletter
\providecommand \@ifxundefined [1]{%
 \@ifx{#1\undefined}
}%
\providecommand \@ifnum [1]{%
 \ifnum #1\expandafter \@firstoftwo
 \else \expandafter \@secondoftwo
 \fi
}%
\providecommand \@ifx [1]{%
 \ifx #1\expandafter \@firstoftwo
 \else \expandafter \@secondoftwo
 \fi
}%
\providecommand \natexlab [1]{#1}%
\providecommand \enquote  [1]{``#1''}%
\providecommand \bibnamefont  [1]{#1}%
\providecommand \bibfnamefont [1]{#1}%
\providecommand \citenamefont [1]{#1}%
\providecommand \href@noop [0]{\@secondoftwo}%
\providecommand \href [0]{\begingroup \@sanitize@url \@href}%
\providecommand \@href[1]{\@@startlink{#1}\@@href}%
\providecommand \@@href[1]{\endgroup#1\@@endlink}%
\providecommand \@sanitize@url [0]{\catcode `\\12\catcode `\$12\catcode
  `\&12\catcode `\#12\catcode `\^12\catcode `\_12\catcode `\%12\relax}%
\providecommand \@@startlink[1]{}%
\providecommand \@@endlink[0]{}%
\providecommand \url  [0]{\begingroup\@sanitize@url \@url }%
\providecommand \@url [1]{\endgroup\@href {#1}{\urlprefix }}%
\providecommand \urlprefix  [0]{URL }%
\providecommand \Eprint [0]{\href }%
\providecommand \doibase [0]{http://dx.doi.org/}%
\providecommand \selectlanguage [0]{\@gobble}%
\providecommand \bibinfo  [0]{\@secondoftwo}%
\providecommand \bibfield  [0]{\@secondoftwo}%
\providecommand \translation [1]{[#1]}%
\providecommand \BibitemOpen [0]{}%
\providecommand \bibitemStop [0]{}%
\providecommand \bibitemNoStop [0]{.\EOS\space}%
\providecommand \EOS [0]{\spacefactor3000\relax}%
\providecommand \BibitemShut  [1]{\csname bibitem#1\endcsname}%
\let\auto@bib@innerbib\@empty
\bibitem [{\citenamefont {{Somerville}}\ and\ \citenamefont
  {{Dav{\'e}}}(2015)}]{GalFormation}%
  \BibitemOpen
  \bibfield  {author} {\bibinfo {author} {\bibfnamefont {R.~S.}\ \bibnamefont
  {{Somerville}}}\ and\ \bibinfo {author} {\bibfnamefont {R.}~\bibnamefont
  {{Dav{\'e}}}},\ }\href {\doibase 10.1146/annurev-astro-082812-140951}
  {\bibfield  {journal} {\bibinfo  {journal} {araa}\ }\textbf {\bibinfo
  {volume} {53}},\ \bibinfo {pages} {51} (\bibinfo {year} {2015})},\ \Eprint
  {http://arxiv.org/abs/1412.2712} {arXiv:1412.2712 [astro-ph.GA]} \BibitemShut
  {NoStop}%
\bibitem [{\citenamefont {Hopkins}\ \emph {et~al.}(2019)\citenamefont
  {Hopkins}, \citenamefont {Chan}, \citenamefont {Garrison-Kimmel},
  \citenamefont {Ji}, \citenamefont {Su}, \citenamefont {Hummels},
  \citenamefont {KereÅ¡}, \citenamefont {Quataert},\ and\ \citenamefont
  {Faucher-GiguÃ¨re}}]{Hopkins}%
  \BibitemOpen
  \bibfield  {author} {\bibinfo {author} {\bibfnamefont {P.~F.}\ \bibnamefont
  {Hopkins}}, \bibinfo {author} {\bibfnamefont {T.~K.}\ \bibnamefont {Chan}},
  \bibinfo {author} {\bibfnamefont {S.}~\bibnamefont {Garrison-Kimmel}},
  \bibinfo {author} {\bibfnamefont {S.}~\bibnamefont {Ji}}, \bibinfo {author}
  {\bibfnamefont {K.-Y.}\ \bibnamefont {Su}}, \bibinfo {author} {\bibfnamefont
  {C.~B.}\ \bibnamefont {Hummels}}, \bibinfo {author} {\bibfnamefont
  {D.}~\bibnamefont {KereÅ¡}}, \bibinfo {author} {\bibfnamefont
  {E.}~\bibnamefont {Quataert}}, \ and\ \bibinfo {author} {\bibfnamefont
  {C.-A.}\ \bibnamefont {Faucher-GiguÃ¨re}},\ }\href {\doibase
  10.1093/mnras/stz3321} {\bibfield  {journal} {\bibinfo  {journal} {Monthly
  Notices of the Royal Astronomical Society}\ }\textbf {\bibinfo {volume}
  {492}},\ \bibinfo {pages} {3465} (\bibinfo {year} {2019})},\ \Eprint
  {http://arxiv.org/abs/https://academic.oup.com/mnras/article-pdf/492/3/3465/31952706/stz3321.pdf}
  {https://academic.oup.com/mnras/article-pdf/492/3/3465/31952706/stz3321.pdf}
  \BibitemShut {NoStop}%
\bibitem [{\citenamefont {Grenier}\ \emph {et~al.}(2015)\citenamefont
  {Grenier}, \citenamefont {Black},\ and\ \citenamefont {Strong}}]{CR-Galaxy}%
  \BibitemOpen
  \bibfield  {author} {\bibinfo {author} {\bibfnamefont {I.~A.}\ \bibnamefont
  {Grenier}}, \bibinfo {author} {\bibfnamefont {J.~H.}\ \bibnamefont {Black}},
  \ and\ \bibinfo {author} {\bibfnamefont {A.~W.}\ \bibnamefont {Strong}},\
  }\href {\doibase 10.1146/annurev-astro-082214-122457} {\bibfield  {journal}
  {\bibinfo  {journal} {Annual Review of Astronomy and Astrophysics}\ }\textbf
  {\bibinfo {volume} {53}},\ \bibinfo {pages} {199} (\bibinfo {year} {2015})},\
  \Eprint
  {http://arxiv.org/abs/https://doi.org/10.1146/annurev-astro-082214-122457}
  {https://doi.org/10.1146/annurev-astro-082214-122457} \BibitemShut {NoStop}%
\bibitem [{\citenamefont {Farber}\ \emph {et~al.}(2018)\citenamefont {Farber},
  \citenamefont {Ruszkowski}, \citenamefont {Yang},\ and\ \citenamefont
  {Zweibel}}]{Farber_2018}%
  \BibitemOpen
  \bibfield  {author} {\bibinfo {author} {\bibfnamefont {R.}~\bibnamefont
  {Farber}}, \bibinfo {author} {\bibfnamefont {M.}~\bibnamefont {Ruszkowski}},
  \bibinfo {author} {\bibfnamefont {H.-Y.~K.}\ \bibnamefont {Yang}}, \ and\
  \bibinfo {author} {\bibfnamefont {E.~G.}\ \bibnamefont {Zweibel}},\ }\href
  {\doibase 10.3847/1538-4357/aab26d} {\bibfield  {journal} {\bibinfo
  {journal} {The Astrophysical Journal}\ }\textbf {\bibinfo {volume} {856}},\
  \bibinfo {pages} {112} (\bibinfo {year} {2018})}\BibitemShut {NoStop}%
\bibitem [{\citenamefont {Gabici}\ \emph {et~al.}(2019)\citenamefont {Gabici},
  \citenamefont {Evoli}, \citenamefont {Gaggero}, \citenamefont {Lipari},
  \citenamefont {Mertsch}, \citenamefont {Orlando}, \citenamefont {Strong},\
  and\ \citenamefont {Vittino}}]{Gabici:2019jvz}%
  \BibitemOpen
  \bibfield  {author} {\bibinfo {author} {\bibfnamefont {S.}~\bibnamefont
  {Gabici}}, \bibinfo {author} {\bibfnamefont {C.}~\bibnamefont {Evoli}},
  \bibinfo {author} {\bibfnamefont {D.}~\bibnamefont {Gaggero}}, \bibinfo
  {author} {\bibfnamefont {P.}~\bibnamefont {Lipari}}, \bibinfo {author}
  {\bibfnamefont {P.}~\bibnamefont {Mertsch}}, \bibinfo {author} {\bibfnamefont
  {E.}~\bibnamefont {Orlando}}, \bibinfo {author} {\bibfnamefont
  {A.}~\bibnamefont {Strong}}, \ and\ \bibinfo {author} {\bibfnamefont
  {A.}~\bibnamefont {Vittino}},\ }\href {\doibase 10.1142/S0218271819300222}
  {\bibfield  {journal} {\bibinfo  {journal} {Int. J. Mod. Phys. D}\ }\textbf
  {\bibinfo {volume} {28}},\ \bibinfo {pages} {1930022} (\bibinfo {year}
  {2019})},\ \Eprint {http://arxiv.org/abs/1903.11584} {arXiv:1903.11584
  [astro-ph.HE]} \BibitemShut {NoStop}%
\bibitem [{\citenamefont {{Ginzburg}}\ and\ \citenamefont
  {{Syrovatskii}}(1969)}]{Ginz&Syr}%
  \BibitemOpen
  \bibfield  {author} {\bibinfo {author} {\bibfnamefont {V.~L.}\ \bibnamefont
  {{Ginzburg}}}\ and\ \bibinfo {author} {\bibfnamefont {S.~I.}\ \bibnamefont
  {{Syrovatskii}}},\ }\href@noop {} {\emph {\bibinfo {title} {Topics in
  Astrophysics and Space Physics, New York: Gordon and Breach, 1969}}}\
  (\bibinfo  {publisher} {Gordon \& Breach Publishing Group},\ \bibinfo {year}
  {1969})\BibitemShut {NoStop}%
\bibitem [{\citenamefont {Berezinskii}\ \emph {et~al.}(1990)\citenamefont
  {Berezinskii}, \citenamefont {Dogiel}, \citenamefont {Bulanov},\ and\
  \citenamefont {Ginzburg}}]{berezinskii1990astrophysics}%
  \BibitemOpen
  \bibfield  {author} {\bibinfo {author} {\bibfnamefont {V.}~\bibnamefont
  {Berezinskii}}, \bibinfo {author} {\bibfnamefont {V.}~\bibnamefont {Dogiel}},
  \bibinfo {author} {\bibfnamefont {S.}~\bibnamefont {Bulanov}}, \ and\
  \bibinfo {author} {\bibfnamefont {V.~L.}\ \bibnamefont {Ginzburg}},\
  }\href@noop {} {\emph {\bibinfo {title} {Astrophysics of cosmic rays}}},\
  Vol.\ \bibinfo {volume} {393}\ (\bibinfo  {publisher} {North-Holland
  Amsterdam},\ \bibinfo {year} {1990})\BibitemShut {NoStop}%
\bibitem [{\citenamefont {De~La Torre~Luque}\ \emph {et~al.}(2021)\citenamefont
  {De~La Torre~Luque}, \citenamefont {Mazziotta}, \citenamefont {Loparco},
  \citenamefont {Gargano},\ and\ \citenamefont {Serini}}]{Luque:2021joz}%
  \BibitemOpen
  \bibfield  {author} {\bibinfo {author} {\bibfnamefont {P.}~\bibnamefont
  {De~La Torre~Luque}}, \bibinfo {author} {\bibfnamefont {M.~N.}\ \bibnamefont
  {Mazziotta}}, \bibinfo {author} {\bibfnamefont {F.}~\bibnamefont {Loparco}},
  \bibinfo {author} {\bibfnamefont {F.}~\bibnamefont {Gargano}}, \ and\
  \bibinfo {author} {\bibfnamefont {D.}~\bibnamefont {Serini}},\ }\href
  {\doibase 10.1088/1475-7516/2021/03/099} {\bibfield  {journal} {\bibinfo
  {journal} {JCAP}\ }\textbf {\bibinfo {volume} {03}},\ \bibinfo {pages} {099}
  (\bibinfo {year} {2021})},\ \Eprint {http://arxiv.org/abs/2101.01547}
  {arXiv:2101.01547 [astro-ph.HE]} \BibitemShut {NoStop}%
\bibitem [{\citenamefont {Evoli}\ \emph {et~al.}(2020)\citenamefont {Evoli},
  \citenamefont {Morlino}, \citenamefont {Blasi},\ and\ \citenamefont
  {Aloisio}}]{CarmeloBeB}%
  \BibitemOpen
  \bibfield  {author} {\bibinfo {author} {\bibfnamefont {C.}~\bibnamefont
  {Evoli}}, \bibinfo {author} {\bibfnamefont {G.}~\bibnamefont {Morlino}},
  \bibinfo {author} {\bibfnamefont {P.}~\bibnamefont {Blasi}}, \ and\ \bibinfo
  {author} {\bibfnamefont {R.}~\bibnamefont {Aloisio}},\ }\href {\doibase
  10.1103/PhysRevD.101.023013} {\bibfield  {journal} {\bibinfo  {journal}
  {Phys.\ Rev.\ D}\ }\textbf {\bibinfo {volume} {101}},\ \bibinfo {pages}
  {023013} (\bibinfo {year} {2020})},\ \Eprint
  {http://arxiv.org/abs/1910.04113} {arXiv:1910.04113 [astro-ph.HE]}
  \BibitemShut {NoStop}%
\bibitem [{\citenamefont {{Weinrich}}\ \emph
  {et~al.}(2020{\natexlab{a}})\citenamefont {{Weinrich}}, \citenamefont
  {{Boudaud}}, \citenamefont {{Derome}}, \citenamefont {{G{\'e}nolini}},
  \citenamefont {{Lavalle}}, \citenamefont {{Maurin}}, \citenamefont
  {{Salati}}, \citenamefont {{Serpico}},\ and\ \citenamefont
  {{Weymann-Despres}}}]{Weinrich_halo}%
  \BibitemOpen
  \bibfield  {author} {\bibinfo {author} {\bibfnamefont {N.}~\bibnamefont
  {{Weinrich}}}, \bibinfo {author} {\bibfnamefont {M.}~\bibnamefont
  {{Boudaud}}}, \bibinfo {author} {\bibfnamefont {L.}~\bibnamefont {{Derome}}},
  \bibinfo {author} {\bibfnamefont {Y.}~\bibnamefont {{G{\'e}nolini}}},
  \bibinfo {author} {\bibfnamefont {J.}~\bibnamefont {{Lavalle}}}, \bibinfo
  {author} {\bibfnamefont {D.}~\bibnamefont {{Maurin}}}, \bibinfo {author}
  {\bibfnamefont {P.}~\bibnamefont {{Salati}}}, \bibinfo {author}
  {\bibfnamefont {P.}~\bibnamefont {{Serpico}}}, \ and\ \bibinfo {author}
  {\bibfnamefont {G.}~\bibnamefont {{Weymann-Despres}}},\ }\href {\doibase
  10.1051/0004-6361/202038064} {\bibfield  {journal} {\bibinfo  {journal} {The
  Astrophysical Journal}\ }\textbf {\bibinfo {volume} {639}},\ \bibinfo {eid}
  {A74} (\bibinfo {year} {2020}{\natexlab{a}})},\ \Eprint
  {http://arxiv.org/abs/2004.00441} {arXiv:2004.00441 [astro-ph.HE]}
  \BibitemShut {NoStop}%
\bibitem [{\citenamefont {Biswas}\ and\ \citenamefont
  {Gupta}(2018)}]{biswas2018constraining}%
  \BibitemOpen
  \bibfield  {author} {\bibinfo {author} {\bibfnamefont {S.}~\bibnamefont
  {Biswas}}\ and\ \bibinfo {author} {\bibfnamefont {N.}~\bibnamefont {Gupta}},\
  }\href {\doibase 10.1088/1475-7516/2018/07/063} {\bibfield  {journal}
  {\bibinfo  {journal} {JCAP}\ }\textbf {\bibinfo {volume} {07}},\ \bibinfo
  {pages} {063} (\bibinfo {year} {2018})},\ \Eprint
  {http://arxiv.org/abs/1802.03538} {arXiv:1802.03538 [astro-ph.HE]}
  \BibitemShut {NoStop}%
\bibitem [{\citenamefont {Moskalenko}\ and\ \citenamefont
  {Strong}(2000)}]{moskalenko2000diffuse}%
  \BibitemOpen
  \bibfield  {author} {\bibinfo {author} {\bibfnamefont {I.}~\bibnamefont
  {Moskalenko}}\ and\ \bibinfo {author} {\bibfnamefont {A.}~\bibnamefont
  {Strong}},\ }\href {\doibase 10.1023/A:1002604831398} {\bibfield  {journal}
  {\bibinfo  {journal} {Astrophys. Space Sci.}\ }\textbf {\bibinfo {volume}
  {272}},\ \bibinfo {pages} {247} (\bibinfo {year} {2000})},\ \Eprint
  {http://arxiv.org/abs/astro-ph/9908032} {arXiv:astro-ph/9908032} \BibitemShut
  {NoStop}%
\bibitem [{\citenamefont {Bringmann}\ \emph {et~al.}(2012)\citenamefont
  {Bringmann}, \citenamefont {Donato},\ and\ \citenamefont
  {Lineros}}]{bringmann2012radio}%
  \BibitemOpen
  \bibfield  {author} {\bibinfo {author} {\bibfnamefont {T.}~\bibnamefont
  {Bringmann}}, \bibinfo {author} {\bibfnamefont {F.}~\bibnamefont {Donato}}, \
  and\ \bibinfo {author} {\bibfnamefont {R.~A.}\ \bibnamefont {Lineros}},\
  }\href {\doibase 10.1088/1475-7516/2012/01/049} {\bibfield  {journal}
  {\bibinfo  {journal} {JCAP}\ }\textbf {\bibinfo {volume} {01}},\ \bibinfo
  {pages} {049} (\bibinfo {year} {2012})},\ \Eprint
  {http://arxiv.org/abs/1106.4821} {arXiv:1106.4821 [astro-ph.GA]} \BibitemShut
  {NoStop}%
\bibitem [{\citenamefont {Stephens}\ and\ \citenamefont
  {Streitmatter}(1998)}]{stephens1998cosmic}%
  \BibitemOpen
  \bibfield  {author} {\bibinfo {author} {\bibfnamefont {S.}~\bibnamefont
  {Stephens}}\ and\ \bibinfo {author} {\bibfnamefont {R.}~\bibnamefont
  {Streitmatter}},\ }\href {\doibase 10.1086/306142} {\bibfield  {journal}
  {\bibinfo  {journal} {The Astrophysical Journal}\ }\textbf {\bibinfo {volume}
  {505}},\ \bibinfo {pages} {266} (\bibinfo {year} {1998})}\BibitemShut
  {NoStop}%
\bibitem [{\citenamefont {Tjus}\ and\ \citenamefont
  {Merten}(2020)}]{tjus2020closing}%
  \BibitemOpen
  \bibfield  {author} {\bibinfo {author} {\bibfnamefont {J.~B.}\ \bibnamefont
  {Tjus}}\ and\ \bibinfo {author} {\bibfnamefont {L.}~\bibnamefont {Merten}},\
  }\href@noop {} {\enquote {\bibinfo {title} {Closing in on the origin of
  galactic cosmic rays using multimessenger information},}\ } (\bibinfo {year}
  {2020}),\ \Eprint {http://arxiv.org/abs/2002.00964} {arXiv:2002.00964
  [astro-ph.HE]} \BibitemShut {NoStop}%
\bibitem [{\citenamefont {Reinert}\ and\ \citenamefont
  {Winkler}(2018)}]{reinert2018precision}%
  \BibitemOpen
  \bibfield  {author} {\bibinfo {author} {\bibfnamefont {A.}~\bibnamefont
  {Reinert}}\ and\ \bibinfo {author} {\bibfnamefont {M.~W.}\ \bibnamefont
  {Winkler}},\ }\href {\doibase 10.1088/1475-7516/2018/01/055} {\bibfield
  {journal} {\bibinfo  {journal} {JCAP}\ }\textbf {\bibinfo {volume} {01}},\
  \bibinfo {pages} {055} (\bibinfo {year} {2018})},\ \Eprint
  {http://arxiv.org/abs/1712.00002} {arXiv:1712.00002 [astro-ph.HE]}
  \BibitemShut {NoStop}%
\bibitem [{\citenamefont {Derome}\ \emph {et~al.}(2020)\citenamefont {Derome},
  \citenamefont {Maurin}, \citenamefont {Salati}, \citenamefont {Boudaud},
  \citenamefont {Génolini},\ and\ \citenamefont {Kunzé}}]{derome2019fitting}%
  \BibitemOpen
  \bibfield  {author} {\bibinfo {author} {\bibfnamefont {L.}~\bibnamefont
  {Derome}}, \bibinfo {author} {\bibfnamefont {D.}~\bibnamefont {Maurin}},
  \bibinfo {author} {\bibfnamefont {P.}~\bibnamefont {Salati}}, \bibinfo
  {author} {\bibfnamefont {M.}~\bibnamefont {Boudaud}}, \bibinfo {author}
  {\bibfnamefont {Y.}~\bibnamefont {Génolini}}, \ and\ \bibinfo {author}
  {\bibfnamefont {P.}~\bibnamefont {Kunzé}},\ }\href {\doibase
  10.22323/1.358.0054} {\bibfield  {journal} {\bibinfo  {journal} {PoS}\
  }\textbf {\bibinfo {volume} {ICRC2019}},\ \bibinfo {pages} {054} (\bibinfo
  {year} {2020})}\BibitemShut {NoStop}%
\bibitem [{\citenamefont {{Weinrich}}\ \emph
  {et~al.}(2020{\natexlab{b}})\citenamefont {{Weinrich}}, \citenamefont
  {{G{\'e}nolini}}, \citenamefont {{Boudaud}}, \citenamefont {{Derome}},\ and\
  \citenamefont {{Maurin}}}]{Weinrich_combined}%
  \BibitemOpen
  \bibfield  {author} {\bibinfo {author} {\bibfnamefont {N.}~\bibnamefont
  {{Weinrich}}}, \bibinfo {author} {\bibfnamefont {Y.}~\bibnamefont
  {{G{\'e}nolini}}}, \bibinfo {author} {\bibfnamefont {M.}~\bibnamefont
  {{Boudaud}}}, \bibinfo {author} {\bibfnamefont {L.}~\bibnamefont {{Derome}}},
  \ and\ \bibinfo {author} {\bibfnamefont {D.}~\bibnamefont {{Maurin}}},\
  }\href {\doibase 10.1051/0004-6361/202037875} {\bibfield  {journal} {\bibinfo
   {journal} {The Astrophysical Journal}\ }\textbf {\bibinfo {volume} {639}},\
  \bibinfo {eid} {A131} (\bibinfo {year} {2020}{\natexlab{b}})},\ \Eprint
  {http://arxiv.org/abs/2002.11406} {arXiv:2002.11406 [astro-ph.HE]}
  \BibitemShut {NoStop}%
\bibitem [{\citenamefont {Korsmeier}\ and\ \citenamefont
  {Cuoco}(2021{\natexlab{a}})}]{Korsmeier:2021brc}%
  \BibitemOpen
  \bibfield  {author} {\bibinfo {author} {\bibfnamefont {M.}~\bibnamefont
  {Korsmeier}}\ and\ \bibinfo {author} {\bibfnamefont {A.}~\bibnamefont
  {Cuoco}},\ }\href {\doibase 10.1103/PhysRevD.103.103016} {\bibfield
  {journal} {\bibinfo  {journal} {Phys. Rev. D}\ }\textbf {\bibinfo {volume}
  {103}},\ \bibinfo {pages} {103016} (\bibinfo {year} {2021}{\natexlab{a}})},\
  \Eprint {http://arxiv.org/abs/2103.09824} {arXiv:2103.09824 [astro-ph.HE]}
  \BibitemShut {NoStop}%
\bibitem [{\citenamefont {la~Torre~Luque}\ \emph {et~al.}(2020)\citenamefont
  {la~Torre~Luque}, \citenamefont {Gargano}, \citenamefont {Loparco},
  \citenamefont {Mazziotta},\ and\ \citenamefont {Serini}}]{ICPPA_Pedro}%
  \BibitemOpen
  \bibfield  {author} {\bibinfo {author} {\bibfnamefont {P.~D.}\ \bibnamefont
  {la~Torre~Luque}}, \bibinfo {author} {\bibfnamefont {F.}~\bibnamefont
  {Gargano}}, \bibinfo {author} {\bibfnamefont {F.}~\bibnamefont {Loparco}},
  \bibinfo {author} {\bibfnamefont {M.~N.}\ \bibnamefont {Mazziotta}}, \ and\
  \bibinfo {author} {\bibfnamefont {D.}~\bibnamefont {Serini}},\ }\href
  {\doibase 10.1088/1742-6596/1690/1/012010} {\bibfield  {journal} {\bibinfo
  {journal} {Journal of Physics: Conference Series}\ }\textbf {\bibinfo
  {volume} {1690}},\ \bibinfo {pages} {012010} (\bibinfo {year}
  {2020})}\BibitemShut {NoStop}%
\bibitem [{\citenamefont {Gondolo}(2014)}]{GONDOLO2014175}%
  \BibitemOpen
  \bibfield  {author} {\bibinfo {author} {\bibfnamefont {P.}~\bibnamefont
  {Gondolo}},\ }\href {\doibase https://doi.org/10.1016/j.nds.2014.07.039}
  {\bibfield  {journal} {\bibinfo  {journal} {Nuclear Data Sheets}\ }\textbf
  {\bibinfo {volume} {120}},\ \bibinfo {pages} {175} (\bibinfo {year}
  {2014})}\BibitemShut {NoStop}%
\bibitem [{\citenamefont {Genolini}\ \emph {et~al.}(2018)\citenamefont
  {Genolini}, \citenamefont {Maurin}, \citenamefont {Moskalenko},\ and\
  \citenamefont {Unger}}]{GenoliniRanking}%
  \BibitemOpen
  \bibfield  {author} {\bibinfo {author} {\bibfnamefont {Y.}~\bibnamefont
  {Genolini}}, \bibinfo {author} {\bibfnamefont {D.}~\bibnamefont {Maurin}},
  \bibinfo {author} {\bibfnamefont {I.~V.}\ \bibnamefont {Moskalenko}}, \ and\
  \bibinfo {author} {\bibfnamefont {M.}~\bibnamefont {Unger}},\ }\href
  {\doibase 10.1103/PhysRevC.98.034611} {\bibfield  {journal} {\bibinfo
  {journal} {Phys. Rev.}\ }\textbf {\bibinfo {volume} {C98}},\ \bibinfo {pages}
  {034611} (\bibinfo {year} {2018})},\ \Eprint
  {http://arxiv.org/abs/1803.04686} {arXiv:1803.04686 [astro-ph.HE]}
  \BibitemShut {NoStop}%
\bibitem [{\citenamefont {Tomassetti}(2015)}]{Tomassetti:2015nha}%
  \BibitemOpen
  \bibfield  {author} {\bibinfo {author} {\bibfnamefont {N.}~\bibnamefont
  {Tomassetti}},\ }\href {\doibase 10.1103/PhysRevC.92.045808} {\bibfield
  {journal} {\bibinfo  {journal} {Phys. Rev. C}\ }\textbf {\bibinfo {volume}
  {92}},\ \bibinfo {pages} {045808} (\bibinfo {year} {2015})},\ \Eprint
  {http://arxiv.org/abs/1509.05776} {arXiv:1509.05776 [astro-ph.HE]}
  \BibitemShut {NoStop}%
\bibitem [{\citenamefont {Mashnik}(2011)}]{Mashnik:2010xh}%
  \BibitemOpen
  \bibfield  {author} {\bibinfo {author} {\bibfnamefont {S.~G.}\ \bibnamefont
  {Mashnik}},\ }\href {\doibase 10.1140/epjp/i2011-11049-1} {\bibfield
  {journal} {\bibinfo  {journal} {Eur. Phys. J. Plus}\ }\textbf {\bibinfo
  {volume} {126}},\ \bibinfo {pages} {49} (\bibinfo {year} {2011})},\ \Eprint
  {http://arxiv.org/abs/1011.4978} {arXiv:1011.4978 [nucl-th]} \BibitemShut
  {NoStop}%
\bibitem [{\citenamefont {Mashnik}\ \emph {et~al.}(2017)\citenamefont
  {Mashnik}, \citenamefont {Kerby}, \citenamefont {Gudima}, \citenamefont
  {Sierk}, \citenamefont {Bull},\ and\ \citenamefont
  {James}}]{Mashnik:2016dmf}%
  \BibitemOpen
  \bibfield  {author} {\bibinfo {author} {\bibfnamefont {S.~G.}\ \bibnamefont
  {Mashnik}}, \bibinfo {author} {\bibfnamefont {L.~M.}\ \bibnamefont {Kerby}},
  \bibinfo {author} {\bibfnamefont {K.~K.}\ \bibnamefont {Gudima}}, \bibinfo
  {author} {\bibfnamefont {A.~J.}\ \bibnamefont {Sierk}}, \bibinfo {author}
  {\bibfnamefont {J.~S.}\ \bibnamefont {Bull}}, \ and\ \bibinfo {author}
  {\bibfnamefont {M.~R.}\ \bibnamefont {James}},\ }\href {\doibase
  10.1103/PhysRevC.95.034613} {\bibfield  {journal} {\bibinfo  {journal} {Phys.
  Rev. C}\ }\textbf {\bibinfo {volume} {95}},\ \bibinfo {pages} {034613}
  (\bibinfo {year} {2017})},\ \Eprint {http://arxiv.org/abs/1607.02506}
  {arXiv:1607.02506 [nucl-th]} \BibitemShut {NoStop}%
\bibitem [{\citenamefont {Kerby}\ and\ \citenamefont
  {Mashnik}(2015)}]{Kerby:2015qaa}%
  \BibitemOpen
  \bibfield  {author} {\bibinfo {author} {\bibfnamefont {L.~M.}\ \bibnamefont
  {Kerby}}\ and\ \bibinfo {author} {\bibfnamefont {S.~G.}\ \bibnamefont
  {Mashnik}},\ }\href {\doibase 10.1016/j.nimb.2015.04.057} {\bibfield
  {journal} {\bibinfo  {journal} {Nucl. Instrum. Meth. B}\ }\textbf {\bibinfo
  {volume} {356-357}},\ \bibinfo {pages} {135} (\bibinfo {year} {2015})},\
  \Eprint {http://arxiv.org/abs/1505.00842} {arXiv:1505.00842 [nucl-th]}
  \BibitemShut {NoStop}%
\bibitem [{\citenamefont {Pshenichnov}\ \emph {et~al.}(2010)\citenamefont
  {Pshenichnov}, \citenamefont {Botvina}, \citenamefont {Mishustin},\ and\
  \citenamefont {Greiner}}]{PSHENICHNOV2010604}%
  \BibitemOpen
  \bibfield  {author} {\bibinfo {author} {\bibfnamefont {I.}~\bibnamefont
  {Pshenichnov}}, \bibinfo {author} {\bibfnamefont {A.}~\bibnamefont
  {Botvina}}, \bibinfo {author} {\bibfnamefont {I.}~\bibnamefont {Mishustin}},
  \ and\ \bibinfo {author} {\bibfnamefont {W.}~\bibnamefont {Greiner}},\ }\href
  {\doibase https://doi.org/10.1016/j.nimb.2009.12.023} {\bibfield  {journal}
  {\bibinfo  {journal} {Nuclear Instruments and Methods in Physics Research
  Section B: Beam Interactions with Materials and Atoms}\ }\textbf {\bibinfo
  {volume} {268}},\ \bibinfo {pages} {604} (\bibinfo {year}
  {2010})}\BibitemShut {NoStop}%
\bibitem [{\citenamefont {Niita}\ \emph {et~al.}(1995)\citenamefont {Niita},
  \citenamefont {Chiba}, \citenamefont {Maruyama}, \citenamefont {Maruyama},
  \citenamefont {Takada}, \citenamefont {Fukahori}, \citenamefont {Nakahara},\
  and\ \citenamefont {Iwamoto}}]{PHIS}%
  \BibitemOpen
  \bibfield  {author} {\bibinfo {author} {\bibfnamefont {K.}~\bibnamefont
  {Niita}}, \bibinfo {author} {\bibfnamefont {S.}~\bibnamefont {Chiba}},
  \bibinfo {author} {\bibfnamefont {T.}~\bibnamefont {Maruyama}}, \bibinfo
  {author} {\bibfnamefont {T.}~\bibnamefont {Maruyama}}, \bibinfo {author}
  {\bibfnamefont {H.}~\bibnamefont {Takada}}, \bibinfo {author} {\bibfnamefont
  {T.}~\bibnamefont {Fukahori}}, \bibinfo {author} {\bibfnamefont
  {Y.}~\bibnamefont {Nakahara}}, \ and\ \bibinfo {author} {\bibfnamefont
  {A.}~\bibnamefont {Iwamoto}},\ }\href {\doibase 10.1103/PhysRevC.52.2620}
  {\bibfield  {journal} {\bibinfo  {journal} {Phys. Rev. C}\ }\textbf {\bibinfo
  {volume} {52}},\ \bibinfo {pages} {2620} (\bibinfo {year}
  {1995})}\BibitemShut {NoStop}%
\bibitem [{\citenamefont {Sihver}\ \emph {et~al.}(2010)\citenamefont {Sihver},
  \citenamefont {Sato}, \citenamefont {Gustafsson}, \citenamefont {Mancusi},
  \citenamefont {Iwase}, \citenamefont {Niita}, \citenamefont {Nakashima},
  \citenamefont {Sakamoto}, \citenamefont {Iwamoto},\ and\ \citenamefont
  {Matsuda}}]{SIHVER2010892}%
  \BibitemOpen
  \bibfield  {author} {\bibinfo {author} {\bibfnamefont {L.}~\bibnamefont
  {Sihver}}, \bibinfo {author} {\bibfnamefont {T.}~\bibnamefont {Sato}},
  \bibinfo {author} {\bibfnamefont {K.}~\bibnamefont {Gustafsson}}, \bibinfo
  {author} {\bibfnamefont {D.}~\bibnamefont {Mancusi}}, \bibinfo {author}
  {\bibfnamefont {H.}~\bibnamefont {Iwase}}, \bibinfo {author} {\bibfnamefont
  {K.}~\bibnamefont {Niita}}, \bibinfo {author} {\bibfnamefont
  {H.}~\bibnamefont {Nakashima}}, \bibinfo {author} {\bibfnamefont
  {Y.}~\bibnamefont {Sakamoto}}, \bibinfo {author} {\bibfnamefont
  {Y.}~\bibnamefont {Iwamoto}}, \ and\ \bibinfo {author} {\bibfnamefont
  {N.}~\bibnamefont {Matsuda}},\ }\href {\doibase
  https://doi.org/10.1016/j.asr.2010.01.002} {\bibfield  {journal} {\bibinfo
  {journal} {Advances in Space Research}\ }\textbf {\bibinfo {volume} {45}},\
  \bibinfo {pages} {892} (\bibinfo {year} {2010})},\ \bibinfo {note} {life
  Sciences in Space}\BibitemShut {NoStop}%
\bibitem [{\citenamefont {Hultqvist}\ \emph {et~al.}(2012)\citenamefont
  {Hultqvist}, \citenamefont {Lazzeroni}, \citenamefont {Botvina},
  \citenamefont {Gudowska}, \citenamefont {Sobolevsky},\ and\ \citenamefont
  {Brahme}}]{Hultqvist_2012}%
  \BibitemOpen
  \bibfield  {author} {\bibinfo {author} {\bibfnamefont {M.}~\bibnamefont
  {Hultqvist}}, \bibinfo {author} {\bibfnamefont {M.}~\bibnamefont
  {Lazzeroni}}, \bibinfo {author} {\bibfnamefont {A.}~\bibnamefont {Botvina}},
  \bibinfo {author} {\bibfnamefont {I.}~\bibnamefont {Gudowska}}, \bibinfo
  {author} {\bibfnamefont {N.}~\bibnamefont {Sobolevsky}}, \ and\ \bibinfo
  {author} {\bibfnamefont {A.}~\bibnamefont {Brahme}},\ }\href {\doibase
  10.1088/0031-9155/57/13/4369} {\bibfield  {journal} {\bibinfo  {journal}
  {Physics in Medicine and Biology}\ }\textbf {\bibinfo {volume} {57}},\
  \bibinfo {pages} {4369} (\bibinfo {year} {2012})}\BibitemShut {NoStop}%
\bibitem [{\citenamefont {Ferrari}\ \emph {et~al.}(2005)\citenamefont
  {Ferrari}, \citenamefont {Sala}, \citenamefont {Fasso},\ and\ \citenamefont
  {Ranft}}]{Ferrari:2005zk}%
  \BibitemOpen
  \bibfield  {author} {\bibinfo {author} {\bibfnamefont {A.}~\bibnamefont
  {Ferrari}}, \bibinfo {author} {\bibfnamefont {P.~R.}\ \bibnamefont {Sala}},
  \bibinfo {author} {\bibfnamefont {A.}~\bibnamefont {Fasso}}, \ and\ \bibinfo
  {author} {\bibfnamefont {J.}~\bibnamefont {Ranft}},\ }\href {\doibase
  10.2172/877507} {\enquote {\bibinfo {title} {{FLUKA: A multi-particle
  transport code (Program version 2005)}},}\ } (\bibinfo {year}
  {2005})\BibitemShut {NoStop}%
\bibitem [{\citenamefont {Fasso}\ \emph {et~al.}(1993)\citenamefont {Fasso},
  \citenamefont {Ferrari}, \citenamefont {Ranft},\ and\ \citenamefont
  {Sala}}]{FLUKA1}%
  \BibitemOpen
  \bibfield  {author} {\bibinfo {author} {\bibfnamefont {A.}~\bibnamefont
  {Fasso}}, \bibinfo {author} {\bibfnamefont {A.}~\bibnamefont {Ferrari}},
  \bibinfo {author} {\bibfnamefont {J.}~\bibnamefont {Ranft}}, \ and\ \bibinfo
  {author} {\bibfnamefont {P.~R.}\ \bibnamefont {Sala}},\ }\href@noop {}
  {\bibfield  {journal} {\bibinfo  {journal} {Conf. Proc. C}\ }\textbf
  {\bibinfo {volume} {9309194}},\ \bibinfo {pages} {493} (\bibinfo {year}
  {1993})}\BibitemShut {NoStop}%
\bibitem [{\citenamefont {Aricò}\ \emph {et~al.}(2019)\citenamefont {Aricò},
  \citenamefont {Ferrari}, \citenamefont {Horst}, \citenamefont {Mairani},
  \citenamefont {Reidel}, \citenamefont {Schuy},\ and\ \citenamefont
  {Weber}}]{Arico:2019pcz}%
  \BibitemOpen
  \bibfield  {author} {\bibinfo {author} {\bibfnamefont {G.}~\bibnamefont
  {Aricò}}, \bibinfo {author} {\bibfnamefont {A.}~\bibnamefont {Ferrari}},
  \bibinfo {author} {\bibfnamefont {F.}~\bibnamefont {Horst}}, \bibinfo
  {author} {\bibfnamefont {A.}~\bibnamefont {Mairani}}, \bibinfo {author}
  {\bibfnamefont {C.}~\bibnamefont {Reidel}}, \bibinfo {author} {\bibfnamefont
  {C.}~\bibnamefont {Schuy}}, \ and\ \bibinfo {author} {\bibfnamefont
  {U.}~\bibnamefont {Weber}},\ }\href@noop {} {\bibfield  {journal} {\bibinfo
  {journal} {CERN Proc.}\ }\textbf {\bibinfo {volume} {1}},\ \bibinfo {pages}
  {321} (\bibinfo {year} {2019})}\BibitemShut {NoStop}%
\bibitem [{\citenamefont {Battistoni}(2008)}]{Battistoni:2008hga}%
  \BibitemOpen
  \bibfield  {author} {\bibinfo {author} {\bibfnamefont {G.}~\bibnamefont
  {Battistoni}} (\bibinfo {collaboration} {FLUKA}),\ }in\ \href {\doibase
  10.1109/NSSMIC.2008.4774716} {\emph {\bibinfo {booktitle} {{2008 IEEE Nuclear
  Science Symposium and Medical Imaging Conference}}}}\ (\bibinfo {year}
  {2008})\ pp.\ \bibinfo {pages} {1609--1615}\BibitemShut {NoStop}%
\bibitem [{\citenamefont {Andersen}\ \emph {et~al.}(2004)\citenamefont
  {Andersen}, \citenamefont {Ballarini}, \citenamefont {Battistoni},
  \citenamefont {Campanella}, \citenamefont {Carboni}, \citenamefont {Cerutti},
  \citenamefont {Empl}, \citenamefont {Fasso}, \citenamefont {Ferrari},
  \citenamefont {Gadioli}, \citenamefont {Garzelli}, \citenamefont {Lee},
  \citenamefont {Ottolenghi}, \citenamefont {Pelliccioni}, \citenamefont
  {Pinsky}, \citenamefont {Ranft}, \citenamefont {Roesler}, \citenamefont
  {Sala},\ and\ \citenamefont {Wilson}}]{Andersenarticle}%
  \BibitemOpen
  \bibfield  {author} {\bibinfo {author} {\bibfnamefont {V.}~\bibnamefont
  {Andersen}}, \bibinfo {author} {\bibfnamefont {F.}~\bibnamefont {Ballarini}},
  \bibinfo {author} {\bibfnamefont {G.}~\bibnamefont {Battistoni}}, \bibinfo
  {author} {\bibfnamefont {M.}~\bibnamefont {Campanella}}, \bibinfo {author}
  {\bibfnamefont {M.}~\bibnamefont {Carboni}}, \bibinfo {author} {\bibfnamefont
  {F.}~\bibnamefont {Cerutti}}, \bibinfo {author} {\bibfnamefont
  {A.}~\bibnamefont {Empl}}, \bibinfo {author} {\bibfnamefont {A.}~\bibnamefont
  {Fasso}}, \bibinfo {author} {\bibfnamefont {A.}~\bibnamefont {Ferrari}},
  \bibinfo {author} {\bibfnamefont {E.}~\bibnamefont {Gadioli}}, \bibinfo
  {author} {\bibfnamefont {M.~V.}\ \bibnamefont {Garzelli}}, \bibinfo {author}
  {\bibfnamefont {K.}~\bibnamefont {Lee}}, \bibinfo {author} {\bibfnamefont
  {A.}~\bibnamefont {Ottolenghi}}, \bibinfo {author} {\bibfnamefont
  {M.}~\bibnamefont {Pelliccioni}}, \bibinfo {author} {\bibfnamefont
  {L.}~\bibnamefont {Pinsky}}, \bibinfo {author} {\bibfnamefont
  {J.}~\bibnamefont {Ranft}}, \bibinfo {author} {\bibfnamefont
  {S.}~\bibnamefont {Roesler}}, \bibinfo {author} {\bibfnamefont
  {P.}~\bibnamefont {Sala}}, \ and\ \bibinfo {author} {\bibfnamefont
  {T.}~\bibnamefont {Wilson}},\ }\href {\doibase 10.1016/j.asr.2003.03.045}
  {\bibfield  {journal} {\bibinfo  {journal} {Advances in space research : the
  official journal of the Committee on Space Research (COSPAR)}\ }\textbf
  {\bibinfo {volume} {34}},\ \bibinfo {pages} {1302} (\bibinfo {year}
  {2004})}\BibitemShut {NoStop}%
\bibitem [{\citenamefont {Heinbockel}\ \emph {et~al.}(2011)\citenamefont
  {Heinbockel} \emph {et~al.}}]{Heinbockel:2011zz}%
  \BibitemOpen
  \bibfield  {author} {\bibinfo {author} {\bibfnamefont {J.~H.}\ \bibnamefont
  {Heinbockel}} \emph {et~al.},\ }\href {\doibase 10.1016/j.asr.2010.11.013}
  {\bibfield  {journal} {\bibinfo  {journal} {Adv. Space Res.}\ }\textbf
  {\bibinfo {volume} {47}},\ \bibinfo {pages} {1089} (\bibinfo {year}
  {2011})}\BibitemShut {NoStop}%
\bibitem [{\citenamefont {Tusnski}\ \emph {et~al.}(2019)\citenamefont
  {Tusnski}, \citenamefont {Szpigel}, \citenamefont {Giménez~de Castro},
  \citenamefont {MacKinnon},\ and\ \citenamefont {Simões}}]{Tusnski:2019rpd}%
  \BibitemOpen
  \bibfield  {author} {\bibinfo {author} {\bibfnamefont {D.~S.}\ \bibnamefont
  {Tusnski}}, \bibinfo {author} {\bibfnamefont {S.}~\bibnamefont {Szpigel}},
  \bibinfo {author} {\bibfnamefont {C.~G.}\ \bibnamefont {Giménez~de Castro}},
  \bibinfo {author} {\bibfnamefont {A.~L.}\ \bibnamefont {MacKinnon}}, \ and\
  \bibinfo {author} {\bibfnamefont {P.~J.~A.}\ \bibnamefont {Simões}},\ }\href
  {\doibase 10.1007/s11207-019-1499-2} {\bibfield  {journal} {\bibinfo
  {journal} {Solar Phys.}\ }\textbf {\bibinfo {volume} {294}},\ \bibinfo
  {pages} {103} (\bibinfo {year} {2019})},\ \Eprint
  {http://arxiv.org/abs/1907.11575} {arXiv:1907.11575 [astro-ph.HE]}
  \BibitemShut {NoStop}%
\bibitem [{\citenamefont {Mazziotta}\ \emph {et~al.}(2020)\citenamefont
  {Mazziotta}, \citenamefont {De~La Torre~Luque}, \citenamefont {Di~Venere},
  \citenamefont {Fassò}, \citenamefont {Ferrari}, \citenamefont {Loparco},
  \citenamefont {Sala},\ and\ \citenamefont {Serini}}]{FlukaSun}%
  \BibitemOpen
  \bibfield  {author} {\bibinfo {author} {\bibfnamefont {M.~N.}\ \bibnamefont
  {Mazziotta}}, \bibinfo {author} {\bibfnamefont {P.}~\bibnamefont {De~La
  Torre~Luque}}, \bibinfo {author} {\bibfnamefont {L.}~\bibnamefont
  {Di~Venere}}, \bibinfo {author} {\bibfnamefont {A.}~\bibnamefont {Fassò}},
  \bibinfo {author} {\bibfnamefont {A.}~\bibnamefont {Ferrari}}, \bibinfo
  {author} {\bibfnamefont {F.}~\bibnamefont {Loparco}}, \bibinfo {author}
  {\bibfnamefont {P.}~\bibnamefont {Sala}}, \ and\ \bibinfo {author}
  {\bibfnamefont {D.}~\bibnamefont {Serini}},\ }\href {\doibase
  10.1103/PhysRevD.101.083011} {\bibfield  {journal} {\bibinfo  {journal}
  {Phys. Rev. D}\ }\textbf {\bibinfo {volume} {101}},\ \bibinfo {pages}
  {083011} (\bibinfo {year} {2020})},\ \Eprint
  {http://arxiv.org/abs/2001.09933} {arXiv:2001.09933 [astro-ph.HE]}
  \BibitemShut {NoStop}%
\bibitem [{\citenamefont {Mazziotta}\ \emph {et~al.}(2016)\citenamefont
  {Mazziotta}, \citenamefont {Cerutti}, \citenamefont {Ferrari}, \citenamefont
  {Gaggero}, \citenamefont {Loparco},\ and\ \citenamefont {Sala}}]{Mazziot}%
  \BibitemOpen
  \bibfield  {author} {\bibinfo {author} {\bibfnamefont {M.~N.}\ \bibnamefont
  {Mazziotta}}, \bibinfo {author} {\bibfnamefont {F.}~\bibnamefont {Cerutti}},
  \bibinfo {author} {\bibfnamefont {A.}~\bibnamefont {Ferrari}}, \bibinfo
  {author} {\bibfnamefont {D.}~\bibnamefont {Gaggero}}, \bibinfo {author}
  {\bibfnamefont {F.}~\bibnamefont {Loparco}}, \ and\ \bibinfo {author}
  {\bibfnamefont {P.~R.}\ \bibnamefont {Sala}},\ }\href {\doibase
  10.1016/j.astropartphys.2016.04.005} {\bibfield  {journal} {\bibinfo
  {journal} {Astropart. Phys.}\ }\textbf {\bibinfo {volume} {81}},\ \bibinfo
  {pages} {21} (\bibinfo {year} {2016})},\ \Eprint
  {http://arxiv.org/abs/1510.04623} {arXiv:1510.04623 [astro-ph.HE]}
  \BibitemShut {NoStop}%
\bibitem [{\citenamefont {Evoli}\ \emph {et~al.}(2017)\citenamefont {Evoli},
  \citenamefont {Gaggero}, \citenamefont {Vittino}, \citenamefont
  {Di~Bernardo}, \citenamefont {Di~Mauro}, \citenamefont {Ligorini},
  \citenamefont {Ullio},\ and\ \citenamefont {Grasso}}]{DRAGON2-1}%
  \BibitemOpen
  \bibfield  {author} {\bibinfo {author} {\bibfnamefont {C.}~\bibnamefont
  {Evoli}}, \bibinfo {author} {\bibfnamefont {D.}~\bibnamefont {Gaggero}},
  \bibinfo {author} {\bibfnamefont {A.}~\bibnamefont {Vittino}}, \bibinfo
  {author} {\bibfnamefont {G.}~\bibnamefont {Di~Bernardo}}, \bibinfo {author}
  {\bibfnamefont {M.}~\bibnamefont {Di~Mauro}}, \bibinfo {author}
  {\bibfnamefont {A.}~\bibnamefont {Ligorini}}, \bibinfo {author}
  {\bibfnamefont {P.}~\bibnamefont {Ullio}}, \ and\ \bibinfo {author}
  {\bibfnamefont {D.}~\bibnamefont {Grasso}},\ }\href {\doibase
  10.1088/1475-7516/2017/02/015} {\bibfield  {journal} {\bibinfo  {journal}
  {JCAP}\ }\textbf {\bibinfo {volume} {02}},\ \bibinfo {pages} {015} (\bibinfo
  {year} {2017})},\ \Eprint {http://arxiv.org/abs/1607.07886} {arXiv:1607.07886
  [astro-ph.HE]} \BibitemShut {NoStop}%
\bibitem [{\citenamefont {Evoli}\ \emph {et~al.}(2018)\citenamefont {Evoli},
  \citenamefont {Gaggero}, \citenamefont {Vittino}, \citenamefont {Di~Mauro},
  \citenamefont {Grasso},\ and\ \citenamefont {Mazziotta}}]{DRAGON2-2}%
  \BibitemOpen
  \bibfield  {author} {\bibinfo {author} {\bibfnamefont {C.}~\bibnamefont
  {Evoli}}, \bibinfo {author} {\bibfnamefont {D.}~\bibnamefont {Gaggero}},
  \bibinfo {author} {\bibfnamefont {A.}~\bibnamefont {Vittino}}, \bibinfo
  {author} {\bibfnamefont {M.}~\bibnamefont {Di~Mauro}}, \bibinfo {author}
  {\bibfnamefont {D.}~\bibnamefont {Grasso}}, \ and\ \bibinfo {author}
  {\bibfnamefont {M.~N.}\ \bibnamefont {Mazziotta}},\ }\href {\doibase
  10.1088/1475-7516/2018/07/006} {\bibfield  {journal} {\bibinfo  {journal}
  {JCAP}\ }\textbf {\bibinfo {volume} {07}},\ \bibinfo {pages} {006} (\bibinfo
  {year} {2018})},\ \Eprint {http://arxiv.org/abs/1711.09616} {arXiv:1711.09616
  [astro-ph.HE]} \BibitemShut {NoStop}%
\bibitem [{\citenamefont {Evoli}\ \emph {et~al.}(2019)\citenamefont {Evoli},
  \citenamefont {Aloisio},\ and\ \citenamefont {Blasi}}]{Evoli:2019wwu}%
  \BibitemOpen
  \bibfield  {author} {\bibinfo {author} {\bibfnamefont {C.}~\bibnamefont
  {Evoli}}, \bibinfo {author} {\bibfnamefont {R.}~\bibnamefont {Aloisio}}, \
  and\ \bibinfo {author} {\bibfnamefont {P.}~\bibnamefont {Blasi}},\ }\href
  {\doibase 10.1103/PhysRevD.99.103023} {\bibfield  {journal} {\bibinfo
  {journal} {Phys. Rev. D}\ }\textbf {\bibinfo {volume} {99}},\ \bibinfo
  {pages} {103023} (\bibinfo {year} {2019})},\ \Eprint
  {http://arxiv.org/abs/1904.10220} {arXiv:1904.10220 [astro-ph.HE]}
  \BibitemShut {NoStop}%
\bibitem [{\citenamefont {{Moskalenko}}\ and\ \citenamefont
  {{Mashnik}}(2003)}]{GALPROPXS}%
  \BibitemOpen
  \bibfield  {author} {\bibinfo {author} {\bibfnamefont {I.~V.}\ \bibnamefont
  {{Moskalenko}}}\ and\ \bibinfo {author} {\bibfnamefont {S.~G.}\ \bibnamefont
  {{Mashnik}}},\ }in\ \href@noop {} {\emph {\bibinfo {booktitle} {International
  Cosmic Ray Conference}}},\ \bibinfo {series} {International Cosmic Ray
  Conference}, Vol.~\bibinfo {volume} {4}\ (\bibinfo {year} {2003})\ p.\
  \bibinfo {pages} {1969},\ \Eprint {http://arxiv.org/abs/astro-ph/0306367}
  {arXiv:astro-ph/0306367 [astro-ph]} \BibitemShut {NoStop}%
\bibitem [{\citenamefont {Moskalenko}\ \emph {et~al.}(2004)\citenamefont
  {Moskalenko}, \citenamefont {Strong},\ and\ \citenamefont
  {Mashnik}}]{GALPROPXS1}%
  \BibitemOpen
  \bibfield  {author} {\bibinfo {author} {\bibfnamefont {I.}~\bibnamefont
  {Moskalenko}}, \bibinfo {author} {\bibfnamefont {A.}~\bibnamefont {Strong}},
  \ and\ \bibinfo {author} {\bibfnamefont {S.}~\bibnamefont {Mashnik}},\ }\href
  {\doibase 10.1063/1.1945315} {\bibfield  {journal} {\bibinfo  {journal} {AIP
  Conference Proceedings}\ }\textbf {\bibinfo {volume} {769}} (\bibinfo {year}
  {2004}),\ 10.1063/1.1945315}\BibitemShut {NoStop}%
\bibitem [{\citenamefont {Luque}\ \emph {et~al.}(2021)\citenamefont {Luque},
  \citenamefont {Mazziotta}, \citenamefont {Loparco}, \citenamefont {Gargano},\
  and\ \citenamefont {Serini}}]{Luque_MCMC}%
  \BibitemOpen
  \bibfield  {author} {\bibinfo {author} {\bibfnamefont {P.~D. L.~T.}\
  \bibnamefont {Luque}}, \bibinfo {author} {\bibfnamefont {M.}~\bibnamefont
  {Mazziotta}}, \bibinfo {author} {\bibfnamefont {F.}~\bibnamefont {Loparco}},
  \bibinfo {author} {\bibfnamefont {F.}~\bibnamefont {Gargano}}, \ and\
  \bibinfo {author} {\bibfnamefont {D.}~\bibnamefont {Serini}},\ }\href
  {\doibase 10.1088/1475-7516/2021/07/010} {\bibfield  {journal} {\bibinfo
  {journal} {Journal of Cosmology and Astroparticle Physics}\ }\textbf
  {\bibinfo {volume} {2021}},\ \bibinfo {pages} {010} (\bibinfo {year}
  {2021})}\BibitemShut {NoStop}%
\bibitem [{\citenamefont {Luque}(2021)}]{Luque_Ap}%
  \BibitemOpen
  \bibfield  {author} {\bibinfo {author} {\bibfnamefont {P.~D. L.~T.}\
  \bibnamefont {Luque}},\ }\href {\doibase 10.1088/1475-7516/2021/11/018}
  {\bibfield  {journal} {\bibinfo  {journal} {Journal of Cosmology and
  Astroparticle Physics}\ }\textbf {\bibinfo {volume} {2021}},\ \bibinfo
  {pages} {018} (\bibinfo {year} {2021})}\BibitemShut {NoStop}%
\bibitem [{\citenamefont {Aguilar}\ \emph
  {et~al.}(2013{\natexlab{a}})\citenamefont {Aguilar} \emph
  {et~al.}}]{AMS_gen}%
  \BibitemOpen
  \bibfield  {author} {\bibinfo {author} {\bibfnamefont {M.}~\bibnamefont
  {Aguilar}} \emph {et~al.} (\bibinfo {collaboration} {AMS Collaboration}),\
  }\href {\doibase 10.1103/PhysRevLett.110.141102} {\bibfield  {journal}
  {\bibinfo  {journal} {Phys. Rev. Lett.}\ }\textbf {\bibinfo {volume} {110}},\
  \bibinfo {pages} {141102} (\bibinfo {year} {2013}{\natexlab{a}})}\BibitemShut
  {NoStop}%
\bibitem [{\citenamefont {Aguilar}\ \emph {et~al.}(2015)\citenamefont {Aguilar}
  \emph {et~al.}}]{Aguilar:2015ooa}%
  \BibitemOpen
  \bibfield  {author} {\bibinfo {author} {\bibfnamefont {M.}~\bibnamefont
  {Aguilar}} \emph {et~al.} (\bibinfo {collaboration} {AMS Collaboration}),\
  }\href {\doibase 10.1103/PhysRevLett.114.171103} {\bibfield  {journal}
  {\bibinfo  {journal} {Phys. Rev. Lett.}\ }\textbf {\bibinfo {volume} {114}},\
  \bibinfo {pages} {171103} (\bibinfo {year} {2015})}\BibitemShut {NoStop}%
\bibitem [{\citenamefont {Aguilar}\ \emph
  {et~al.}(2018{\natexlab{a}})\citenamefont {Aguilar} \emph
  {et~al.}}]{Aguilar:2018keu}%
  \BibitemOpen
  \bibfield  {author} {\bibinfo {author} {\bibfnamefont {M.}~\bibnamefont
  {Aguilar}} \emph {et~al.} (\bibinfo {collaboration} {AMS Collaboration}),\
  }\href {\doibase 10.1103/PhysRevLett.121.051103} {\bibfield  {journal}
  {\bibinfo  {journal} {Phys. Rev. Lett.}\ }\textbf {\bibinfo {volume} {121}},\
  \bibinfo {pages} {051103} (\bibinfo {year} {2018}{\natexlab{a}})}\BibitemShut
  {NoStop}%
\bibitem [{\citenamefont {Aguilar}\ \emph {et~al.}(2020)\citenamefont {Aguilar}
  \emph {et~al.}}]{Aguilar:2020ohx}%
  \BibitemOpen
  \bibfield  {author} {\bibinfo {author} {\bibfnamefont {M.}~\bibnamefont
  {Aguilar}} \emph {et~al.} (\bibinfo {collaboration} {AMS Collaboration}),\
  }\href {\doibase 10.1103/PhysRevLett.124.211102} {\bibfield  {journal}
  {\bibinfo  {journal} {Phys. Rev. Lett.}\ }\textbf {\bibinfo {volume} {124}},\
  \bibinfo {pages} {211102} (\bibinfo {year} {2020})}\BibitemShut {NoStop}%
\bibitem [{\citenamefont {Aguilar}\ \emph {et~al.}(2017)\citenamefont {Aguilar}
  \emph {et~al.}}]{aguilar2017observation}%
  \BibitemOpen
  \bibfield  {author} {\bibinfo {author} {\bibfnamefont {M.}~\bibnamefont
  {Aguilar}} \emph {et~al.} (\bibinfo {collaboration} {AMS Collaboration}),\
  }\href {\doibase 10.1103/PhysRevLett.119.251101} {\bibfield  {journal}
  {\bibinfo  {journal} {Phys. Rev. Lett.}\ }\textbf {\bibinfo {volume} {119}},\
  \bibinfo {pages} {251101} (\bibinfo {year} {2017})}\BibitemShut {NoStop}%
\bibitem [{\citenamefont {Aguilar}\ \emph
  {et~al.}(2018{\natexlab{b}})\citenamefont {Aguilar} \emph
  {et~al.}}]{aguilar2018observation}%
  \BibitemOpen
  \bibfield  {author} {\bibinfo {author} {\bibfnamefont {M.}~\bibnamefont
  {Aguilar}} \emph {et~al.} (\bibinfo {collaboration} {AMS Collaboration}),\
  }\href {\doibase 10.1103/PhysRevLett.120.021101} {\bibfield  {journal}
  {\bibinfo  {journal} {Phys. Rev. Lett.}\ }\textbf {\bibinfo {volume} {120}},\
  \bibinfo {pages} {021101} (\bibinfo {year} {2018}{\natexlab{b}})}\BibitemShut
  {NoStop}%
\bibitem [{\citenamefont {Aguilar}\ \emph
  {et~al.}(2019{\natexlab{a}})\citenamefont {Aguilar} \emph
  {et~al.}}]{aguilar2019towards}%
  \BibitemOpen
  \bibfield  {author} {\bibinfo {author} {\bibfnamefont {M.}~\bibnamefont
  {Aguilar}} \emph {et~al.} (\bibinfo {collaboration} {AMS Collaboration}),\
  }\href {\doibase 10.1103/PhysRevLett.122.101101} {\bibfield  {journal}
  {\bibinfo  {journal} {Phys. Rev. Lett.}\ }\textbf {\bibinfo {volume} {122}},\
  \bibinfo {pages} {101101} (\bibinfo {year} {2019}{\natexlab{a}})}\BibitemShut
  {NoStop}%
\bibitem [{\citenamefont {Aguilar}\ \emph
  {et~al.}(2021{\natexlab{a}})\citenamefont {Aguilar} \emph
  {et~al.}}]{AMS_NNaAl}%
  \BibitemOpen
  \bibfield  {author} {\bibinfo {author} {\bibfnamefont {M.}~\bibnamefont
  {Aguilar}} \emph {et~al.} (\bibinfo {collaboration} {AMS Collaboration}),\
  }\href {\doibase 10.1103/PhysRevLett.127.021101} {\bibfield  {journal}
  {\bibinfo  {journal} {Phys. Rev. Lett.}\ }\textbf {\bibinfo {volume} {127}},\
  \bibinfo {pages} {021101} (\bibinfo {year} {2021}{\natexlab{a}})}\BibitemShut
  {NoStop}%
\bibitem [{\citenamefont {Aguilar}\ \emph
  {et~al.}(2021{\natexlab{b}})\citenamefont {Aguilar} \emph {et~al.}}]{AMS_Fe}%
  \BibitemOpen
  \bibfield  {author} {\bibinfo {author} {\bibfnamefont {M.}~\bibnamefont
  {Aguilar}} \emph {et~al.} (\bibinfo {collaboration} {AMS Collaboration}),\
  }\href {\doibase 10.1103/PhysRevLett.126.041104} {\bibfield  {journal}
  {\bibinfo  {journal} {Phys. Rev. Lett.}\ }\textbf {\bibinfo {volume} {126}},\
  \bibinfo {pages} {041104} (\bibinfo {year} {2021}{\natexlab{b}})}\BibitemShut
  {NoStop}%
\bibitem [{\citenamefont {Aguilar}\ \emph
  {et~al.}(2019{\natexlab{b}})\citenamefont {Aguilar} \emph
  {et~al.}}]{AMS_Pos2019}%
  \BibitemOpen
  \bibfield  {author} {\bibinfo {author} {\bibfnamefont {M.}~\bibnamefont
  {Aguilar}} \emph {et~al.} (\bibinfo {collaboration} {AMS Collaboration}),\
  }\href {\doibase 10.1103/PhysRevLett.122.041102} {\bibfield  {journal}
  {\bibinfo  {journal} {Phys. Rev. Lett.}\ }\textbf {\bibinfo {volume} {122}},\
  \bibinfo {pages} {041102} (\bibinfo {year} {2019}{\natexlab{b}})}\BibitemShut
  {NoStop}%
\bibitem [{\citenamefont {Aguilar}\ \emph
  {et~al.}(2021{\natexlab{c}})\citenamefont {Aguilar} \emph {et~al.}}]{AMS_F}%
  \BibitemOpen
  \bibfield  {author} {\bibinfo {author} {\bibfnamefont {M.}~\bibnamefont
  {Aguilar}} \emph {et~al.} (\bibinfo {collaboration} {AMS Collaboration}),\
  }\href {\doibase 10.1103/PhysRevLett.126.081102} {\bibfield  {journal}
  {\bibinfo  {journal} {Phys. Rev. Lett.}\ }\textbf {\bibinfo {volume} {126}},\
  \bibinfo {pages} {081102} (\bibinfo {year} {2021}{\natexlab{c}})}\BibitemShut
  {NoStop}%
\bibitem [{\citenamefont {Cirelli}\ \emph
  {et~al.}(2014{\natexlab{a}})\citenamefont {Cirelli}, \citenamefont {Gaggero},
  \citenamefont {Giesen}, \citenamefont {Taoso},\ and\ \citenamefont
  {Urbano}}]{Cirelli_2014}%
  \BibitemOpen
  \bibfield  {author} {\bibinfo {author} {\bibfnamefont {M.}~\bibnamefont
  {Cirelli}}, \bibinfo {author} {\bibfnamefont {D.}~\bibnamefont {Gaggero}},
  \bibinfo {author} {\bibfnamefont {G.}~\bibnamefont {Giesen}}, \bibinfo
  {author} {\bibfnamefont {M.}~\bibnamefont {Taoso}}, \ and\ \bibinfo {author}
  {\bibfnamefont {A.}~\bibnamefont {Urbano}},\ }\href {\doibase
  10.1088/1475-7516/2014/12/045} {\bibfield  {journal} {\bibinfo  {journal}
  {Journal of Cosmology and Astroparticle Physics}\ }\textbf {\bibinfo {volume}
  {2014}},\ \bibinfo {pages} {045} (\bibinfo {year}
  {2014}{\natexlab{a}})}\BibitemShut {NoStop}%
\bibitem [{\citenamefont {Evoli}\ \emph {et~al.}(2012)\citenamefont {Evoli},
  \citenamefont {Gaggero}, \citenamefont {Grasso},\ and\ \citenamefont
  {Maccione}}]{GSky_diffuse}%
  \BibitemOpen
  \bibfield  {author} {\bibinfo {author} {\bibfnamefont {C.}~\bibnamefont
  {Evoli}}, \bibinfo {author} {\bibfnamefont {D.}~\bibnamefont {Gaggero}},
  \bibinfo {author} {\bibfnamefont {D.}~\bibnamefont {Grasso}}, \ and\ \bibinfo
  {author} {\bibfnamefont {L.}~\bibnamefont {Maccione}},\ }\href {\doibase
  10.1103/PhysRevLett.108.211102} {\bibfield  {journal} {\bibinfo  {journal}
  {Phys. Rev. Lett.}\ }\textbf {\bibinfo {volume} {108}},\ \bibinfo {pages}
  {211102} (\bibinfo {year} {2012})}\BibitemShut {NoStop}%
\bibitem [{\citenamefont {Tavakoli}\ \emph {et~al.}(2011)\citenamefont
  {Tavakoli}, \citenamefont {Cholis}, \citenamefont {Evoli},\ and\
  \citenamefont {Ullio}}]{Tavakoli:2011wz}%
  \BibitemOpen
  \bibfield  {author} {\bibinfo {author} {\bibfnamefont {M.}~\bibnamefont
  {Tavakoli}}, \bibinfo {author} {\bibfnamefont {I.}~\bibnamefont {Cholis}},
  \bibinfo {author} {\bibfnamefont {C.}~\bibnamefont {Evoli}}, \ and\ \bibinfo
  {author} {\bibfnamefont {P.}~\bibnamefont {Ullio}},\ }in\ \href@noop {}
  {\emph {\bibinfo {booktitle} {{3rd International Fermi Symposium}}}}\
  (\bibinfo {year} {2011})\ \Eprint {http://arxiv.org/abs/1110.5922}
  {arXiv:1110.5922 [astro-ph.HE]} \BibitemShut {NoStop}%
\bibitem [{\citenamefont {Casandjian}(2015)}]{Casandjian:2015hja}%
  \BibitemOpen
  \bibfield  {author} {\bibinfo {author} {\bibfnamefont {J.-M.}\ \bibnamefont
  {Casandjian}},\ }\href {\doibase 10.1088/0004-637X/806/2/240} {\bibfield
  {journal} {\bibinfo  {journal} {Astrophys. J.}\ }\textbf {\bibinfo {volume}
  {806}},\ \bibinfo {pages} {240} (\bibinfo {year} {2015})},\ \Eprint
  {http://arxiv.org/abs/1506.00047} {arXiv:1506.00047 [astro-ph.HE]}
  \BibitemShut {NoStop}%
\bibitem [{\citenamefont {Orlando}(2018)}]{Orlando:2017mvd}%
  \BibitemOpen
  \bibfield  {author} {\bibinfo {author} {\bibfnamefont {E.}~\bibnamefont
  {Orlando}},\ }\href {\doibase 10.1093/mnras/stx3280} {\bibfield  {journal}
  {\bibinfo  {journal} {Mon. Not. Roy. Astron. Soc.}\ }\textbf {\bibinfo
  {volume} {475}},\ \bibinfo {pages} {2724} (\bibinfo {year} {2018})},\ \Eprint
  {http://arxiv.org/abs/1712.07127} {arXiv:1712.07127 [astro-ph.HE]}
  \BibitemShut {NoStop}%
\bibitem [{\citenamefont {Bobchenko}\ \emph {et~al.}(1979)\citenamefont
  {Bobchenko} \emph {et~al.}}]{Bobchenko:1979hp}%
  \BibitemOpen
  \bibfield  {author} {\bibinfo {author} {\bibfnamefont {B.}~\bibnamefont
  {Bobchenko}} \emph {et~al.},\ }\href@noop {} {\bibfield  {journal} {\bibinfo
  {journal} {Sov. J. Nucl. Phys.}\ }\textbf {\bibinfo {volume} {30}},\ \bibinfo
  {pages} {805} (\bibinfo {year} {1979})}\BibitemShut {NoStop}%
\bibitem [{\citenamefont {Otuka}\ \emph {et~al.}(2014)\citenamefont {Otuka}
  \emph {et~al.}}]{OTUKA2014272}%
  \BibitemOpen
  \bibfield  {author} {\bibinfo {author} {\bibfnamefont {N.}~\bibnamefont
  {Otuka}} \emph {et~al.},\ }\href {\doibase
  https://doi.org/10.1016/j.nds.2014.07.065} {\bibfield  {journal} {\bibinfo
  {journal} {Nuclear Data Sheets}\ }\textbf {\bibinfo {volume} {120}},\
  \bibinfo {pages} {272 } (\bibinfo {year} {2014})}\BibitemShut {NoStop}%
\bibitem [{\citenamefont {Zerkin}\ and\ \citenamefont
  {Pritychenko}(2018)}]{Zerkin_2018}%
  \BibitemOpen
  \bibfield  {author} {\bibinfo {author} {\bibfnamefont {V.}~\bibnamefont
  {Zerkin}}\ and\ \bibinfo {author} {\bibfnamefont {B.}~\bibnamefont
  {Pritychenko}},\ }\href {\doibase 10.1016/j.nima.2018.01.045} {\bibfield
  {journal} {\bibinfo  {journal} {Nuclear Instruments and Methods in Physics
  Research Section A: Accelerators, Spectrometers, Detectors and Associated
  Equipment}\ }\textbf {\bibinfo {volume} {888}},\ \bibinfo {pages} {31–43}
  (\bibinfo {year} {2018})}\BibitemShut {NoStop}%
\bibitem [{\citenamefont {Barashenkov}\ and\ \citenamefont
  {Polanski}(1994)}]{Barashenkov:1994cp}%
  \BibitemOpen
  \bibfield  {author} {\bibinfo {author} {\bibfnamefont {V.~S.}\ \bibnamefont
  {Barashenkov}}\ and\ \bibinfo {author} {\bibfnamefont {A.}~\bibnamefont
  {Polanski}},\ }\href {http://lt-jds.jinr.ru/record/5725/export/hx?ln=en}
  {\emph {\bibinfo {title} {Electronic guide for nuclear cross-sections:
  version 1994}}},\ \bibinfo {type} {Tech. Rep.}\ \bibinfo {number} {E2-94-417.
  JINR-E2-94-417}\ (\bibinfo  {institution} {Joint Inst. Nucl. Res.},\ \bibinfo
  {address} {Dubna},\ \bibinfo {year} {1994})\BibitemShut {NoStop}%
\bibitem [{\citenamefont {Kafexhiu}\ \emph {et~al.}(2014)\citenamefont
  {Kafexhiu}, \citenamefont {Aharonian}, \citenamefont {Taylor},\ and\
  \citenamefont {Vila}}]{Kafexhiu:2014cua}%
  \BibitemOpen
  \bibfield  {author} {\bibinfo {author} {\bibfnamefont {E.}~\bibnamefont
  {Kafexhiu}}, \bibinfo {author} {\bibfnamefont {F.}~\bibnamefont {Aharonian}},
  \bibinfo {author} {\bibfnamefont {A.~M.}\ \bibnamefont {Taylor}}, \ and\
  \bibinfo {author} {\bibfnamefont {G.~S.}\ \bibnamefont {Vila}},\ }\href
  {\doibase 10.1103/PhysRevD.90.123014} {\bibfield  {journal} {\bibinfo
  {journal} {Phys. Rev. D}\ }\textbf {\bibinfo {volume} {90}},\ \bibinfo
  {pages} {123014} (\bibinfo {year} {2014})},\ \Eprint
  {http://arxiv.org/abs/1406.7369} {arXiv:1406.7369 [astro-ph.HE]} \BibitemShut
  {NoStop}%
\bibitem [{\citenamefont {Block}\ \emph
  {et~al.}(2015{\natexlab{a}})\citenamefont {Block}, \citenamefont {Durand},
  \citenamefont {Ha},\ and\ \citenamefont {Halzen}}]{RiseofXS}%
  \BibitemOpen
  \bibfield  {author} {\bibinfo {author} {\bibfnamefont {M.~M.}\ \bibnamefont
  {Block}}, \bibinfo {author} {\bibfnamefont {L.}~\bibnamefont {Durand}},
  \bibinfo {author} {\bibfnamefont {P.}~\bibnamefont {Ha}}, \ and\ \bibinfo
  {author} {\bibfnamefont {F.}~\bibnamefont {Halzen}},\ }\href {\doibase
  10.1103/PhysRevD.92.114021} {\bibfield  {journal} {\bibinfo  {journal} {Phys.
  Rev. D}\ }\textbf {\bibinfo {volume} {92}},\ \bibinfo {pages} {114021}
  (\bibinfo {year} {2015}{\natexlab{a}})}\BibitemShut {NoStop}%
\bibitem [{\citenamefont {Block}\ and\ \citenamefont
  {Halzen}(2011)}]{Block_2011}%
  \BibitemOpen
  \bibfield  {author} {\bibinfo {author} {\bibfnamefont {M.~M.}\ \bibnamefont
  {Block}}\ and\ \bibinfo {author} {\bibfnamefont {F.}~\bibnamefont {Halzen}},\
  }\href {\doibase 10.1103/physrevlett.107.212002} {\bibfield  {journal}
  {\bibinfo  {journal} {Physical Review Letters}\ }\textbf {\bibinfo {volume}
  {107}} (\bibinfo {year} {2011}),\ 10.1103/physrevlett.107.212002}\BibitemShut
  {NoStop}%
\bibitem [{\citenamefont {Block}\ \emph
  {et~al.}(2015{\natexlab{b}})\citenamefont {Block}, \citenamefont {Durand},
  \citenamefont {Ha},\ and\ \citenamefont {Halzen}}]{PhysRevD_rise}%
  \BibitemOpen
  \bibfield  {author} {\bibinfo {author} {\bibfnamefont {M.~M.}\ \bibnamefont
  {Block}}, \bibinfo {author} {\bibfnamefont {L.}~\bibnamefont {Durand}},
  \bibinfo {author} {\bibfnamefont {P.}~\bibnamefont {Ha}}, \ and\ \bibinfo
  {author} {\bibfnamefont {F.}~\bibnamefont {Halzen}},\ }\href {\doibase
  10.1103/PhysRevD.92.114021} {\bibfield  {journal} {\bibinfo  {journal} {Phys.
  Rev. D}\ }\textbf {\bibinfo {volume} {92}},\ \bibinfo {pages} {114021}
  (\bibinfo {year} {2015}{\natexlab{b}})}\BibitemShut {NoStop}%
\bibitem [{\citenamefont {Webber}\ \emph {et~al.}(2003)\citenamefont {Webber},
  \citenamefont {Soutoul}, \citenamefont {Kish},\ and\ \citenamefont
  {Rockstroh}}]{webber2003updated}%
  \BibitemOpen
  \bibfield  {author} {\bibinfo {author} {\bibfnamefont {W.}~\bibnamefont
  {Webber}}, \bibinfo {author} {\bibfnamefont {A.}~\bibnamefont {Soutoul}},
  \bibinfo {author} {\bibfnamefont {J.}~\bibnamefont {Kish}}, \ and\ \bibinfo
  {author} {\bibfnamefont {J.}~\bibnamefont {Rockstroh}},\ }\href@noop {}
  {\bibfield  {journal} {\bibinfo  {journal} {ApJ Supplement Series}\ }\textbf
  {\bibinfo {volume} {144}},\ \bibinfo {pages} {153} (\bibinfo {year}
  {2003})}\BibitemShut {NoStop}%
\bibitem [{\citenamefont {Webber}\ \emph {et~al.}(1990)\citenamefont {Webber},
  \citenamefont {Kish},\ and\ \citenamefont {Schrier}}]{webber1990formula}%
  \BibitemOpen
  \bibfield  {author} {\bibinfo {author} {\bibfnamefont {W.}~\bibnamefont
  {Webber}}, \bibinfo {author} {\bibfnamefont {J.}~\bibnamefont {Kish}}, \ and\
  \bibinfo {author} {\bibfnamefont {D.}~\bibnamefont {Schrier}},\ }\href@noop
  {} {\bibfield  {journal} {\bibinfo  {journal} {Physical Review C}\ }\textbf
  {\bibinfo {volume} {41}},\ \bibinfo {pages} {566} (\bibinfo {year}
  {1990})}\BibitemShut {NoStop}%
\bibitem [{\citenamefont {Tsao}\ \emph {et~al.}(1998)\citenamefont {Tsao},
  \citenamefont {Silberberg},\ and\ \citenamefont
  {Barghouty}}]{tsao1998partial}%
  \BibitemOpen
  \bibfield  {author} {\bibinfo {author} {\bibfnamefont {C.}~\bibnamefont
  {Tsao}}, \bibinfo {author} {\bibfnamefont {R.}~\bibnamefont {Silberberg}}, \
  and\ \bibinfo {author} {\bibfnamefont {A.}~\bibnamefont {Barghouty}},\
  }\href@noop {} {\bibfield  {journal} {\bibinfo  {journal} {ApJ}\ }\textbf
  {\bibinfo {volume} {501}},\ \bibinfo {pages} {920} (\bibinfo {year}
  {1998})}\BibitemShut {NoStop}%
\bibitem [{\citenamefont {Silberberg}\ \emph {et~al.}(1998)\citenamefont
  {Silberberg}, \citenamefont {Tsao},\ and\ \citenamefont
  {Barghouty}}]{silberberg1998updated}%
  \BibitemOpen
  \bibfield  {author} {\bibinfo {author} {\bibfnamefont {R.}~\bibnamefont
  {Silberberg}}, \bibinfo {author} {\bibfnamefont {C.}~\bibnamefont {Tsao}}, \
  and\ \bibinfo {author} {\bibfnamefont {A.}~\bibnamefont {Barghouty}},\
  }\href@noop {} {\bibfield  {journal} {\bibinfo  {journal} {ApJ}\ }\textbf
  {\bibinfo {volume} {501}},\ \bibinfo {pages} {911} (\bibinfo {year}
  {1998})}\BibitemShut {NoStop}%
\bibitem [{\citenamefont {G\'enolini}\ \emph {et~al.}(2017)\citenamefont
  {G\'enolini} \emph {et~al.}}]{genolini2017indications}%
  \BibitemOpen
  \bibfield  {author} {\bibinfo {author} {\bibfnamefont {Y.}~\bibnamefont
  {G\'enolini}} \emph {et~al.},\ }\href {\doibase
  10.1103/PhysRevLett.119.241101} {\bibfield  {journal} {\bibinfo  {journal}
  {Phys. Rev. Lett.}\ }\textbf {\bibinfo {volume} {119}},\ \bibinfo {pages}
  {241101} (\bibinfo {year} {2017})},\ \Eprint
  {http://arxiv.org/abs/1706.09812} {arXiv:1706.09812 [astro-ph.HE]}
  \BibitemShut {NoStop}%
\bibitem [{\citenamefont {Weinrich}\ \emph {et~al.}(2020)\citenamefont
  {Weinrich}, \citenamefont {G\'enolini}, \citenamefont {Boudaud},
  \citenamefont {Derome},\ and\ \citenamefont {Maurin}}]{Weinrich:2020cmw}%
  \BibitemOpen
  \bibfield  {author} {\bibinfo {author} {\bibfnamefont {N.}~\bibnamefont
  {Weinrich}}, \bibinfo {author} {\bibfnamefont {Y.}~\bibnamefont
  {G\'enolini}}, \bibinfo {author} {\bibfnamefont {M.}~\bibnamefont {Boudaud}},
  \bibinfo {author} {\bibfnamefont {L.}~\bibnamefont {Derome}}, \ and\ \bibinfo
  {author} {\bibfnamefont {D.}~\bibnamefont {Maurin}},\ }\href {\doibase
  10.1051/0004-6361/202037875} {\bibfield  {journal} {\bibinfo  {journal}
  {Astron. Astrophys.}\ }\textbf {\bibinfo {volume} {639}},\ \bibinfo {pages}
  {A131} (\bibinfo {year} {2020})},\ \Eprint {http://arxiv.org/abs/2002.11406}
  {arXiv:2002.11406 [astro-ph.HE]} \BibitemShut {NoStop}%
\bibitem [{\citenamefont {Adriani}\ and\ \citenamefont
  {thers}(2011)}]{Adriani69}%
  \BibitemOpen
  \bibfield  {author} {\bibinfo {author} {\bibfnamefont {O.}~\bibnamefont
  {Adriani}}\ and\ \bibinfo {author} {\bibnamefont {thers}},\ }\href {\doibase
  10.1126/science.1199172} {\bibfield  {journal} {\bibinfo  {journal}
  {Science}\ }\textbf {\bibinfo {volume} {332}},\ \bibinfo {pages} {69}
  (\bibinfo {year} {2011})},\ \Eprint
  {http://arxiv.org/abs/https://science.sciencemag.org/content/332/6025/69.full.pdf}
  {https://science.sciencemag.org/content/332/6025/69.full.pdf} \BibitemShut
  {NoStop}%
\bibitem [{\citenamefont {Ahn}\ \emph {et~al.}(2010)\citenamefont {Ahn} \emph
  {et~al.}}]{Ahn_2010}%
  \BibitemOpen
  \bibfield  {author} {\bibinfo {author} {\bibfnamefont {H.~S.}\ \bibnamefont
  {Ahn}} \emph {et~al.},\ }\href {\doibase 10.1088/2041-8205/714/1/l89}
  {\bibfield  {journal} {\bibinfo  {journal} {The Astrophysical Journal}\
  }\textbf {\bibinfo {volume} {714}},\ \bibinfo {pages} {L89} (\bibinfo {year}
  {2010})}\BibitemShut {NoStop}%
\bibitem [{\citenamefont {Aguilar}\ \emph
  {et~al.}(2013{\natexlab{b}})\citenamefont {Aguilar} \emph
  {et~al.}}]{AMS_hardening}%
  \BibitemOpen
  \bibfield  {author} {\bibinfo {author} {\bibfnamefont {M.}~\bibnamefont
  {Aguilar}} \emph {et~al.} (\bibinfo {collaboration} {AMS Collaboration}),\
  }\href {\doibase 10.1103/PhysRevLett.110.141102} {\bibfield  {journal}
  {\bibinfo  {journal} {Phys. Rev. Lett.}\ }\textbf {\bibinfo {volume} {110}},\
  \bibinfo {pages} {141102} (\bibinfo {year} {2013}{\natexlab{b}})}\BibitemShut
  {NoStop}%
\bibitem [{\citenamefont {Panov}\ \emph {et~al.}(2007)\citenamefont {Panov}
  \emph {et~al.}}]{Panov:2006kf}%
  \BibitemOpen
  \bibfield  {author} {\bibinfo {author} {\bibfnamefont {A.~D.}\ \bibnamefont
  {Panov}} \emph {et~al.},\ }\href {\doibase 10.3103/S1062873807040168}
  {\bibfield  {journal} {\bibinfo  {journal} {Bull. Russ. Acad. Sci. Phys.}\
  }\textbf {\bibinfo {volume} {71}},\ \bibinfo {pages} {494} (\bibinfo {year}
  {2007})},\ \Eprint {http://arxiv.org/abs/astro-ph/0612377}
  {arXiv:astro-ph/0612377} \BibitemShut {NoStop}%
\bibitem [{\citenamefont {{Maurin}}\ \emph {et~al.}(2014)\citenamefont
  {{Maurin}}, \citenamefont {{Melot}},\ and\ \citenamefont
  {{Taillet}}}]{Maurin_db1}%
  \BibitemOpen
  \bibfield  {author} {\bibinfo {author} {\bibfnamefont {D.}~\bibnamefont
  {{Maurin}}}, \bibinfo {author} {\bibfnamefont {F.}~\bibnamefont {{Melot}}}, \
  and\ \bibinfo {author} {\bibfnamefont {R.}~\bibnamefont {{Taillet}}},\ }\href
  {\doibase 10.1051/0004-6361/201321344} {\bibfield  {journal} {\bibinfo
  {journal} {ApJ}\ }\textbf {\bibinfo {volume} {569}},\ \bibinfo {eid} {A32}
  (\bibinfo {year} {2014})},\ \Eprint {http://arxiv.org/abs/1302.5525}
  {arXiv:1302.5525 [astro-ph.HE]} \BibitemShut {NoStop}%
\bibitem [{\citenamefont {{Maurin}}\ \emph {et~al.}(2020)\citenamefont
  {{Maurin}}, \citenamefont {{Dembinski}}, \citenamefont {{Gonzalez}},
  \citenamefont {{Mari{\textcommabelow s}}},\ and\ \citenamefont
  {{Melot}}}]{Maurin_db2}%
  \BibitemOpen
  \bibfield  {author} {\bibinfo {author} {\bibfnamefont {D.}~\bibnamefont
  {{Maurin}}}, \bibinfo {author} {\bibfnamefont {H.~P.}\ \bibnamefont
  {{Dembinski}}}, \bibinfo {author} {\bibfnamefont {J.}~\bibnamefont
  {{Gonzalez}}}, \bibinfo {author} {\bibfnamefont {I.~C.}\ \bibnamefont
  {{Mari{\textcommabelow s}}}}, \ and\ \bibinfo {author} {\bibfnamefont
  {F.}~\bibnamefont {{Melot}}},\ }\href {\doibase 10.3390/universe6080102}
  {\bibfield  {journal} {\bibinfo  {journal} {Universe}\ }\textbf {\bibinfo
  {volume} {6}},\ \bibinfo {pages} {102} (\bibinfo {year} {2020})},\ \Eprint
  {http://arxiv.org/abs/2005.14663} {arXiv:2005.14663 [astro-ph.HE]}
  \BibitemShut {NoStop}%
\bibitem [{\citenamefont {Pizzolotto}\ \emph {et~al.}(2017)\citenamefont
  {Pizzolotto} \emph {et~al.}}]{ssdc}%
  \BibitemOpen
  \bibfield  {author} {\bibinfo {author} {\bibfnamefont {C.}~\bibnamefont
  {Pizzolotto}} \emph {et~al.},\ }\href {\doibase 10.22323/1.301.0227}
  {\bibfield  {journal} {\bibinfo  {journal} {PoS}\ }\textbf {\bibinfo {volume}
  {ICRC2017}},\ \bibinfo {pages} {227} (\bibinfo {year} {2017})}\BibitemShut
  {NoStop}%
\bibitem [{\citenamefont {Korsmeier}\ and\ \citenamefont
  {Cuoco}(2021{\natexlab{b}})}]{Korsmeier:2021bkw}%
  \BibitemOpen
  \bibfield  {author} {\bibinfo {author} {\bibfnamefont {M.}~\bibnamefont
  {Korsmeier}}\ and\ \bibinfo {author} {\bibfnamefont {A.}~\bibnamefont
  {Cuoco}},\ }\href@noop {} {\  (\bibinfo {year} {2021}{\natexlab{b}})},\
  \Eprint {http://arxiv.org/abs/2112.08381} {arXiv:2112.08381 [astro-ph.HE]}
  \BibitemShut {NoStop}%
\bibitem [{\citenamefont {Chmeleff}\ and\ \citenamefont
  {et~al.}(2010)}]{chmeleff2010determination}%
  \BibitemOpen
  \bibfield  {author} {\bibinfo {author} {\bibfnamefont {J.}~\bibnamefont
  {Chmeleff}}\ and\ \bibinfo {author} {\bibnamefont {et~al.}},\ }\href@noop {}
  {\bibfield  {journal} {\bibinfo  {journal} {Nuclear Instruments and Methods
  in Physics Research Section B: Beam Interactions with Materials and Atoms}\
  }\textbf {\bibinfo {volume} {268}},\ \bibinfo {pages} {192} (\bibinfo {year}
  {2010})}\BibitemShut {NoStop}%
\bibitem [{\citenamefont {{Yanasak}}\ \emph {et~al.}(2001)\citenamefont
  {{Yanasak}} \emph {et~al.}}]{ACEBe}%
  \BibitemOpen
  \bibfield  {author} {\bibinfo {author} {\bibfnamefont {N.~E.}\ \bibnamefont
  {{Yanasak}}} \emph {et~al.},\ }\href {\doibase 10.1086/323842} {\bibfield
  {journal} {\bibinfo  {journal} {apj}\ }\textbf {\bibinfo {volume} {563}},\
  \bibinfo {pages} {768} (\bibinfo {year} {2001})}\BibitemShut {NoStop}%
\bibitem [{\citenamefont {{Garcia-Munoz}}\ \emph {et~al.}(1977)\citenamefont
  {{Garcia-Munoz}}, \citenamefont {{Mason}},\ and\ \citenamefont
  {{Simpson}}}]{IMP1}%
  \BibitemOpen
  \bibfield  {author} {\bibinfo {author} {\bibfnamefont {M.}~\bibnamefont
  {{Garcia-Munoz}}}, \bibinfo {author} {\bibfnamefont {G.~M.}\ \bibnamefont
  {{Mason}}}, \ and\ \bibinfo {author} {\bibfnamefont {J.~A.}\ \bibnamefont
  {{Simpson}}},\ }\href {\doibase 10.1086/155632} {\bibfield  {journal}
  {\bibinfo  {journal} {apj}\ }\textbf {\bibinfo {volume} {217}},\ \bibinfo
  {pages} {859} (\bibinfo {year} {1977})}\BibitemShut {NoStop}%
\bibitem [{\citenamefont {{Garcia-Munoz}}\ \emph {et~al.}(1981)\citenamefont
  {{Garcia-Munoz}}, \citenamefont {{Simpson}},\ and\ \citenamefont
  {{Wefel}}}]{IMP2}%
  \BibitemOpen
  \bibfield  {author} {\bibinfo {author} {\bibfnamefont {M.}~\bibnamefont
  {{Garcia-Munoz}}}, \bibinfo {author} {\bibfnamefont {J.~A.}\ \bibnamefont
  {{Simpson}}}, \ and\ \bibinfo {author} {\bibfnamefont {J.~P.}\ \bibnamefont
  {{Wefel}}},\ }in\ \href@noop {} {\emph {\bibinfo {booktitle} {International
  Cosmic Ray Conference}}},\ \bibinfo {series} {International Cosmic Ray
  Conference}, Vol.~\bibinfo {volume} {2}\ (\bibinfo {year} {1981})\
  p.~\bibinfo {pages} {72}\BibitemShut {NoStop}%
\bibitem [{\citenamefont {{Wiedenbeck}}\ and\ \citenamefont
  {{Greiner}}(1980)}]{ISEE}%
  \BibitemOpen
  \bibfield  {author} {\bibinfo {author} {\bibfnamefont {M.~E.}\ \bibnamefont
  {{Wiedenbeck}}}\ and\ \bibinfo {author} {\bibfnamefont {D.~E.}\ \bibnamefont
  {{Greiner}}},\ }\href {\doibase 10.1086/183310} {\bibfield  {journal}
  {\bibinfo  {journal} {apjl}\ }\textbf {\bibinfo {volume} {239}},\ \bibinfo
  {pages} {L139} (\bibinfo {year} {1980})}\BibitemShut {NoStop}%
\bibitem [{\citenamefont {Hams}\ \emph {et~al.}(2004)\citenamefont {Hams} \emph
  {et~al.}}]{Hams_2004}%
  \BibitemOpen
  \bibfield  {author} {\bibinfo {author} {\bibfnamefont {T.}~\bibnamefont
  {Hams}} \emph {et~al.},\ }\href {\doibase 10.1086/422384} {\bibfield
  {journal} {\bibinfo  {journal} {ApJ}\ }\textbf {\bibinfo {volume} {611}},\
  \bibinfo {pages} {892} (\bibinfo {year} {2004})}\BibitemShut {NoStop}%
\bibitem [{\citenamefont {{Connell}}(1998)}]{Ulysses}%
  \BibitemOpen
  \bibfield  {author} {\bibinfo {author} {\bibfnamefont {J.~J.}\ \bibnamefont
  {{Connell}}},\ }\href {\doibase 10.1086/311437} {\bibfield  {journal}
  {\bibinfo  {journal} {apjl}\ }\textbf {\bibinfo {volume} {501}},\ \bibinfo
  {pages} {L59} (\bibinfo {year} {1998})}\BibitemShut {NoStop}%
\bibitem [{\citenamefont {Webber}\ \emph {et~al.}(2002)\citenamefont {Webber},
  \citenamefont {Lukasiak},\ and\ \citenamefont {Mcdonald}}]{VoyagerMO}%
  \BibitemOpen
  \bibfield  {author} {\bibinfo {author} {\bibfnamefont {W.}~\bibnamefont
  {Webber}}, \bibinfo {author} {\bibfnamefont {A.}~\bibnamefont {Lukasiak}}, \
  and\ \bibinfo {author} {\bibfnamefont {F.}~\bibnamefont {Mcdonald}},\
  }\href@noop {} {\bibfield  {journal} {\bibinfo  {journal} {ApJ}\ }\textbf
  {\bibinfo {volume} {568}},\ \bibinfo {pages} {210} (\bibinfo {year}
  {2002})}\BibitemShut {NoStop}%
\bibitem [{\citenamefont {Zaharijas}\ \emph {et~al.}(2013)\citenamefont
  {Zaharijas}, \citenamefont {Conrad}, \citenamefont {Cuoco},\ and\
  \citenamefont {Yang}}]{zaharijas2012fermi}%
  \BibitemOpen
  \bibfield  {author} {\bibinfo {author} {\bibfnamefont {G.}~\bibnamefont
  {Zaharijas}}, \bibinfo {author} {\bibfnamefont {J.}~\bibnamefont {Conrad}},
  \bibinfo {author} {\bibfnamefont {A.}~\bibnamefont {Cuoco}}, \ and\ \bibinfo
  {author} {\bibfnamefont {Z.}~\bibnamefont {Yang}} (\bibinfo {collaboration}
  {Fermi-LAT}),\ }\href {\doibase 10.1016/j.nuclphysbps.2013.05.014} {\bibfield
   {journal} {\bibinfo  {journal} {Nucl.\ Phys.\ B Proc.\ Suppl.}\ }\textbf
  {\bibinfo {volume} {239-240}},\ \bibinfo {pages} {88} (\bibinfo {year}
  {2013})},\ \Eprint {http://arxiv.org/abs/1212.6755} {arXiv:1212.6755
  [astro-ph.HE]} \BibitemShut {NoStop}%
\bibitem [{\citenamefont {Di~Bernardo}\ \emph {et~al.}(2013)\citenamefont
  {Di~Bernardo}, \citenamefont {Evoli}, \citenamefont {Gaggero}, \citenamefont
  {Grasso},\ and\ \citenamefont {Maccione}}]{di2013cosmic}%
  \BibitemOpen
  \bibfield  {author} {\bibinfo {author} {\bibfnamefont {G.}~\bibnamefont
  {Di~Bernardo}}, \bibinfo {author} {\bibfnamefont {C.}~\bibnamefont {Evoli}},
  \bibinfo {author} {\bibfnamefont {D.}~\bibnamefont {Gaggero}}, \bibinfo
  {author} {\bibfnamefont {D.}~\bibnamefont {Grasso}}, \ and\ \bibinfo {author}
  {\bibfnamefont {L.}~\bibnamefont {Maccione}},\ }\href {\doibase
  10.1088/1475-7516/2013/03/036} {\bibfield  {journal} {\bibinfo  {journal}
  {JCAP}\ }\textbf {\bibinfo {volume} {03}},\ \bibinfo {pages} {036} (\bibinfo
  {year} {2013})},\ \Eprint {http://arxiv.org/abs/1210.4546} {arXiv:1210.4546
  [astro-ph.HE]} \BibitemShut {NoStop}%
\bibitem [{\citenamefont {Beuermann}\ \emph {et~al.}(1985)\citenamefont
  {Beuermann}, \citenamefont {Kanbach},\ and\ \citenamefont
  {Berkhuijsen}}]{beuermann1985radio}%
  \BibitemOpen
  \bibfield  {author} {\bibinfo {author} {\bibfnamefont {K.}~\bibnamefont
  {Beuermann}}, \bibinfo {author} {\bibfnamefont {G.}~\bibnamefont {Kanbach}},
  \ and\ \bibinfo {author} {\bibfnamefont {E.}~\bibnamefont {Berkhuijsen}},\
  }\href@noop {} {\bibfield  {journal} {\bibinfo  {journal} {Astronomy and
  Astrophysics}\ }\textbf {\bibinfo {volume} {153}},\ \bibinfo {pages} {17}
  (\bibinfo {year} {1985})}\BibitemShut {NoStop}%
\bibitem [{\citenamefont {Orlando}\ and\ \citenamefont
  {Strong}(2013)}]{orlando2013galactic}%
  \BibitemOpen
  \bibfield  {author} {\bibinfo {author} {\bibfnamefont {E.}~\bibnamefont
  {Orlando}}\ and\ \bibinfo {author} {\bibfnamefont {A.}~\bibnamefont
  {Strong}},\ }\href {\doibase 10.1093/mnras/stt1718} {\bibfield  {journal}
  {\bibinfo  {journal} {MNRAS}\ }\textbf {\bibinfo {volume} {436}},\ \bibinfo
  {pages} {2127} (\bibinfo {year} {2013})},\ \Eprint
  {http://arxiv.org/abs/1309.2947} {arXiv:1309.2947 [astro-ph.GA]} \BibitemShut
  {NoStop}%
\bibitem [{\citenamefont {{Spanier}}\ and\ \citenamefont
  {{Schlickeiser}}(2005)}]{Spanier_2005}%
  \BibitemOpen
  \bibfield  {author} {\bibinfo {author} {\bibfnamefont {F.}~\bibnamefont
  {{Spanier}}}\ and\ \bibinfo {author} {\bibfnamefont {R.}~\bibnamefont
  {{Schlickeiser}}},\ }\href {\doibase 10.1051/0004-6361:20040364} {\bibfield
  {journal} {\bibinfo  {journal} {aap}\ }\textbf {\bibinfo {volume} {436}},\
  \bibinfo {pages} {9} (\bibinfo {year} {2005})}\BibitemShut {NoStop}%
\bibitem [{\citenamefont {Spangler}\ \emph {et~al.}(2011)\citenamefont
  {Spangler}, \citenamefont {Savage},\ and\ \citenamefont
  {Redfield}}]{Spangler:2010nu}%
  \BibitemOpen
  \bibfield  {author} {\bibinfo {author} {\bibfnamefont {S.~R.}\ \bibnamefont
  {Spangler}}, \bibinfo {author} {\bibfnamefont {A.~H.}\ \bibnamefont
  {Savage}}, \ and\ \bibinfo {author} {\bibfnamefont {S.}~\bibnamefont
  {Redfield}},\ }\href {\doibase 10.1063/1.3625594} {\bibfield  {journal}
  {\bibinfo  {journal} {AIP Conf. Proc.}\ }\textbf {\bibinfo {volume} {1366}},\
  \bibinfo {pages} {97} (\bibinfo {year} {2011})},\ \Eprint
  {http://arxiv.org/abs/1012.4121} {arXiv:1012.4121 [astro-ph.GA]} \BibitemShut
  {NoStop}%
\bibitem [{\citenamefont {{Lerche}}\ and\ \citenamefont
  {{Schlickeiser}}(2001)}]{Lerche_2001}%
  \BibitemOpen
  \bibfield  {author} {\bibinfo {author} {\bibfnamefont {I.}~\bibnamefont
  {{Lerche}}}\ and\ \bibinfo {author} {\bibfnamefont {R.}~\bibnamefont
  {{Schlickeiser}}},\ }\href {\doibase 10.1051/0004-6361:20000242} {\bibfield
  {journal} {\bibinfo  {journal} {aap}\ }\textbf {\bibinfo {volume} {366}},\
  \bibinfo {pages} {1008} (\bibinfo {year} {2001})}\BibitemShut {NoStop}%
\bibitem [{\citenamefont {{Drury, Luke O\'{}C.}}\ and\ \citenamefont {{Strong,
  Andrew W.}}(2017)}]{Drudy_VA_Energetics}%
  \BibitemOpen
  \bibfield  {author} {\bibinfo {author} {\bibnamefont {{Drury, Luke
  O\'{}C.}}}\ and\ \bibinfo {author} {\bibnamefont {{Strong, Andrew W.}}},\
  }\href {\doibase 10.1051/0004-6361/201629526} {\bibfield  {journal} {\bibinfo
   {journal} {A\&A}\ }\textbf {\bibinfo {volume} {597}},\ \bibinfo {pages}
  {A117} (\bibinfo {year} {2017})}\BibitemShut {NoStop}%
\bibitem [{\citenamefont {Ptuskin}\ \emph {et~al.}(2006)\citenamefont
  {Ptuskin}, \citenamefont {Moskalenko}, \citenamefont {Jones}, \citenamefont
  {Strong},\ and\ \citenamefont {Zirakashvili}}]{Ptuskin_2006}%
  \BibitemOpen
  \bibfield  {author} {\bibinfo {author} {\bibfnamefont {V.~S.}\ \bibnamefont
  {Ptuskin}}, \bibinfo {author} {\bibfnamefont {I.~V.}\ \bibnamefont
  {Moskalenko}}, \bibinfo {author} {\bibfnamefont {F.~C.}\ \bibnamefont
  {Jones}}, \bibinfo {author} {\bibfnamefont {A.~W.}\ \bibnamefont {Strong}}, \
  and\ \bibinfo {author} {\bibfnamefont {V.~N.}\ \bibnamefont {Zirakashvili}},\
  }\href {\doibase 10.1086/501117} {\bibfield  {journal} {\bibinfo  {journal}
  {The Astrophysical Journal}\ }\textbf {\bibinfo {volume} {642}},\ \bibinfo
  {pages} {902} (\bibinfo {year} {2006})}\BibitemShut {NoStop}%
\bibitem [{\citenamefont {Fornieri}\ \emph {et~al.}(2021)\citenamefont
  {Fornieri}, \citenamefont {Gaggero}, \citenamefont {Cerri}, \citenamefont
  {De~la Torre~Luque},\ and\ \citenamefont {Gabici}}]{Fornieri:2020wrr}%
  \BibitemOpen
  \bibfield  {author} {\bibinfo {author} {\bibfnamefont {O.}~\bibnamefont
  {Fornieri}}, \bibinfo {author} {\bibfnamefont {D.}~\bibnamefont {Gaggero}},
  \bibinfo {author} {\bibfnamefont {S.~S.}\ \bibnamefont {Cerri}}, \bibinfo
  {author} {\bibfnamefont {P.}~\bibnamefont {De~la Torre~Luque}}, \ and\
  \bibinfo {author} {\bibfnamefont {S.}~\bibnamefont {Gabici}},\ }\href
  {\doibase 10.1093/mnras/stab355} {\bibfield  {journal} {\bibinfo  {journal}
  {Mon. Not. Roy. Astron. Soc.}\ }\textbf {\bibinfo {volume} {502}},\ \bibinfo
  {pages} {5821} (\bibinfo {year} {2021})},\ \Eprint
  {http://arxiv.org/abs/2011.09197} {arXiv:2011.09197 [astro-ph.HE]}
  \BibitemShut {NoStop}%
\bibitem [{\citenamefont {Ackermann}\ \emph
  {et~al.}(2012{\natexlab{a}})\citenamefont {Ackermann} \emph
  {et~al.}}]{Ackermann_2012}%
  \BibitemOpen
  \bibfield  {author} {\bibinfo {author} {\bibfnamefont {M.}~\bibnamefont
  {Ackermann}} \emph {et~al.},\ }\href {\doibase 10.1088/0004-637x/750/1/3}
  {\bibfield  {journal} {\bibinfo  {journal} {The Astrophysical Journal}\
  }\textbf {\bibinfo {volume} {750}},\ \bibinfo {pages} {3} (\bibinfo {year}
  {2012}{\natexlab{a}})}\BibitemShut {NoStop}%
\bibitem [{\citenamefont {Acero}\ \emph {et~al.}(2016)\citenamefont {Acero}
  \emph {et~al.}}]{Acero_2016}%
  \BibitemOpen
  \bibfield  {author} {\bibinfo {author} {\bibfnamefont {F.}~\bibnamefont
  {Acero}} \emph {et~al.},\ }\href {\doibase 10.3847/0067-0049/223/2/26}
  {\bibfield  {journal} {\bibinfo  {journal} {The Astrophysical Journal
  Supplement Series}\ }\textbf {\bibinfo {volume} {223}},\ \bibinfo {pages}
  {26} (\bibinfo {year} {2016})}\BibitemShut {NoStop}%
\bibitem [{\citenamefont {Amato}(2014)}]{Amato:2013fua}%
  \BibitemOpen
  \bibfield  {author} {\bibinfo {author} {\bibfnamefont {E.}~\bibnamefont
  {Amato}},\ }\href {\doibase 10.1142/S2010194514601604} {\bibfield  {journal}
  {\bibinfo  {journal} {Int. J. Mod. Phys. Conf. Ser.}\ }\textbf {\bibinfo
  {volume} {28}},\ \bibinfo {pages} {1460160} (\bibinfo {year} {2014})},\
  \Eprint {http://arxiv.org/abs/1312.5945} {arXiv:1312.5945 [astro-ph.HE]}
  \BibitemShut {NoStop}%
\bibitem [{\citenamefont {Serpico}(2009)}]{Serpico_efraction}%
  \BibitemOpen
  \bibfield  {author} {\bibinfo {author} {\bibfnamefont {P.~D.}\ \bibnamefont
  {Serpico}},\ }\href {\doibase 10.1103/PhysRevD.79.021302} {\bibfield
  {journal} {\bibinfo  {journal} {Phys. Rev. D}\ }\textbf {\bibinfo {volume}
  {79}},\ \bibinfo {pages} {021302} (\bibinfo {year} {2009})}\BibitemShut
  {NoStop}%
\bibitem [{\citenamefont {Cirelli}\ \emph
  {et~al.}(2014{\natexlab{b}})\citenamefont {Cirelli}, \citenamefont {Gaggero},
  \citenamefont {Giesen}, \citenamefont {Taoso},\ and\ \citenamefont
  {Urbano}}]{Cirelli:2014lwa}%
  \BibitemOpen
  \bibfield  {author} {\bibinfo {author} {\bibfnamefont {M.}~\bibnamefont
  {Cirelli}}, \bibinfo {author} {\bibfnamefont {D.}~\bibnamefont {Gaggero}},
  \bibinfo {author} {\bibfnamefont {G.}~\bibnamefont {Giesen}}, \bibinfo
  {author} {\bibfnamefont {M.}~\bibnamefont {Taoso}}, \ and\ \bibinfo {author}
  {\bibfnamefont {A.}~\bibnamefont {Urbano}},\ }\href {\doibase
  10.1088/1475-7516/2014/12/045} {\bibfield  {journal} {\bibinfo  {journal}
  {JCAP}\ }\textbf {\bibinfo {volume} {12}},\ \bibinfo {pages} {045} (\bibinfo
  {year} {2014}{\natexlab{b}})},\ \Eprint {http://arxiv.org/abs/1407.2173}
  {arXiv:1407.2173 [hep-ph]} \BibitemShut {NoStop}%
\bibitem [{\citenamefont {Gaggero}\ \emph {et~al.}(2015)\citenamefont
  {Gaggero}, \citenamefont {Grasso}, \citenamefont {Marinelli}, \citenamefont
  {Urbano},\ and\ \citenamefont {Valli}}]{Gaggero:2015xza}%
  \BibitemOpen
  \bibfield  {author} {\bibinfo {author} {\bibfnamefont {D.}~\bibnamefont
  {Gaggero}}, \bibinfo {author} {\bibfnamefont {D.}~\bibnamefont {Grasso}},
  \bibinfo {author} {\bibfnamefont {A.}~\bibnamefont {Marinelli}}, \bibinfo
  {author} {\bibfnamefont {A.}~\bibnamefont {Urbano}}, \ and\ \bibinfo {author}
  {\bibfnamefont {M.}~\bibnamefont {Valli}},\ }\href {\doibase
  10.1088/2041-8205/815/2/L25} {\bibfield  {journal} {\bibinfo  {journal}
  {Astrophys. J. Lett.}\ }\textbf {\bibinfo {volume} {815}},\ \bibinfo {pages}
  {L25} (\bibinfo {year} {2015})},\ \Eprint {http://arxiv.org/abs/1504.00227}
  {arXiv:1504.00227 [astro-ph.HE]} \BibitemShut {NoStop}%
\bibitem [{\citenamefont {Moskalenko}\ \emph {et~al.}(2002)\citenamefont
  {Moskalenko}, \citenamefont {Strong}, \citenamefont {Ormes},\ and\
  \citenamefont {Potgieter}}]{Moskalenko:2001ya}%
  \BibitemOpen
  \bibfield  {author} {\bibinfo {author} {\bibfnamefont {I.~V.}\ \bibnamefont
  {Moskalenko}}, \bibinfo {author} {\bibfnamefont {A.~W.}\ \bibnamefont
  {Strong}}, \bibinfo {author} {\bibfnamefont {J.~F.}\ \bibnamefont {Ormes}}, \
  and\ \bibinfo {author} {\bibfnamefont {M.~S.}\ \bibnamefont {Potgieter}},\
  }\href {\doibase 10.1086/324402} {\bibfield  {journal} {\bibinfo  {journal}
  {Astrophys. J.}\ }\textbf {\bibinfo {volume} {565}},\ \bibinfo {pages} {280}
  (\bibinfo {year} {2002})},\ \Eprint {http://arxiv.org/abs/astro-ph/0106567}
  {arXiv:astro-ph/0106567} \BibitemShut {NoStop}%
\bibitem [{\citenamefont {Ackermann}\ \emph
  {et~al.}(2012{\natexlab{b}})\citenamefont {Ackermann} \emph
  {et~al.}}]{Ackermann:2012pya}%
  \BibitemOpen
  \bibfield  {author} {\bibinfo {author} {\bibfnamefont {M.}~\bibnamefont
  {Ackermann}} \emph {et~al.} (\bibinfo {collaboration} {Fermi-LAT}),\ }\href
  {\doibase 10.1088/0004-637X/750/1/3} {\bibfield  {journal} {\bibinfo
  {journal} {Astrophys. J.}\ }\textbf {\bibinfo {volume} {750}},\ \bibinfo
  {pages} {3} (\bibinfo {year} {2012}{\natexlab{b}})},\ \Eprint
  {http://arxiv.org/abs/1202.4039} {arXiv:1202.4039 [astro-ph.HE]} \BibitemShut
  {NoStop}%
\bibitem [{\citenamefont {Casandjian}\ and\ \citenamefont
  {Grenier}(2009)}]{Casandjian:2009wq}%
  \BibitemOpen
  \bibfield  {author} {\bibinfo {author} {\bibfnamefont {J.-M.}\ \bibnamefont
  {Casandjian}}\ and\ \bibinfo {author} {\bibfnamefont {I.}~\bibnamefont
  {Grenier}} (\bibinfo {collaboration} {Fermi-LAT}),\ }\href@noop {} {\enquote
  {\bibinfo {title} {{High Energy Gamma-Ray Emission from the Loop I
  region}},}\ } (\bibinfo {year} {2009}),\ \Eprint
  {http://arxiv.org/abs/0912.3478} {arXiv:0912.3478 [astro-ph.HE]} \BibitemShut
  {NoStop}%
\bibitem [{\citenamefont {Ackermann}\ \emph {et~al.}(2014)\citenamefont
  {Ackermann} \emph {et~al.}}]{Fermi-LAT:2014sfa}%
  \BibitemOpen
  \bibfield  {author} {\bibinfo {author} {\bibfnamefont {M.}~\bibnamefont
  {Ackermann}} \emph {et~al.} (\bibinfo {collaboration} {Fermi-LAT}),\ }\href
  {\doibase 10.1088/0004-637X/793/1/64} {\bibfield  {journal} {\bibinfo
  {journal} {Astrophys. J.}\ }\textbf {\bibinfo {volume} {793}},\ \bibinfo
  {pages} {64} (\bibinfo {year} {2014})},\ \Eprint
  {http://arxiv.org/abs/1407.7905} {arXiv:1407.7905 [astro-ph.HE]} \BibitemShut
  {NoStop}%
\bibitem [{\citenamefont {Fornieri}\ \emph {et~al.}(2020)\citenamefont
  {Fornieri}, \citenamefont {Gaggero},\ and\ \citenamefont
  {Grasso}}]{Fornieri:2019ddi}%
  \BibitemOpen
  \bibfield  {author} {\bibinfo {author} {\bibfnamefont {O.}~\bibnamefont
  {Fornieri}}, \bibinfo {author} {\bibfnamefont {D.}~\bibnamefont {Gaggero}}, \
  and\ \bibinfo {author} {\bibfnamefont {D.}~\bibnamefont {Grasso}},\ }\href
  {\doibase 10.1088/1475-7516/2020/02/009} {\bibfield  {journal} {\bibinfo
  {journal} {JCAP}\ }\textbf {\bibinfo {volume} {02}},\ \bibinfo {pages} {009}
  (\bibinfo {year} {2020})},\ \Eprint {http://arxiv.org/abs/1907.03696}
  {arXiv:1907.03696 [astro-ph.HE]} \BibitemShut {NoStop}%
\bibitem [{\citenamefont {{De Angelis}}\ \emph {et~al.}(2018)\citenamefont {{De
  Angelis}} \emph {et~al.}}]{DEANGELIS20181}%
  \BibitemOpen
  \bibfield  {author} {\bibinfo {author} {\bibfnamefont {A.}~\bibnamefont {{De
  Angelis}}} \emph {et~al.},\ }\href {\doibase
  https://doi.org/10.1016/j.jheap.2018.07.001} {\bibfield  {journal} {\bibinfo
  {journal} {Journal of High Energy Astrophysics}\ }\textbf {\bibinfo {volume}
  {19}},\ \bibinfo {pages} {1 } (\bibinfo {year} {2018})}\BibitemShut {NoStop}%
\bibitem [{\citenamefont {Benhabiles-Mezhoud}\ \emph
  {et~al.}(2012)\citenamefont {Benhabiles-Mezhoud}, \citenamefont {Kiener},
  \citenamefont {Tatischeff},\ and\ \citenamefont {Strong}}]{gamma_lines}%
  \BibitemOpen
  \bibfield  {author} {\bibinfo {author} {\bibfnamefont {H.}~\bibnamefont
  {Benhabiles-Mezhoud}}, \bibinfo {author} {\bibfnamefont {J.}~\bibnamefont
  {Kiener}}, \bibinfo {author} {\bibfnamefont {V.}~\bibnamefont {Tatischeff}},
  \ and\ \bibinfo {author} {\bibfnamefont {A.}~\bibnamefont {Strong}},\ }\href
  {\doibase 10.1088/0004-637X/763/2/98} {\bibfield  {journal} {\bibinfo
  {journal} {The Astrophysical Journal}\ }\textbf {\bibinfo {volume} {763}}
  (\bibinfo {year} {2012}),\ 10.1088/0004-637X/763/2/98}\BibitemShut {NoStop}%
\bibitem [{\citenamefont {Adriani}\ \emph {et~al.}(2011)\citenamefont {Adriani}
  \emph {et~al.}}]{PamHeData}%
  \BibitemOpen
  \bibfield  {author} {\bibinfo {author} {\bibfnamefont {O.}~\bibnamefont
  {Adriani}} \emph {et~al.},\ }\href {\doibase 10.1126/science.1199172}
  {\bibfield  {journal} {\bibinfo  {journal} {Science}\ }\textbf {\bibinfo
  {volume} {332}},\ \bibinfo {pages} {69} (\bibinfo {year} {2011})},\ \Eprint
  {http://arxiv.org/abs/https://www.science.org/doi/pdf/10.1126/science.1199172}
  {https://www.science.org/doi/pdf/10.1126/science.1199172} \BibitemShut
  {NoStop}%
\bibitem [{\citenamefont {Pshirkov}\ \emph {et~al.}(2011)\citenamefont
  {Pshirkov}, \citenamefont {Tinyakov}, \citenamefont {Kronberg},\ and\
  \citenamefont {Newton-McGee}}]{pshirkov2011deriving}%
  \BibitemOpen
  \bibfield  {author} {\bibinfo {author} {\bibfnamefont {M.}~\bibnamefont
  {Pshirkov}}, \bibinfo {author} {\bibfnamefont {P.}~\bibnamefont {Tinyakov}},
  \bibinfo {author} {\bibfnamefont {P.}~\bibnamefont {Kronberg}}, \ and\
  \bibinfo {author} {\bibfnamefont {K.}~\bibnamefont {Newton-McGee}},\
  }\href@noop {} {\bibfield  {journal} {\bibinfo  {journal} {The Astrophysical
  Journal}\ }\textbf {\bibinfo {volume} {738}},\ \bibinfo {pages} {192}
  (\bibinfo {year} {2011})}\BibitemShut {NoStop}%
\bibitem [{\citenamefont {Porter}\ \emph {et~al.}(2008)\citenamefont {Porter},
  \citenamefont {Moskalenko}, \citenamefont {Strong}, \citenamefont {Orlando},\
  and\ \citenamefont {Bouchet}}]{porter2008inverse}%
  \BibitemOpen
  \bibfield  {author} {\bibinfo {author} {\bibfnamefont {T.~A.}\ \bibnamefont
  {Porter}}, \bibinfo {author} {\bibfnamefont {I.~V.}\ \bibnamefont
  {Moskalenko}}, \bibinfo {author} {\bibfnamefont {A.~W.}\ \bibnamefont
  {Strong}}, \bibinfo {author} {\bibfnamefont {E.}~\bibnamefont {Orlando}}, \
  and\ \bibinfo {author} {\bibfnamefont {L.}~\bibnamefont {Bouchet}},\
  }\href@noop {} {\bibfield  {journal} {\bibinfo  {journal} {The Astrophysical
  Journal}\ }\textbf {\bibinfo {volume} {682}},\ \bibinfo {pages} {400}
  (\bibinfo {year} {2008})}\BibitemShut {NoStop}%
\bibitem [{\citenamefont {Stone}\ \emph {et~al.}(2013)\citenamefont {Stone},
  \citenamefont {Cummings}, \citenamefont {McDonald}, \citenamefont {Heikkila},
  \citenamefont {Lal},\ and\ \citenamefont {Webber}}]{Stone150}%
  \BibitemOpen
  \bibfield  {author} {\bibinfo {author} {\bibfnamefont {E.~C.}\ \bibnamefont
  {Stone}}, \bibinfo {author} {\bibfnamefont {A.~C.}\ \bibnamefont {Cummings}},
  \bibinfo {author} {\bibfnamefont {F.~B.}\ \bibnamefont {McDonald}}, \bibinfo
  {author} {\bibfnamefont {B.~C.}\ \bibnamefont {Heikkila}}, \bibinfo {author}
  {\bibfnamefont {N.}~\bibnamefont {Lal}}, \ and\ \bibinfo {author}
  {\bibfnamefont {W.~R.}\ \bibnamefont {Webber}},\ }\href {\doibase
  10.1126/science.1236408} {\bibfield  {journal} {\bibinfo  {journal}
  {Science}\ }\textbf {\bibinfo {volume} {341}},\ \bibinfo {pages} {150}
  (\bibinfo {year} {2013})},\ \Eprint
  {http://arxiv.org/abs/https://science.sciencemag.org/content/341/6142/150.full.pdf}
  {https://science.sciencemag.org/content/341/6142/150.full.pdf} \BibitemShut
  {NoStop}%
\bibitem [{\citenamefont {Ackermann}\ \emph {et~al.}(2010)\citenamefont
  {Ackermann} \emph {et~al.}}]{Ackermann:2010ij}%
  \BibitemOpen
  \bibfield  {author} {\bibinfo {author} {\bibfnamefont {M.}~\bibnamefont
  {Ackermann}} \emph {et~al.} (\bibinfo {collaboration} {Fermi-LAT}),\ }\href
  {\doibase 10.1103/PhysRevD.82.092004} {\bibfield  {journal} {\bibinfo
  {journal} {Phys. Rev. D}\ }\textbf {\bibinfo {volume} {82}},\ \bibinfo
  {pages} {092004} (\bibinfo {year} {2010})},\ \Eprint
  {http://arxiv.org/abs/1008.3999} {arXiv:1008.3999 [astro-ph.HE]} \BibitemShut
  {NoStop}%
\bibitem [{\citenamefont {Ambrosi}\ \emph {et~al.}(2017)\citenamefont {Ambrosi}
  \emph {et~al.}}]{Ambrosi:2017wek}%
  \BibitemOpen
  \bibfield  {author} {\bibinfo {author} {\bibfnamefont {G.}~\bibnamefont
  {Ambrosi}} \emph {et~al.} (\bibinfo {collaboration} {DAMPE}),\ }\href
  {\doibase 10.1038/nature24475} {\bibfield  {journal} {\bibinfo  {journal}
  {Nature}\ }\textbf {\bibinfo {volume} {552}},\ \bibinfo {pages} {63}
  (\bibinfo {year} {2017})},\ \Eprint {http://arxiv.org/abs/1711.10981}
  {arXiv:1711.10981 [astro-ph.HE]} \BibitemShut {NoStop}%
\bibitem [{\citenamefont {Adriani}\ \emph {et~al.}(2018)\citenamefont {Adriani}
  \emph {et~al.}}]{CALET_electrons}%
  \BibitemOpen
  \bibfield  {author} {\bibinfo {author} {\bibfnamefont {O.}~\bibnamefont
  {Adriani}} \emph {et~al.} (\bibinfo {collaboration} {CALET Collaboration}),\
  }\href {\doibase 10.1103/PhysRevLett.120.261102} {\bibfield  {journal}
  {\bibinfo  {journal} {Phys. Rev. Lett.}\ }\textbf {\bibinfo {volume} {120}},\
  \bibinfo {pages} {261102} (\bibinfo {year} {2018})}\BibitemShut {NoStop}%
\bibitem [{\citenamefont {Aguilar}\ \emph {et~al.}(2014)\citenamefont {Aguilar}
  \emph {et~al.}}]{Aguilar:2014mma}%
  \BibitemOpen
  \bibfield  {author} {\bibinfo {author} {\bibfnamefont {M.}~\bibnamefont
  {Aguilar}} \emph {et~al.} (\bibinfo {collaboration} {AMS}),\ }\href {\doibase
  10.1103/PhysRevLett.113.121102} {\bibfield  {journal} {\bibinfo  {journal}
  {Phys. Rev. Lett.}\ }\textbf {\bibinfo {volume} {113}},\ \bibinfo {pages}
  {121102} (\bibinfo {year} {2014})}\BibitemShut {NoStop}%
\end{thebibliography}%

\newpage
\appendix
\section{Summary of the MCMC results}
\label{sec:appendixA}

In this appendix, the results obtained from the MCMC algorithm of each secondary-to-primary ratio (B/C, B/O, Be/C, Be/O, Li/C, Li/O), in table~\ref{tab:MCMC_table}, and their combined analysis, in table~\ref{tab:MCMC_Combined}, for both diffusion parameterisations used ('Source' and 'Diffusion' hypotheses) are summarized. They contain the median value (coinciding with the maximum posterior probability value) $\pm$ the $1\sigma$ uncertainty related to its determination, assuming their PDFS to be Gaussian, and the actual range of values contained in the 95\% probability of the distribution. 

\vspace{0.65cm}
\textbf{\large{Independent analyses}}

\begin{table}[htb!]
\centering
\resizebox*{0.81\columnwidth}{0.17\textheight}{

\begin{tabular}{|c|c|c|c|c|c|c|c|}
\hline
\multicolumn{7}{|c|}{\textbf{\large{Source hypothesis}}} \\
\hline &\textbf{B/C} & \textbf{B/O} & \textbf{Be/C} & \textbf{Be/O} & \textbf{Li/C} & \textbf{Li/O} \\ 
\hline

\multirow{2}{7em}{\centering $\frac{D_0 \, (10^{28}\units{cm^{2}/s})}{H\, (kpc)}$} & 0.87 $\pm$ 0.03 & 0.97$\pm$ 0.04   &  1.11$\pm$ 0.02 &  1.06 $\pm$ 0.05 & 1.22  $\pm$ 0.07 &  1.31 $\pm$ 0.06  \\  &  [0.82, 0.93]  &   [0.89, 1.05] &   [1.06, 1.16]  &  [0.96, 1.17] &   [1.1, 1.27] &   [1.15, 1.41] \\ \hline 

\multirow{2}{6em}{\centering\large{$v_A$ (km/s)}}  & 25.86  $\pm$ 2.16 & 32.4 $\pm$ 1.86 &  5.74 $\pm$ 5.6  &  17.02 $\pm$ 6.8 &  24.3 $\pm$ 6.11 &  37.03 $\pm$ 3.73 \\  &  [21.54, 30.16]  &   [29.01, 35.79] &   [0., 16.94]  &  [5.52, 29.14] &   [13.98, 36.52] &   [29.55, 40.83] \\ \hline 

\multirow{2}{6em}{\centering\large{$\eta$}}    & -0.53 $\pm$ 0.13 & -0.51 $\pm$ 0.11 &  -1.88 $\pm$ 0.18 &  -1.85 $\pm$ 0.28 & -1.91 $\pm$ 0.34 &  -1.23 $\pm$ 0.24  \\ &  [-1.29, -0.8]  &   [-0.75, -0.31] &   [-2.19, -1.52]  &  [-2.35, -1.29] &   [-2.41, -1.29] &   [-1.75, -0.92] \\ \hline 

\multirow{2}{6em}{\centering\large{$\delta$}}  & 0.43 $\pm$ 0.01 & 0.40 $\pm$ 0.01 &  0.41 $\pm$ 0.02  &  0.42 $\pm$ 0.01 & 0.39 $\pm$ 0.01 &  0.37 $\pm$ 0.02  \\  &  [0.41, 0.45]  &   [0.38, 0.43] &   [0.39, 0.44]  &  [0.4, 0.44] &   [0.37, 0.41] &   [0.35, 0.4] \\ \hline 
\end{tabular}
}

\vspace{0.5 cm}
\resizebox*{0.81\columnwidth}{0.17\textheight}{

\begin{tabular}{|c|c|c|c|c|c|c|c|}
\hline
\multicolumn{7}{|c|}{\textbf{\large{Diffusion hypothesis}}} \\
\hline &\textbf{B/C} & \textbf{B/O} & \textbf{Be/C} & \textbf{Be/O} & \textbf{Li/C} & \textbf{Li/O} \\ 
\hline

\multirow{2}{7em}{\centering $\frac{D_0 \, (10^{28}\units{cm^{2}/s})}{H\, (kpc)} $} & 0.82 $\pm$ 0.03 & 0.91 $\pm$ 0.04   &  1.1 $\pm$ 0.02 &  1.03 $\pm$ 0.05 & 1.18  $\pm$ 0.04 &  1.23 $\pm$ 0.07  \\  &  [0.88, 0.76]  &   [0.99, 0.83] &   [1.14, 1.06]  &  [0.94, 1.13] &   [1.11, 1.26] &   [1.09, 1.37] \\ \hline 

\multirow{2}{6em}{\centering\large{$v_A$ (km/s)}}  & 23.26  $\pm$ 2.25 & 30.3 $\pm$ 2.01 &  5.06 $\pm$ 5.2  &  13.19 $\pm$ 6.76 &  20.82 $\pm$ 3.82 &  33.1 $\pm$ 4.78 \\  &  [18.6, 27.71]  &   [26.14, 34.16] &   [0., 15.29]  &  [0., 25.87] &   [13.18, 28.44] &   [23.54, 40.64] \\ \hline 

\multirow{2}{6em}{\centering\large{$\eta$}}    & -0.67 $\pm$ 0.13 & -0.66 $\pm$ 0.11 &  -1.97 $\pm$ 0.17 &  -2.02 $\pm$ 0.26 & -2.13 $\pm$ 0.19 &  -1.52 $\pm$ 0.32  \\ &  [-0.93, -0.41]  &   [-0.88, -0.44] &   [-2.26, -1.63]  &  [-2.4, -1.74] &   [-2.37, -1.73] &   [-2.16, -1.04] \\ \hline 

\multirow{2}{6em}{\centering\large{$\delta$}}  & 0.45 $\pm$ 0.01 & 0.42 $\pm$ 0.01 &  0.42 $\pm$ 0.01  &  0.43 $\pm$ 0.02 & 0.40 $\pm$ 0.02 &  0.38 $\pm$ 0.02  \\  &  [0.43, 0.47]  &   [0.40, 0.44] &   [0.40, 0.44]  &  [0.41, 0.45] &  [0.38, 0.43] &   [0.37, 0.43] \\ \hline 
\end{tabular}
}

\caption{Results obtained in the independent analysis of the flux ratios B/C, Be/C, Li/C, B/O, Be/O and Li/O for the diffusion coefficient parameterisations of Eqs.~\ref{eq:sourcehyp} and~\ref{eq:breakhyp} for {\tt FLUKA} the cross sections derived in this work. The halo size value, $H$, obtained from the $^{10}$Be flux ratios was $7.5$~kpc}
\label{tab:MCMC_table}
\end{table}

\vspace{0.4 cm}
\textbf{\large{Combined analysis}}

\begin{table}[htb!]
\centering

\resizebox*{0.66\columnwidth}{0.2\textheight}{

\begin{tabular}{|c|c|c|c|}
\hline
\multicolumn{4}{|c|}{\textbf{\large{Source hypothesis}}} \\
\hline \multirow{2}{0.1 em} \centering \textbf{$\frac{D_0 \, (10^{28}\units{cm^{2}/s})}{H\, (kpc)}$} & \textbf{\large{$v_A$\units{(km/s)}}} & \textbf{\large{$\eta$}} & \textbf{\large{$\delta$}} \\ 
\hline
1.33 $\pm$ 0.02 & 39.8 $\pm$ 0.15   &  -0.72 $\pm$ 0.05 &  0.35 $\pm$ 0.02  \\ 

[1.29, 1.37]  &   [39.59, 40.02] &   [-0.83, -0.62]  &  [0.33, 0.39]  \\    \hline 

\multicolumn{4}{|c|}{} \\ \hline
\multicolumn{4}{|c|}{\textbf{\large{Diffusion hypothesis}}} \\
\hline \multirow{2}{0.1 em} \centering \textbf{$\frac{D_0 \, (10^{28}\units{cm^{2}/s})}{H\, (kpc)}$} & \textbf{\large{$v_A$\units{(km/s)}}} & \textbf{\large{$\eta$}} & \textbf{\large{$\delta$}} \\ 
\hline
1.31 $\pm$ 0.02 & 39.7 $\pm$ 0.33   &   -0.81 $\pm$ 0.07 &  0.36 $\pm$ 0.01  \\ 

[1.27, 1.35] &  [39.98, 40.1]  &   [-0.95, -0.67]  &  [0.34, 0.39]  \\    \hline 

\end{tabular}
}

\caption{Same as in table \ref{tab:MCMC_table} but for the combined analysis of the B, Be and Li flux ratios.}
\label{tab:MCMC_Combined}
\end{table}

\vskip 0.5cm

Here, we report tables with the $\chi^2$ values obtained with the best-fit parameters of the independent and combined analyses FLUKA cross sections. Interestingly, we observe that the difference between the $\chi^2$ values from the source and diffusion hypotheses in the independent analyses are of $\Delta \chi^2 \sim 6$, for the B and Li ratios. This yields a significance of $\sim 3.5 \sigma$ (making use of Wilk's theorem), which is $1.2\sigma$ less significant than for the DRAGON2 and GALPROP parameterisations, as studied in Ref.~\cite{Luque_MCMC}. This is due to the inclusion of the rise of cross sections at high energies which is not taken into account in the parameterisations and implies that the interpretation of the origin of the break in the CR spectra is sensitive to the cross sections used.

\hskip 0.4cm
\begin{table}[htb!]
\centering
\resizebox*{0.75\columnwidth}{0.17\textheight}{
\begin{tabular}{|c|c|c|c|c|c|c|c|}
\hline
\multicolumn{7}{|c|}{\textbf{\large{$\chi^2$ values of best fit parameters}}} \\
\hline &\textbf{B/C} & \textbf{B/O} & \textbf{Be/C} & \textbf{Be/O} & \textbf{Li/C} & \textbf{Li/O} \\ 
\hline
\multirow{2}{10em}{\centering\large{Source hypothesis Indep. analysis}} & \multirow{2}{*}{35.45} & \multirow{2}{*}{27.49}  &  \multirow{2}{*}{59.01} &  \multirow{2}{*}{55.24} & \multirow{2}{*}{57.87} &  \multirow{2}{*}{40.45}  \\ & & & & & & \\  \hline 

\multirow{2}{10em}{\centering\large{Diffusion hypothesis Indep. analysis}}  & \multirow{2}{*}{29.43} & \multirow{2}{*}{22.99}  &  \multirow{2}{*}{44.47} &  \multirow{2}{*}{47.6} & \multirow{2}{*}{51.38} &  \multirow{2}{*}{34.13}  \\ & & & & & & \\  \hline 

\multirow{2}{10em}{\centering\large{Source hypothesis Comb. analysis}}  & \multirow{2}{*}{51.79} & \multirow{2}{*}{32.8}  &  \multirow{2}{*}{72.33} &  \multirow{2}{*}{55.92} & \multirow{2}{*}{63.44} &  \multirow{2}{*}{66.56}  \\ & & & & & & \\  \hline 

\multirow{2}{10em}{\centering\large{Diffusion hypothesis Comb. analysis}} & \multirow{2}{*}{48.17} & \multirow{2}{*}{32.53}  &  \multirow{2}{*}{62.77} &  \multirow{2}{*}{54.3} & \multirow{2}{*}{60.0} &  \multirow{2}{*}{63.52}  \\ & & & & & & \\  \hline 
\end{tabular}}
\caption{$\chi^2$ values of the best-fit diffusion parameters found for the MCMC analyses with the {\tt FLUKA} cross sections.}
\label{tab:Xi_FLUKA}
\end{table}

\newpage
\section{$^3$He FLUKA cross sections}
\label{sec:appendixB}

\begin{figure}[!hb]
	\centering
	\includegraphics[width=0.49\textwidth]{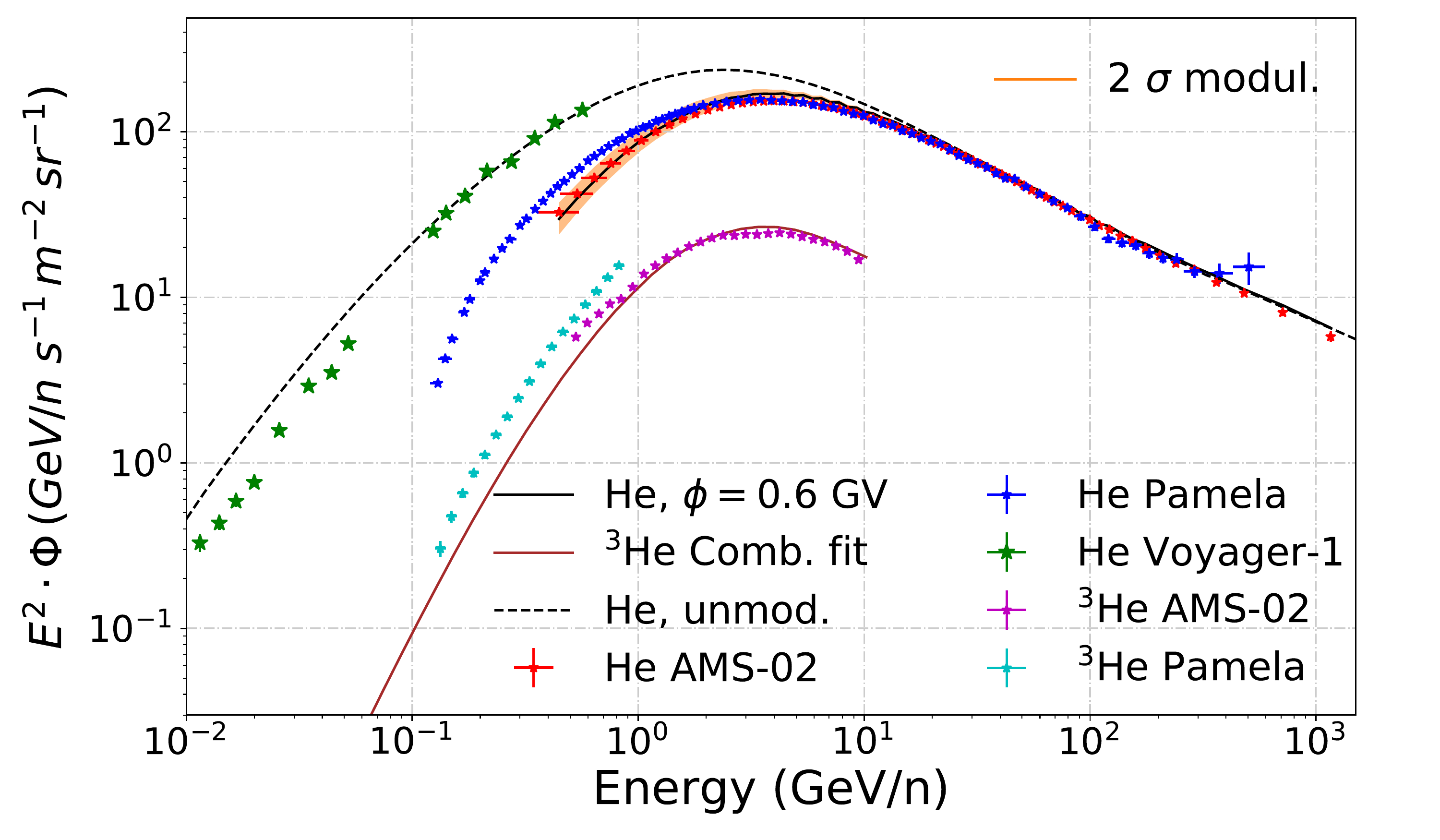} 
	\includegraphics[width=0.48\textwidth]{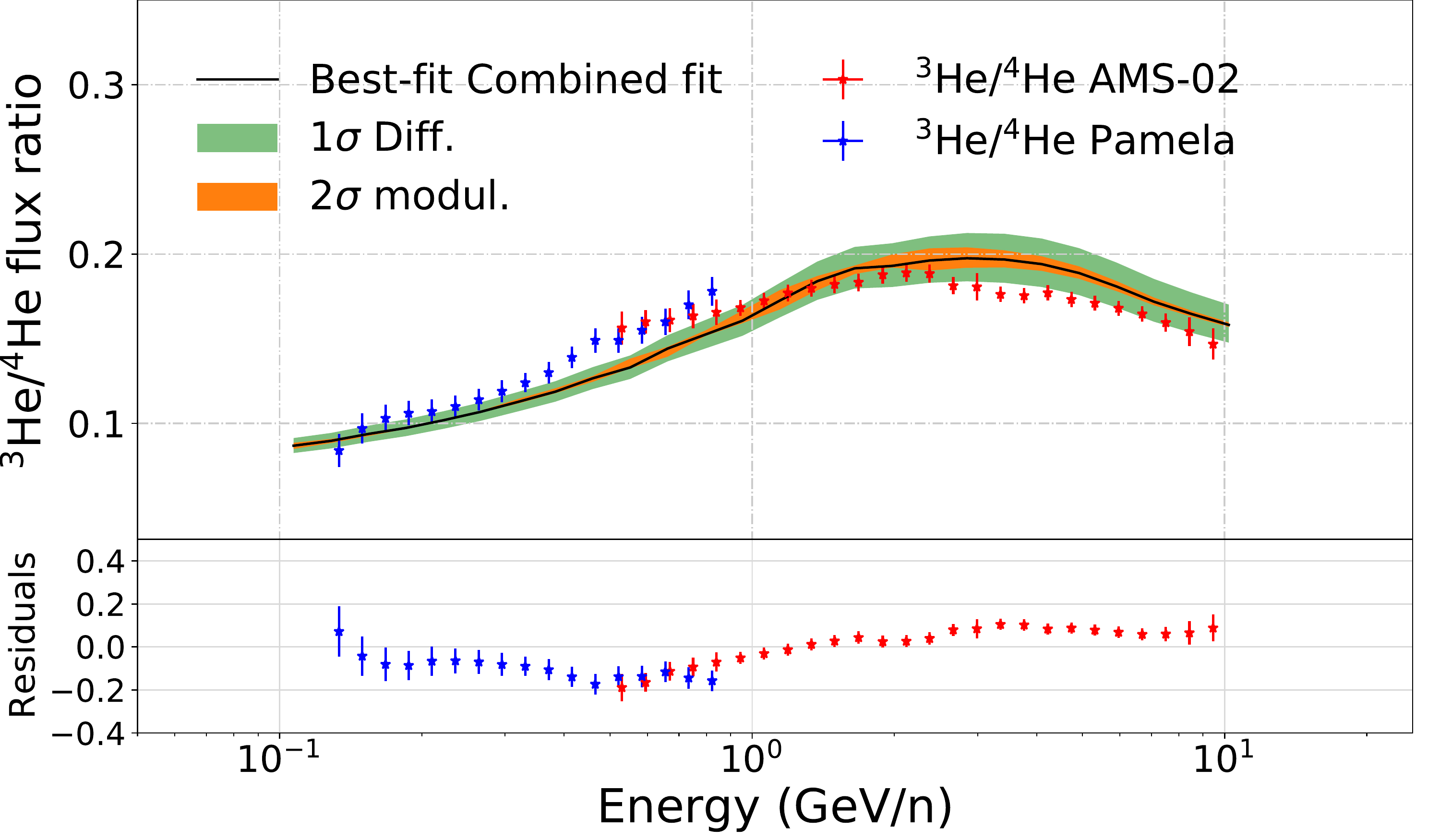} 
	\caption{\textbf{Left panel:} Modulated and unmodulated predicted spectra of He and $^3$He compared to AMS-02 and Voyager 1 and Pamela data. \textbf{Left panel:} Predicted $^3$He/$^4$He flux ratio compared to Pamela and AMS-02 data. The residuals, calculated as model-data/model are also shown. These spectra are calculated with the diffusion coefficient obtained from the combined fit.}
	\label{fig:He3_Fluka}
\end{figure}

In this appendix we report the predicted flux of another of the main light secondary CRs, the $^3$He isotope. We have computed the full $^3$He production cross sections from isotopes up to Iron ($Z=26$). The contribution to the total $^3$He flux from nuclei heavier than He is, in average, $\sim 44\%$. As we observe from Figure~\ref{fig:He3_Fluka}, the predicted $^3$He flux, with the diffusion coefficient obtained in the combined analysis, is reasonably compatible with data in both, energy-dependence and intensity. We also include data from the PAMELA mission~\cite{PamHeData} in this figure, taken in the period from 2006/07 to 2008/12. It is worth mentioning that the predicted $^3$He flux with the diffusion coefficient obtained in the independent B/C analysis is overpredicted in the energy region between $1$~GeV and $10$~GeV by around $20\%$.

\newpage
\section{Electrons and protons spectra}
\label{sec:appendixD}

In this appendix we show the evaluated spectra of proton, helium and leptons. These are obtained by tuning the injection spectra to reproduce experimental data. In the case of protons and helium, we also include the Voyager-01 data. The reproduction of the CR electrons spectrum (CRE) involves also choosing a configuration of the magnetic field and the radiation fields. The {\tt DRAGON2} code allows the parameterisation of three components of the magnetic field (see Ref.~\cite{pshirkov2011deriving}): one for the disc, one for the halo and a turbulent one. We set the magnetic field parameters to $B_{disc}$ = $2\mu G$, $B_{halo}$ = $4\mu G$ and $B_{turb}$ = $7.5\mu G$ as found in Ref.~\cite{Fornieri:2019ddi}.  The energy density of the radiation fields, important for the energy losses related to IC scattering, were taken from Ref.~\cite{porter2008inverse}. 

\begin{figure}[!hb]
	\centering
	\includegraphics[width=0.47\textwidth, height=0.24\textheight]{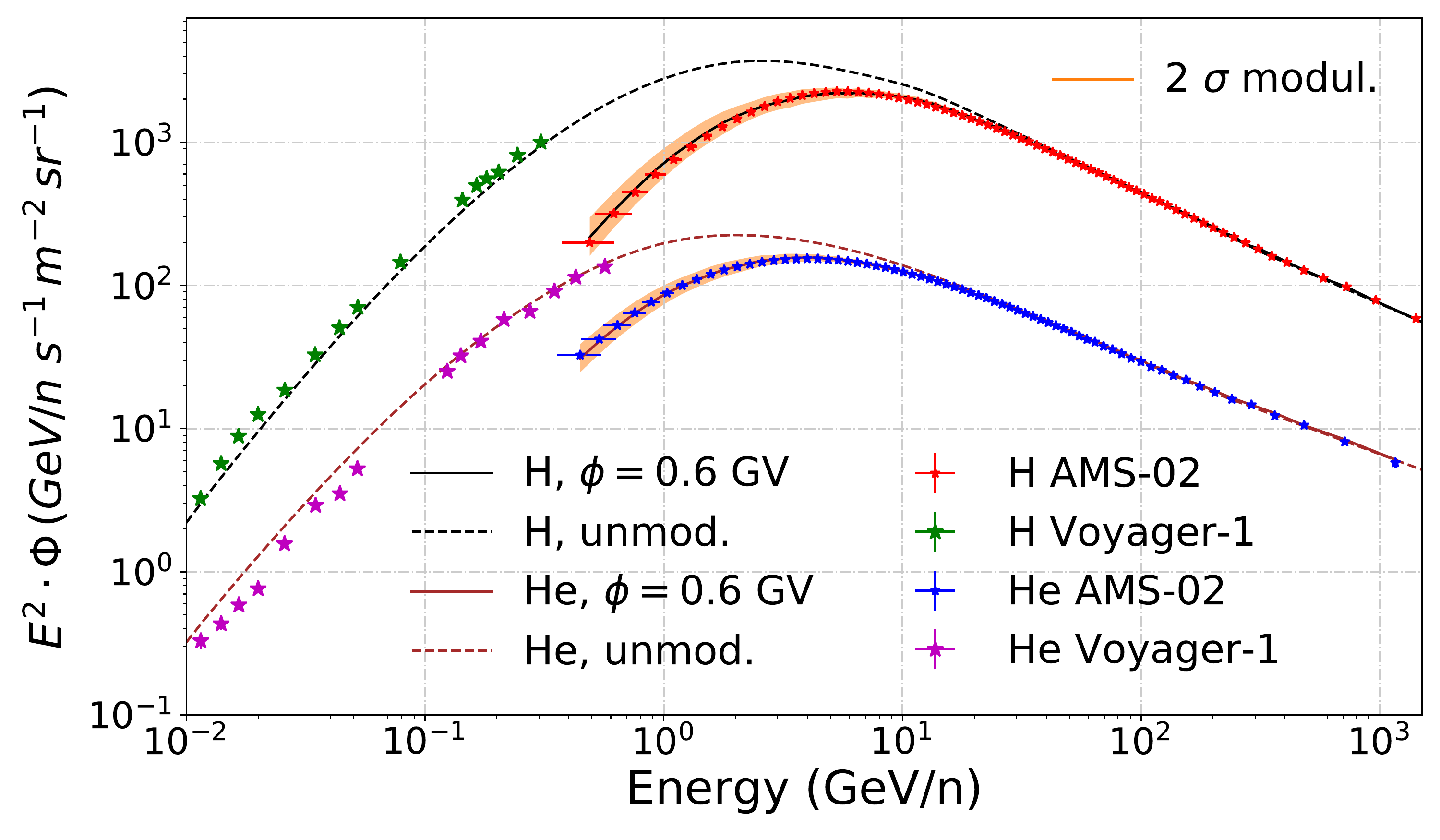} \hspace{0.2 cm}
	\includegraphics[width=0.47\textwidth, height=0.24\textheight]{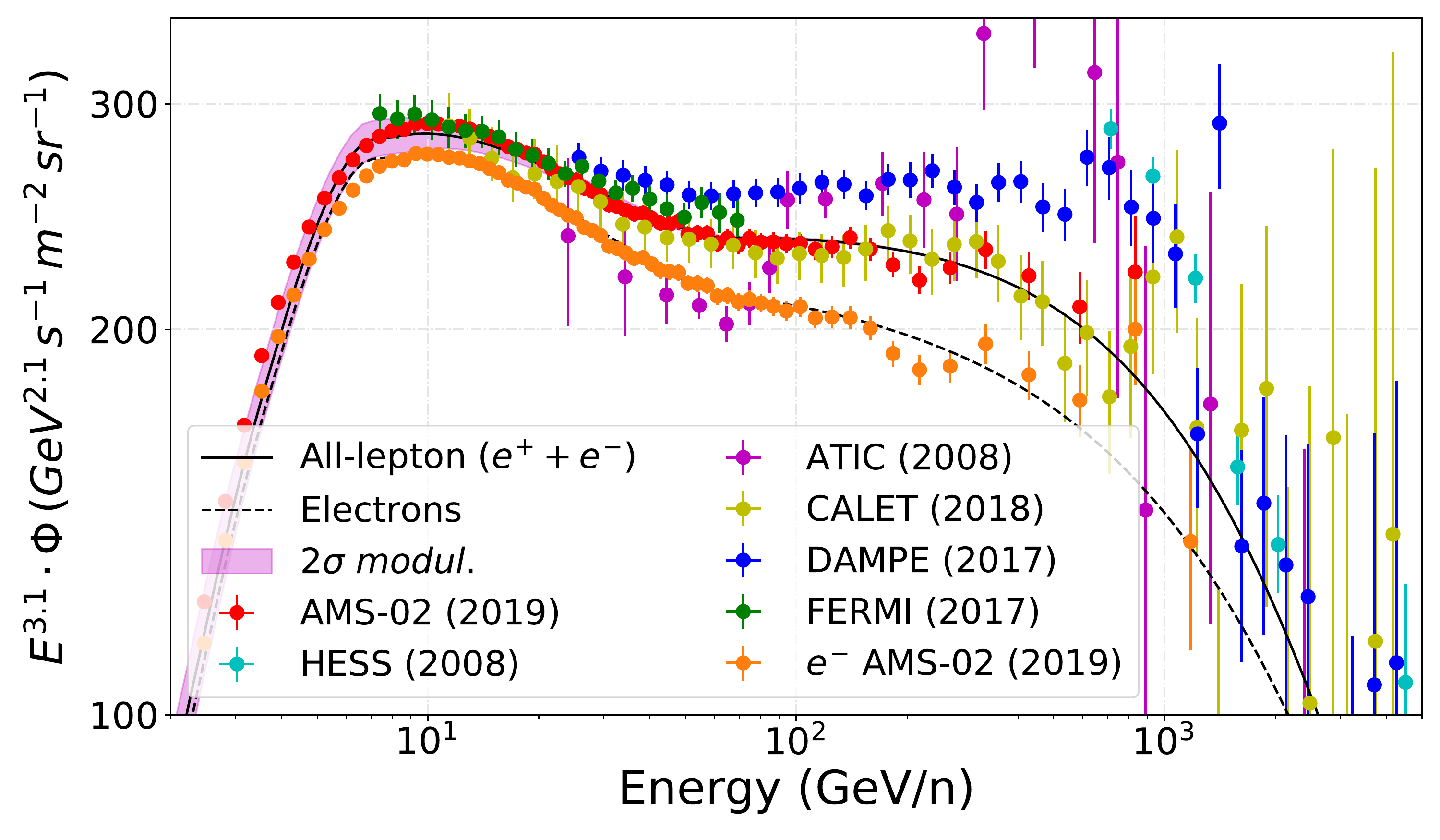}
	
	\caption{\textbf{Left panel}: Modulated and unmodulated predicted spectra of H and He compared to AMS-02 and Voyager 1 data. \textbf{Right panel:} Predicted total CRE spectra compared to AMS-02, Fermi-LAT, DAMPE and CALET data. The dashed line represents the predicted flux of electrons, fitting to the AMS-02 electron flux. The uncertainty band related to a variation of the Fisk potential of $\pm0.1$~GV is shown for the spectra in both figures.}
	\label{fig:Electrons_posit}
\end{figure}

The computed proton and helium spectra are shown in Figure~\ref{fig:Electrons_posit}, left, compared to Voyager-1~\cite{Stone150} (unmodulated) and  AMS-02~\cite{Aguilar:2015ooa}  (using a Fisk potential $\phi = 0.6 \units{GV}$). The electron and total CRE spectra are compared to data, from Fermi-LAT~\cite{Ackermann:2010ij}, DAMPE~\cite{Ambrosi:2017wek}, CALET~\cite{CALET_electrons} and AMS-02~\cite{Aguilar:2014mma}, in the right panel of Figure~\ref{fig:Electrons_posit}. These spectra are derived using the propagation parameters inferred from the B/C analysis with the FLUKA cross sections in the diffusion approximation. The uncertainty band related to a variation of the Fisk potential of $\pm0.1$~GV is included for these spectra. 

\end{document}